%% file: ONcorrelators.tex
\documentclass[11pt,a4wide]{article}
\pdfoutput=1

\usepackage{jheppub}
\input{preamble}

\title{Analytic bootstrap of mixed correlators in the $\boldsymbol{O(n)}$ CFT\hspace{-20pt}\phantom{x}}

\author{Francesco Bertucci,}
\author{Johan Henriksson, \&}
\author{Brian McPeak}
\affiliation{Universit\`a di Pisa and INFN, sezione di Pisa}

\emailAdd{francesco.bertucci@phd.unipi.it} \emailAdd{johan.henriksson@df.unipi.it} \emailAdd{ brian.mcpeak@df.unipi.it}

\abstract{We use large spin perturbation theory and the Lorentzian inversion formula to compute order-$\eps$ corrections to mixed correlators in the $O(n)$ Wilson--Fisher CFT in $4 - \eps$ dimensions. In particular, we find the scaling dimensions and averaged OPE coefficients appearing in all correlators involving the operators $\varphi$ and $\varphi^2$, for $\varphi^2$ in both the singlet and symmetric traceless representations of $O(n)$. We extend some computations to the next order, and find order-$\eps^2$ data for a number of quantities for the Ising case at $n = 1$. Along the way, we discuss several interesting technical aspects which arise, including subleading corrections to mixed conformal blocks, projections onto higher twists in the inversion formula, and multiplet recombination.
}

\begin{document}
\maketitle


\section{Introduction}\label{sec:intro}

One of the most striking features of conformal field theories, realized over many years of study, is just how delicately they fit together. 
We may think of a conformal field theory (CFT) as specified by its spectrum of local operators and the three-point couplings characterizing the interactions of those operators.
By now it is clear that these elements, collectively comprising the ``CFT data,'' are highly constrained by the CFT axioms, such as conformal symmetry, unitarity, crossing symmetry. 

Extracting useful constraints on the CFT data from these general properties is, however, a difficult business. The primary difficulty is that unitarity and crossing symmetry are most naturally expressed as constraints on correlation function, and it is not always easy to translate these into constraints on CFT data. A major breakthrough came with \cite{Rattazzi:2008pe}, which developed a precise numerical algorithm for deriving rigorous bounds on CFT data.
This was enabled in turn by previous progress by Dolan and Osborn \cite{Dolan:2000ut, Dolan:2003hv} in deriving closed-form and otherwise usable expressions for the conformal blocks in $d>2$ dimensions.
Among its many uses, the conformal block expansion turns unitarity into the simple requirement that square OPE coefficients are positive. Subsequently these methods, now called the conformal bootstrap, were expanded and applied to a wide variety of CFTs; see \cite{Poland:2018epd} for an overview.

In this paper, we shall be more interested in a complementary analytic approach to computing CFT data. These methods, which comprise the ``analytic bootstrap'' \cite{Bissi:2022mrs,Hartman:2022zik}, center on identifying and exploiting different perturbative expansions to fix data order-by-order. 
Apart from simplifications in certain limits of ``external'' parameters, such as small coupling constant or large number of degrees of freedom, the analytic bootstrap also considers universal behavior in terms of ``internal'' parameters, such as the charge \cite{Hellerman:2015nra,Monin:2016jmo,Gaume:2020bmp} or -- as in this paper -- the spin of operators in the spectrum.
One benefit of this approach is that it is possible to parametrize the behavior in certain \emph{sectors} of the theory in question -- in our case conformal data of spinning operators -- in terms of just a few parameters, and therefore reduce the complexity of the problem significantly. 

The study of spinning operators is intrinsically tied to CFT in Lorentzian signature and the lightcone limit, defined as $(x_1-x_2)^2\to0$ with $x_1^\mu-x_2^\mu$ finite. 
In this regime, the OPE $\O(x_1)\times \O(x_2)$ is dominated by twist $\tau=\Delta-\ell$ rather than scaling dimension $\Delta$.\footnote{This is reflected within the four-point function by the scaling of the conformal block: $G_{\Delta,\ell}(u,v)\sim u^{\tau/2}$.} 
It is therefore natural to group operators of approximately equal twist into ``twist families,'' and study the dependence of spin for each twist family.\footnote{In fact, interest in families of operators indexed by spin far predates the analytic bootstrap and goes beyond the conformal fixed-point. For instance, such operators have attracted a lot of attention in QCD \cite{Gribov:1972ri,Gribov:1972rt,Lipatov:1974qm,Altarelli:1977zs,Dokshitzer:1977sg,Cornalba:2009ax} and its superconformal cousin $\mathcal N=4$ SYM \cite{Kotikov:2002ab,Brower:2006ea,Costa:2012cb,Gromov:2015wca,Aprile:2017bgs,Chester:2019pvm,Kniehl:2021ysp,Drummond:2022dxw}, where they control the deep inelastic scattering amplitude.} 
Famously, Nachtmann's theorem \cite{Nachtmann:1973mr} shows that positivity of the deep inelastic scattering cross-section implies that for the family of (global symmetry singlet) operators of leading twist, the twists $\tau_\ell$ form a convex function of $\ell$.
Moreover, in the context more relevant to this paper, the $\eps$-expansion for the Ising and $O(n)$ CFT, it was noted perturbatively that ``twist additivity'' holds, namely that for any two operators with twist $\tau_1$ and $\tau_2$, there exists an infinite sequence of spinning operators appearing in their OPE with $\tau_\ell\to\tau_1+\tau_2$ as $\ell\to\infty$ \cite{Kehrein:1995ia}.

A more unified picture emerged through the works of \cite{Fitzpatrick:2012yx,Komargodski:2012ek} (see also \cite{Alday:2007mf}), where it was shown that twist additivity in fact applies to generic CFTs, and more generally, that the large-spin asymptotics are controlled by mean field theory. Consider for example the four-point function of identical scalars $\O$. 
In this case, the considerations of \cite{Fitzpatrick:2012yx,Komargodski:2012ek} imply the existence of operators of the form
\begin{equation}
\label{eq:GFFopsintro}
[\O,\O]_{n,\ell} \sim \O \square^n \partial^\ell \O.
\end{equation}
They are sometimes referred to as GFF operators, since together with the identity operator $\1$, they are the only operators contained in the $\O\times\O$ OPE when $\O$ is a ``generalized free field,'' \emph{i.e.} a non-interacting field with arbitrary $\Delta_\O$. In an interacting theory, the operators $[\O,\O]_{n,\ell}$ acquire anomalous dimensions $\gamma_{n,\ell}$, which tend to zero at large $\ell$. For $n=0$, where we refer to $\O\partial^\ell\O$ as ``double-twist operators,'' this fact was noted a long time ago \cite{Callan:1973pu,Parisi:1973xn}.
Supported by perturbative arguments \cite{Gribov:1972rt,Dokshitzer:1995ev,Basso:2006nk}, and ultimately proven for CFT in \cite{Alday:2015eya}, the anomalous dimensions naturally expand in inverse powers of the ``conformal spin''
\begin{equation}
J^2=
\hb(\hb-1), \qquad \hb=\frac{\Delta+\ell}2
\label{eq:fullconfspin}
,
\end{equation}
which is the eigenvalue of the Casimir on the lightcone.
In fact, \cite{Fitzpatrick:2012yx,Komargodski:2012ek} showed universally that in a generic interacting CFT, the double-twist operators $\O\partial^\ell\O$ have dimensions with leading large spin asymptotics of the form
\begin{equation}
\Delta_{\O\partial^\ell\O} = 2\Delta_\O + \ell + \gamma_{\ell}, \qquad \gamma_{\ell}=- \frac{Ca_{\tau_{\mathrm{min}}}}{(J^2)^{\tau_{\mathrm{min}}/2}}
,
\end{equation}
where $a_{\tau_{\mathrm{min}}}=\lambda^2_{\O\O \O_{\tau_{\mathrm{min}}}}$ is the squared OPE coefficient of the operator with minimal twist $\tau_{\mathrm{min}}$ in the $\O\times\O$ OPE, and $C$ is a positive constant, given explicitly in these papers.
Different methods were developed to determine corrections to this expansion and to the involved OPE coefficients \cite{Alday:2015ota,Alday:2015ewa,Simmons-Duffin:2016wlq}, culminating in the systematic framework of large spin perturbation theory \cite{Alday:2016njk,Alday:2016jfr}.

The proposal and derivation of the Lorentzian inversion formula \cite{Caron-Huot:2017vep} provided several new insights. Firstly, it showed that the anomalous dimensions and OPE coefficients for operators in the same twist family are in fact described by functions which are analytic in spin, putting on a rigorous footing a lot of the previous work.\footnote{In the context of gauge theories, analytic continuation in spin of the leading twist (twist-2) operators is essential when studying the relation between the DGLAP and BFKL regimes \cite{Jaroszewicz:1982gr,Kotikov:2000pm,Brower:2006ea} and in the construction of light-ray operators \cite{Balitsky:1987bk,Braun:2003rp,Balitsky:2013npa}. In fact, the Lorentzian inversion formula has been related to light-ray operators \cite{Kravchuk:2018htv} and provided a proof of Nachtmann's theorem for complex spin \cite{Costa:2017twz}.}
Moreover, boundedness of the four-point correlator in the Regge limit imposes that the range of analyticity extends down to include all operators with spin $\ell>1$ (in the generic case),\footnote{This assumes the general expectation that the correlator should be bounded in the Regge limit by a Regge intercept $\ell_*<2$ \cite{Caron-Huot:2017vep}. In fact, in the $\eps$-expansion, and in the 3d $O(N)$ CFT, we expect $\ell_*<1$ \cite{Liu:2020tpf,Caron-Huot:2020ouj,CaronHuot2022}, implying that the inversion formula should be valid for $\ell\geqslant 1$.
} meaning that the complete spectrum of spinning operators inside a given correlator (and a global symmetry representation) with $\ell>1$ can be captured by a single function $C(\Delta,\ell)$ analytic in $\ell$.
For integer spin $\ell=\ell_0$, this function has poles in $\Delta$ at the position of all spin-$\ell_0$ operators, with a residue that is given by the OPE coefficient,
\begin{equation}
C(\Delta,\ell_0)= \frac{\lambda^2_{\O\O \O_{\Delta_i,\ell_0}}}{\Delta-\Delta_i}, \qquad \text{as $ \Delta\sim \Delta_i$}.
\end{equation}
The work \cite{Caron-Huot:2017vep} also gave a compact formula for computing this function,
\begin{align}
\label{eq:Inversionformulaintro}
C_{1234}(\Delta,\ell)&=\kappa_{\frac{\Delta+\ell}2}
\int\frac{dzd\zb}4\mu(z,\zb)G_{\ell+d-1,\Delta+1-d}(z,\zb)\dDisc\left[\G_{1234}(u,v)\right]
+(-1)^\ell \,\big(\, 1\leftrightarrow 2\, \big)
,
\end{align}
valid in the general case involving the correlator of four potentially distinct scalar operators $\O_i$, $i=1,2,3,4$. Here we have defined the usual cross-ratios,
\begin{equation}u=z\zb=\frac{x_{12}^2x_{34}^2}{x_{13}^2x_{24}^2}, \qquad
v=(1-z)(1-\zb)=\frac{x_{14}^2x_{23}^2}{x_{13}^2x_{24}^2}.
\label{eq:crossratios}
\end{equation}
The precise meaning of all ingredients in the formula \eqref{eq:Inversionformulaintro} will be spelled out in the subsequent sections. Since the formula is valid for $\ell>\ell_*$ with $\ell_*<2$, it must be complemented with finite-spin contributions at spins $\ell\leqslant \ell_*$.

The key property of the inversion formula \eqref{eq:Inversionformulaintro} is that it depends on the correlator $\G_{1234}(u,v)$ only through the so-called double-discontinuity, which is often much easier to evaluate than the entire correlator. It is sensitive only to certain singularities that appear in the (crossed) lightcone limit $(x_2-x_3)^2\to0$, or equivalently $v\to0$ for fixed $u$. This implies, in principle, a complete dispersion relation, where the correlator can be constructed in terms of its double-discontinuity, plus potentially the contribution from operators at spin $0$ and $1$ \cite{Carmi:2019cub}.

The upshot of this is that the double-discontinuity suppresses exactly the part of the correlator that carries the universal spin dependence, namely the contribution from operators that approach the form of the GFF operators \eqref{eq:GFFopsintro}. This behavior is most apparent when one combines the inversion formula with the crossing equation,
\begin{equation}
\G_{1234}(u,v)=\frac{u^{\frac{\Delta_1+\Delta_2}2}}{v^{\frac{\Delta_2+\Delta_3}2}}\G_{3214}(v,u)
.
\end{equation}
Let us look at this in more detail, focusing on the first term in the right-hand side of \eqref{eq:Inversionformulaintro}. Using the crossing equation derived from exchanging operators $\O_1$ and $\O_3$, giving $u\leftrightarrow v$, and writing the crossed channel as a sum over conformal blocks, we get
\begin{equation}
\dDisc[\G_{1234}(u,v)]=\dDisc\left[\frac{u^{\frac{\Delta_1+\Delta_2}2}}{v^{\frac{\Delta_2+\Delta_3}2}}
\sum_{ \O' }\lambda_{32\O'}\lambda_{14\O'} G_{\Delta_{\O'},\ell_{\O'}}(v,u)
\right]
.
\end{equation}
The crossed channel conformal blocks scale as $v^{\tau_{\O'}/2}$ in the limit $v\to0$, where $\tau_{\O'}=\Delta_{\O'}-\ell_{\O'}$. Consider for simplicity the case $\O_2=\O_1$ and $\O_4=\O_3$.\footnote{In full generality, the definition of the double-discontinuity depends on the differences $\Delta_{12}$ and $\Delta_{34}$, where $\Delta_{ij}=\Delta_i-\Delta_j$. See \eqref{eq:dDiscgen} below.} In this case, by an explicit computation that we will review in the main text, one finds that
\begin{equation}
\dDisc[v^{\alpha}] = 2\sin^2(\pi\alpha)v^\alpha
,
\end{equation}
where we expect for $\O'$ that $\alpha=\frac12\left(\tau_{\O'}-\Delta_1-\Delta_3\right)$.
The precise prefactors and the integrations in \eqref{eq:Inversionformulaintro} introduce many additional factors; but what matters for us is a dependence on $\alpha$ of the form $\Gamma(\alpha+1)^2$. So in summary
\begin{equation}
\label{eq:ddiscsuppr}
C_{1234}(\Delta,\ell)|_{O'} \sim \sin^2(\pi\alpha)\Gamma(\alpha+1)^2.
\end{equation}
This factor has a double zero exactly when $\tau_{\O'}=\Delta_1+\Delta_3+2n$ for $n=0,1,2,\ldots$, which is exactly when $\O'$ has a dimension and spin that agrees with the GFF operator $[\O_1,\O_3]_{n,\ell}$.

The fact that GFF operators have suppressed contribution in the inversion formula is particularly useful in the presence of a small perturbative parameter $\delta$, if the spectrum at $\delta\to0$ is similar or identical to the GFF spectrum.\footnote{However, the Lorentzian inversion formula has also been applied in non-perturbative theories, such as three-dimensional CFTs \cite{Albayrak:2019gnz,Caron-Huot:2020ouj,Liu:2020tpf,Atanasov:2022bpi} and 6d $(2,0)$ theory \cite{Lemos:2021azv}. Recently a hybrid bootstrap method was proposed, which incorporates information from the Lorentzian inversion formula in the numerical bootstrap \cite{Su:2022xnj}.} In this case, we can consider the contribution from operators $\O'$ with different properties:
\begin{itemize}
\item $\O'=\1$, the identity operator: this is always present for pairwise identical operators $\O_1=\O_4$, $\O_2=\O_3$, and gives rise to the CFT data that exactly corresponds to the GFF spectrum, with OPE coefficients $a_{n,\ell}^{\mathrm{GFF}}$, which were first found in general $d$ in \cite{Fitzpatrick:2011dm} and are given in \eqref{eq:aGFFgen} below.
\item $\O'$ with generic twist $\tau_{\O'}$, or twist $\tau_{\O'}$ below the double-twist threshold: such an operator gives a contribution to $C(\Delta,\ell)$ proportional to the OPE coefficient $\lambda_{32\O'}\lambda_{14\O'}$.
\item $\O'$ an approximate GFF operator, or any other operator with $\tau_{\O'}=\Delta_3+\Delta_2+2n+ \gamma$, where $\gamma$ denotes a small anomalous dimension ($\gamma=\gamma^{(1)}\delta+\ldots$). Then the factors in \eqref{eq:ddiscsuppr} conspire to give a suppressed contribution
\begin{equation}
C_{1234}(\Delta,\ell)|_{O'} \sim \lambda_{32\O'}\lambda_{14\O'} \gamma^2, \qquad (\tau_{\O'}=\Delta_3+\Delta_2+2n+ \gamma)
.
\end{equation}
\end{itemize}

By clever usage of the above observations, the Lorentzian inversion formula has found applications in various settings, primarily holographic CFTs \cite{Alday:2017vkk,Aharony:2018npf,Alday:2018pdi,Alday:2018kkw,Li:2019zba,Alday:2020tgi,Binder:2021cif,Alday:2021ymb} and the $\langle\phi\phi\phi\phi\rangle$ correlator in critical $\phi^4$ theories \cite{Alday:2017zzv,Henriksson:2018myn,Alday:2019clp,Henriksson:2020fqi,Henriksson:2021lwn}, yielding results that sometimes go beyond other methods.\footnote{Other applications in a perturbative setting include $\phi^3$ theory near six dimensions \cite{Goncalves:2018nlv}, general $\phi^k$ theories \cite{Henriksson:2020jwk}, finite temperature \cite{Iliesiu:2018fao}, defects \cite{Lemos:2017vnx,Liendo:2019jpu,Barrat:2021yvp}.} This is particularly true for corrections to OPE coefficients, which require considerable effort using diagrammatic methods.
The mentioned works involved primarily the four-point function of identical operators, however the inversion formula applies to more general four-point functions, and to investigate these will be the purpose of this study.

In this paper, we will apply large spin perturbation theory and the Lorentzian inversion formula to $\phi^4$ theories, but extend the considerations to other correlators than the $\phi$ four-point function $\langle\phi\phi\phi\phi\rangle$. Specifically, we focus on the $O(n)$-symmetric case, called the $O(n)$ CFT, and work in the $\eps$-expansion.
Since the initial proposal of the $\eps$-expansion by Wilson and Fisher \cite{Wilson:1971dc}, a lot of data has been computed over the years, as reviewed in \cite{Henriksson:2022rnm}. The most recent contributions involve high-order diagrammatic results \cite{Kompaniets:2016hct,Kompaniets:2017yct,Schnetz:2016fhy,SchnetzUnp,Bednyakov:2021ojn} and various bootstrap methods, such as Mellin space \cite{Gopakumar:2016cpb,Gopakumar:2016wkt,Dey:2016mcs} and the related method of analytic functionals \cite{Carmi:2020ekr},\footnote{See \cite{Trinh:2021mll} for a numerical application of the analytic functionals to mixed correlators in the 3d Ising theory.} Lorentzian inversion formula \cite{Alday:2017zzv,Henriksson:2018myn} and multiplet recombination \cite{Rychkov:2015naa,Giombi:2016hkj,Liendo:2017wsn}.

There are several motivations for the extension to mixed correlators. First, the numerical conformal bootstrap involves systems of correlators, and it is instructive to see how the data entering these fit together with the $\eps$-expansion.\footnote{The interpolation between the perturbative and the non-perturbative numerical regime has been investigated in \cite{El-Showk:2013nia,Behan:2016dtz,Cappelli:2018vir,Sirois:2022vth}.} More generally, an increased understanding of perturbative data can give insight about reasonable gap assumptions for future numerical studies. Secondly, conformal perturbation theory \cite{Cardy:1996xt,Komargodski:2016auf,Amoretti:2017aze}, which can be used to find observables in the vicinity of the fixed-point, or at nearby fixed-point, requires the knowledge of four-point correlators to go beyond leading order. Finally, due to the paradigmatic role of the $O(n)$ CFT in the study of conformal field theories, closed-form expressions of correlators and conformal data will be of use for the development and testing of new methods utilizing conformal symmetry.

The goal of this paper, then, will be to pursue this strategy for mixed correlators in the $O(n)$ CFT. Specifically, we shall compute all correlators involving the operators $\varphi$, $s$, and $t$. Here $\varphi$ is the fundamental field transforming in the vector representation of the global symmetry group $O(n)$. The operators $s$ and $t$ are the quadratic operators formed from $\varphi$, transforming in the scalar and symmetric tensor representations respectively. The result of our calculations will be order-$\varepsilon$ corrections to the anomalous dimensions and OPE coefficients of operators exchanged in these correlators, as well as explicit expressions for the correlators and some order-$\eps^2$ results for $n=1$. The reader primarily interested in our results can find them collected in section~\ref{sec:collectedresults}.

\subsection[Example: mixed $\phi$ $\phi^2$ system for $n=1$]{Example: mixed $\boldsymbol \phi$ $\boldsymbol{\phi^2}$ system for $\boldsymbol{n=1}$}

To better illustrate the logic of our paper, let us explain the argument using the simplest possible example, which is the correlators involving $\phi$ and $\phi^2$ when $n=1$.\footnote{For $n=1$, we use the symbol $\phi$ for the single fundamental field, and $\phi^2$ for the operator $s$.} In this case, the $O(n)$ CFT reduces to the Ising CFT, and the global symmetry reduces to $\mathbb Z_2$. The operators $\phi$ and $\phi^2$ are the leading $\Z_2$-odd and $\Z_2$-even operators in the spectrum, and there are three non-vanishing four-point correlators involving these operators:
\begin{equation}
\langle\phi\phi\phi\phi\rangle, \quad \langle\phi\phi\phi^2\phi^2\rangle, \quad \langle\phi^2\phi^2\phi^2\phi^2\rangle.
\end{equation}
The study of these correlators in the numerical bootstrap led to the isolation of a small unitary island in the 3d theory \cite{Kos:2014bka,Simmons-Duffin:2015qma,Kos:2016ysd}, where $\phi$ and $\phi^2$ are conventionally denoted $\sigma$ and $\epsilon$. We will now describe how to study this system using the analytic bootstrap. 
In practice, this involves many technical steps that will be discussed in the main text -- here we give an overview of the method.

\paragraph{Step 1: $\boldsymbol{\langle\phi\phi\phi\phi\rangle}$ correlator} The starting point is the $\langle\phi\phi\phi\phi\rangle $ correlator. It was studied using the Lorentzian inversion formula to order $\eps^4$ in \cite{Alday:2017zzv}. As described above, the CFT data of operators in this correlator is found by computing the double-discontinuity starting from operators $\O'$ in the crossed channel. A number of simplifications occur in the $\langle\phi\phi\phi\phi\rangle $ correlator. The identity operator $\1$ contributes, but all other operators have a contribution that is suppressed by certain powers of $\eps$. In fact, working to order $\eps^3$, the only operator to consider besides the identity is $\O'=\phi^2$. Thus

\begin{equation}
\label{eq:ddiscffffintro}
\dDisc[\G_{\phi\phi\phi\phi}(u,v)]=\dDisc\left[
\frac{u^{\Delta_\phi}}{v^{\Delta_\phi}}\left(1+a_{\phi^2} G_{\Delta_{\phi^2},0}(v,u)\right)
\right]+O(\eps^4)
,
\end{equation}
where $a_{\phi^2}=\lambda^2_{\phi\phi\phi^2}=2+O(\eps)$, computable from the free theory. Note that the contribution from $\phi^2$ will be suppressed by two powers of the anomalous dimension $\gamma = \Delta_{\phi^2}-2\Delta_\phi$. The double-discontinuity in \eqref{eq:ddiscffffintro} gives rise to CFT data for operators at twist 2 and twist 4. The twist-2 operators are exactly the double-twist operators $[\phi,\phi]_{0,\ell}=\phi\partial^\ell\phi$. We will introduce a special name for them: $\mathcal J_\ell=\phi\partial^\ell\phi$. The reason is that they will appear in other correlators, where they are no longer double-twist operators.\footnote{For instance, in the $\phi^2$ four-point function, the double-twist operators will be $[\phi^2,\phi^2]_{0,\ell}$, at twist four, and the operators $\mathcal J_\ell$ at twist two will be below the double-twist threshold.}
One finds that
\begin{equation}
\Delta_{\mathcal J_\ell}=2\Delta_\phi+\ell-\frac{a_{\phi^2}\gamma^2}{J^2} + O(\eps^3)
,
\label{eq:LOdata}
\end{equation}
where the $J^2$ dependence is exact to this order in $\eps$.
Conservation of the stress-tensor $\Delta_{\mathcal J_2}=d=4-\eps$ and an analytic continuation to spin zero fixes the remaining two parameters in this expression: $\gamma_\phi$ at order $\eps^2$ and $\gamma$.\footnote{Specifically, conservation of the stress-tensor sets $\gamma_\phi=\frac16\gamma^2+O(\eps^3)$. Then, by evaluating $J^2$ at the full conformal spin, \eqref{eq:fullconfspin}, it is possible to formally evaluate \eqref{eq:LOdata} at $\ell=0$ and impose $\Delta_{\mathcal J_{\ell}}|_{\ell=0}=\Delta_{\phi^2}$. Solving this equation to order $\eps$ fixes $\gamma=\frac\eps3$ to leading order; see \cite{Alday:2017zzv} for more details. It is not known exactly why such analytic continuation gives the correct result. Finally, a direct crossing analysis shows that the spin-0 OPE coefficient takes the value $\lambda_{\phi\phi\phi^2}^2=2(1-\frac\eps3)+O(\eps^2)$.} Specifically, one arrives at the data
\begin{align}
\Delta_\phi &= 1-\frac\eps2+\frac{\eps^2}{108}+O(\eps^3)
,
\\
\Delta_{\phi^2}&=2-\eps+\frac{\eps}3+O(\eps^2), & \Delta_{\mathcal J_{\ell}}&=2+\ell-\eps+\frac{\eps^2}{54}\left(1-\frac6{\ell(\ell+1)}\right)+O(\eps^3)
\label{eq:deltaJellIntro}
,
\\
\lambda^2_{\phi\phi\phi^2}&=2\left(1-\frac\eps3+O(\eps^2)\right), & \lambda^2_{\phi\phi\mathcal J_\ell}&=\frac{2\Gamma(\ell+1)^2}{\Gamma(2\ell+1)}\Big(1+\eps(S_1(2\ell)-2S_1(\ell))+O(\eps^2)\Big)
,
\label{eq:dataffffintro2}
\end{align}
where $S_1(\ell)$ denotes the harmonic numbers.

\paragraph{Step 2: $\boldsymbol{\langle\phi\phi\phi^2\phi^2\rangle}$ correlator} Next we consider the mixed correlator of $\phi$ and $\phi^2$. Working to order $\eps$, the whole double-discontinuity is given by the operator $\O'=\phi$:
\begin{equation}
\label{eq:ddiscffssintro}
\dDisc[\G_{\phi\phi\phi^2\phi^2}(u,v)]=\dDisc\left[
\frac{u^{\Delta_\phi}}{v^{\frac{\Delta_\phi+\Delta_{\phi^2}}2}}a_{\phi} G_{\Delta_{\phi},0}(v,u)
\right]+O(\eps^2)
.
\end{equation}
Now $a_\phi=\lambda^2_{\phi\phi^2\phi}$, but since OPE coefficients are permutation invariant, this is given by \eqref{eq:dataffffintro2}. Since also the involved scaling dimensions have been previously determined to appropriate order, one proceeds and finds the new CFT data
\begin{equation}
\lambda_{\phi\phi\mathcal J_\ell}\lambda_{\phi^2\phi^2\mathcal J_\ell}=\frac{4\Gamma(\ell+1)^2}{\Gamma(2\ell+1)}\left(1+\eps\left(S_1(2\ell)-\tfrac53S_1(\ell)-\tfrac13\right)+O(\eps^2)\right)
.
\end{equation}
At spin zero, there could potentially be a finite spin contribution, and we write
\begin{equation}
\lambda_{\phi\phi\phi^2}\lambda_{\phi^2\phi^2\phi^2}=4\left(1+\eps\left(-\tfrac13+\tilde\alpha\right)+O(\eps^2)\right)
\label{eq:introducingalpha}
.
\end{equation}
\paragraph{Step 3: $\boldsymbol{\langle\phi^2\phi^2\phi^2\phi^2\rangle}$ correlator} This correlator is the most involved one to compute because the operators that contribute to the double-discontinuity are $\O'=\1$ and the whole sum of twist-two operators $\O'=\phi^2,\,\mathcal J_\ell$:
\begin{equation}
\label{eq:ddiscssssintro}
\dDisc[\G_{\phi^2\phi^2\phi^2\phi^2}(u,v)]=\dDisc\bigg[
\frac{u^{\Delta_{\phi^2}}}{v^{\Delta_{\phi^2}}}\bigg(1+a_{2,0} G_{\Delta_{\phi^2},0}(v,u)+\sum_{\ell=2,4,\ldots}a_{2,\ell}G_{\Delta_{\mathcal J_\ell},\ell}(v,u)\bigg)
\bigg]+O(\eps^2)
,
\end{equation}
where $a_{2,\ell}$ denotes the OPE coefficients of the twist-two operators in the $\phi^2\times\phi^2$ OPE. But they can be determined in terms of the quantities above,
\begin{equation}
a_{2,0}=\frac{\left(\lambda_{\phi\phi\phi^2}\lambda_{\phi^2\phi^2\phi^2}\right)^2}{\lambda_{\phi\phi\phi^2}^2}, \qquad a_{2,\ell}=\frac{\left(\lambda_{\phi\phi\mathcal J_\ell}\lambda_{\phi^2\phi^2\mathcal J_\ell}\right)^2}{\lambda_{\phi\phi\mathcal J_\ell}^2}
,
\end{equation}
and are thus completely fixed, up to the constant $\tilde\alpha$. One can then use the Lorentzian inversion formula to find the order $\eps^0$ and $\eps^1$ CFT data for all operators exchanged in the correlator. This step requires considerable effort and is performed with the help of the projection to subleading twists within the inversion formula, which we discuss in section~\ref{eq:Inversionformulaintro}.

\paragraph{Resumming the correlators}

With all data at hand, we can also determine closed-form expressions of the correlators by computing explicit sums of conformal blocks,
\begin{align}
G_{\phi\phi\phi\phi}(u,v)&=1+\left(u+\frac uv\right)+\eps\left(-\frac{u(1+v)}{2v}\log u+\frac{u}{2v}\log v -\frac u3\Phi(u,v) \right) + O(\eps^2)
,
\\
G_{\phi\phi\phi^2\phi^2}(u,v)&=1+2\left(u+\frac uv\right)+\eps\left(-\frac{u(1+v)}{v}\log u+\frac{u(2+v)}{3v}\log v \right.
\nonumber \\
& \quad \left. -\frac{2u(1+v)}{3v} -\frac{2u}{3}\Phi(u,v) \right)
+4\eps\left(\tilde\alpha+\frac13\right)u\frac{\log(\frac{1-\zb}{1-z})}{z-\bar z}
+ O(\eps^2)
,
\\
G_{\phi^2\phi^2\phi^2\phi^2}(u,v)&=1+u^2+\frac{u^2}{v^2}+4\left( u+\frac uv +\frac{u^2}v \right) +\eps\left(-\frac{8u(u+v+1)}{3v} \right.
\nonumber \\
& \quad -\frac{2u\big( v(v+1)+8v(u+v+1)+u(3v^2 -2v+3) \big)}{9v^2}\log u
\nonumber \\
& \quad -\frac{2u\big(-4v(u+v+1)+(-3u-2v^2+v+uv) \big)}{9v^2}\log v
\nonumber \\
& \quad \left. -\frac{4u(u+v+uv)}{3v}\Phi(u,v) \right)
-16\eps\left(\tilde\alpha+\frac13\right)\tilde\G(u,v)
+ O(\eps^2)
,
\label{eq:phi2final}
\end{align}
where $\Phi(u,v)$ is the standard box integral, given in \eqref{eq:boxfunction}, and 
\begin{align}
	\tilde\G(u,v)&=\left(\frac{2 u^3\Phi(u,v)
	+
\frac{u^2}v\big( (1-v)^2-u(v+1) \big)\log u
 + 
u^2 (u-v+1)\log v}{(u-v+1)^2 -4u } +\frac{u\log ( \frac{1-z}{1-\zb})}{z-\zb} \right).
\end{align}
These expressions still depend on the undetermined constant $\tilde\alpha$, introduced in \eqref{eq:introducingalpha}. However, it is now easy to see that crossing invariance of the final formula \eqref{eq:phi2final} always leaves a piece proportional to $(\tilde\alpha+\frac13)$, which can only vanish if $\tilde\alpha=-\frac13$. Using this value, we have achieved a complete determination of the three correlators at order $\eps$.

\subsection{Structure of this paper}

This paper is structured as follows.
In section~\ref{sec:two} we set up the problem and describe in detail all ingredients, including a review of the $O(n)$ CFT, some generalities about conformal correlators and conformal blocks, and some technical details required to use the inversion formula for mixed correlators at higher-twist. In section~\ref{sec:phi-s-system} we study correlators involving $\varphi$ and $s=\varphi^i\varphi^i$. In this case, the $O(n)$ representation theory simplifies, so we focus on the computations that involve the determination of the double-discontinuity and the inversion integral. The results are the CFT data for all operators appearing in the $\langle \varphi \varphi \varphi \varphi \rangle$, $\langle \varphi \varphi s s \rangle$, and $\langle ssss \rangle$ correlators, and expressions for these correlators. In section~\ref{sec:phi-s-t-system} we expand the set of operators to include $t$, the traceless symmetric operator $t=\varphi^{\{i}\varphi^{j\}}$. 
This introduces some complexity due to the variety of representations that can appear in $t \times t$. 
However the technical manipulations involving the inversion formula are similar to those in the preceding section, so we proceed more quickly towards the results.

In sections~\ref{sec:phi-s-system} and \ref{sec:phi-s-t-system}, we take the viewpoint that all data should be fixed by crossing and analyticity alone. However, in section~\ref{sec:fseps2} we take a less strict approach; we combine bootstrap constraints with previous results in the literature, mainly obtained diagrammatically, to push our method as far as possible. The result is an improvement on the data for $\langle \phi \phi \phi^2 \phi^2 \rangle$ at $n = 1$, and includes a number of new order-$\eps^2$ results.

In section~\ref{sec:collectedresults} we collect our results in one place, and present explicit data for as many non-degenerate and pairwise degenerate operators as possible. Finally, we conclude and discuss future directions in section~\ref{sec:disc}.

\subsubsection{Conventions}

In this paper, we use the following conventions, which are mostly in alignment with \cite{Henriksson:2022rnm}:
\begin{description}
\item[Operators for general $\boldsymbol n$] The fundamental field is denoted $\varphi$,  and composite operators in the $O(n)$ representation $R$ by their form $\square^n\partial^\ell\varphi^k_R$; see \eqref{eq:opform}. Special names are given to the operators
\begin{equation}
s=\varphi^2_S=\varphi^i\varphi^i, \qquad t=\varphi^2_T=\varphi^{\{i}\varphi^{j\}},
\end{equation}
where curly brackets denote symmetrization and removal of traces.
\item[Operators for $\boldsymbol{n=1}$] The fundamental field is denoted $\phi$, and composite operators by their form $\square^n\partial^\ell\phi^k$.
\item[CFT data] Dimensions and OPE coefficients of individual operators are denoted by $\Delta_\O$ and $\lambda_{\O_1\O_2\O_3}$. For nearly degenerate operators at twist $\tau_0+\gamma_i$ (with $\gamma_i=O(\eps)$) and spin $\ell$, we use $a^{1234}_{\tau_0,\ell}$ and $\gamma^{1234}_{\tau_0,\ell}$ to denote the weighted sum of OPE coefficients and anomalous dimensions,
\begin{equation}
a^{1234}_{\tau_0,\ell}=\sum_i \lambda_{\O_1\O_2\O_i}\lambda_{\O_3\O_4\O_i}, \qquad \gamma^{1234}_{\tau_0,\ell}=\frac1{a^{1234}_{\tau_0,\ell}}\sum_i \lambda_{\O_1\O_2\O_i}\lambda_{\O_3\O_4\O_i}\gamma_i.
\end{equation}
We will often use the notation $a^{1234}_{\tau_0,\ell}=a^{(0),1234}_{\tau_0,\ell}\left(1+\alpha^{1234}_{\tau_0,\ell}\eps\right)+O(\eps^2)$. 
When there is no chance of confusion, we will drop the dependence on $1234$ and write $\tau$ instead of $\tau_0$. When different $O(n)$ representations are involved, we have to introduce additional normalization factors, which we define in section~\ref{sec:kinematicconventions}.
\item[Cross-ratios and conformal blocks] We use the standard conformal cross-ratios \eqref{eq:Inversionformulaintro} and, unless specified, evaluate the conformal blocks in $d=4-\eps$ dimensions. The conformal blocks are such that there are no explicit factors of $2^\ell$; see \eqref{eq:4dblock}, \eqref{eq:blockleading}.
\end{description}

\section{Review and setup}
\label{sec:two}

Let us start by reviewing some of the topics that are central to this paper. We will first consider the construction of the $O(n)$ model, focusing specifically on the operators that appear in its CFT fixed point, and some group-theoretic elements that will be relevant later in the paper. We then review some basics of CFTs, which will allow us to state our conventions and discuss some details that arise for mixed correlators. The reader already familiar with the $O(n)$ model and conformal symmetry may want to proceed to the next section and then consult the present one when specific formulas are needed.

\subsection[The $O(n)$ CFT]{The $\boldsymbol{O(n)}$ CFT}\label{sec:ONCFT}

Consider the action of $n$ free scalar fields $\varphi^i$, with a mass term and quartic interaction,
\begin{equation}
\mathcal S=\int d^dx\left(\frac12(\partial_\mu\varphi^i)^2+\frac{m^2}2\varphi^2+\frac{\lambda}{24}(\varphi^2)^2\right)
.
\end{equation}
For $d<4$, the quartic interaction is relevant and triggers an RG flow away from the free theory. For a fine-tuned value of the mass parameter $m^2$, the flow reaches an IR fixed-point. The theory at this fixed-point is scale-invariant and conjecturally also conformal, and is what we will call the $O(n)$ CFT.

In $d=4-\eps$ dimensions and working at small $\eps$, the IR fixed-point can be studied perturbatively in $\eps$, and all the observables can be given as series expansions in $\eps$. We will be interested in local conformal primary operators, which in this expansion are constructed as (normal-ordered) composite operators in terms of the field $\varphi^i$ and the partial derivative $\de^\mu$. We will always work in CFT normalization, where the operators have unit two-point functions. For spacetime scalars, this entails
\begin{equation}
\langle\O_i^I(x_1)\O_j^J(x_2)\rangle = \frac{\delta_{ij}}{(x_{12}^2)^{\Delta_{\O_i}}}\delta^{IJ}
,
\end{equation}
where $x_{12}^\mu=x_1^\mu-x_2^\mu$ and $I$, $J$ are indices belonging to the particular $O(n)$ representations that $\O_i$ and $\O_j$ transform in.

The construction of conformal primaries is complicated because one must consider all possible operators with equal engineering dimension (dimension in the 4d free theory) transforming in a given $O(n)$ and Lorentz representation. The conformal primary operators are the linear combinations of such operators that are annihilated by the generator $K_\mu$ of special conformal transformations, or equivalently those that cannot be written as a total derivative of any other operator.
The explicit expressions for the operators are very complicated, and we will not write them here.

To leading order in $\eps$, however, the construction of primary operators simplifies, and one only needs to take into account those operators with the same number of fields and partial derivatives.\footnote{By the equations of motion, to higher orders in $\eps$, there is a mixture between $\varphi^i\varphi^i$ and $\partial_\mu\partial^\mu$.} 
Within such a subspace, the mixing can be solved in a systematic way, which was first done in \cite{Kehrein:1992fn,Kehrein:1994ff}, and revisited in \cite{Hogervorst:2015akt} and \cite{Henriksson:2022rnm}. Following the notation of \cite{Henriksson:2022rnm}, one can therefore write down the ``form'' of an operator as
\begin{equation}
\label{eq:opform}
\O = \square^n\partial^\ell\varphi^m_R
,
\end{equation}
where $R$ denotes an irreducible representation (irrep) of the $O(n)$ symmetry group, and we consider only traceless-symmetric Lorentz representations of spin $\ell$. 
The operator in \eqref{eq:opform} should be read as ``an operator constructed as a linear combination of operators with $m$ fields in the irrep $R$, and $2n$ contracted and $\ell$ uncontracted derivatives, in such a way that it is a conformal primary.'' Note that many operators, both primaries and descendants, may have the same form, since the derivatives can act on different fields in many different ways. 
When there is more than one \emph{primary} operator of the given form, we say that the operators are degenerate, even if this degeneracy is lifted by anomalous dimensions in the $\eps$-expansion.\footnote{Not all of these degeneracies are broken at leading order in $\eps$; however, it is believed that all degeneracies will ultimately be broken at sufficiently high order in $\eps$. See \cite{Kehrein:1995ia} for a nice discussion and survey at order $\eps^2$ for a subclass of operators.}

We give special names to the bilinear operators $s=\varphi^2_S=\varphi^i\varphi^i$ and $t=\varphi^2_T$, with components $t^{ij}=\varphi^{\{i}\varphi^{j\}}$, which together with $\varphi=\varphi_V$ are the operators whose correlators we study in this paper.
Here $S$ denotes the singlet representation, $V$ the vector and $T=T_2$ the rank 2 traceless symmetric representation. 
The non-trivial tensor products of these representations read
\begin{align}
\label{eq:tensorprodVV}
V\otimes V&=S\oplus T\oplus A =\bullet\oplus \raisebox{0pt}{\tiny \yng(2)}\oplus \raisebox{-3pt}{\tiny \yng(1,1)}
\,
,
\\
V\otimes T&=V\oplus T_3\oplus H_3 = \raisebox{0pt}{\tiny \yng(1)} \oplus \raisebox{0pt}{\tiny \yng(3)}\oplus \raisebox{-3pt}{\tiny \yng(2,1)}
\,
,
\label{eq:tensorprodVT}
\\
T\otimes T &= S\oplus T\oplus A\oplus T_4\oplus H_4\oplus B_4 = \bullet\oplus \raisebox{0pt}{\tiny \yng(2)}\oplus \raisebox{-3pt}{\tiny \yng(1,1)} \oplus \raisebox{0pt}{\tiny \yng(4)}\oplus \raisebox{-3pt}{\tiny \yng(3,1)} \oplus \raisebox{-3pt}{\tiny \yng(2,2)}
\,
,
\label{eq:tensorprodT}
\end{align}
where we have given the Young tableaux labeling the irreps. The abbreviations can be read as antisymmetric ($A$), rank-$m$ traceless symmetric ($T_m$), rank-$m$ hook ($H_m$) and box ($B_4$).

Let us describe the operator content in each of the $O(n)$ representations involved. In the limit $\eps\to0$, this can be found by studying the spectrum of the free theory in $d=4$, and taking into account that some operators which are primaries in the free theory become descendants in the interacting theory by the multiplet recombination effect, to be discussed below.  
It is convenient to organize the spectrum by twist $\tau$ rather than scaling dimension. By using the twist in the free 4d theory as our label, the relevant operators for our considerations can be indexed by four labels,
\begin{equation}
\O=\O_{R,\tau,\ell,i}
,
\end{equation}
where $\tau$ denotes the twist in the free 4d theory ($\tau=m+2n$ for $\O$ in \eqref{eq:opform}), and $i$ denotes an additional label for that distinguishes the operators that have identical scaling dimension in the limit $\eps\to0$.

\paragraph{Singlet $\boldsymbol S$ representation}

\begin{itemize}
\item At $\tau=0$ we have the identity operator $\1$.
\item At $\tau=2$ we have a family of non-degenerate operators at each spin $\ell=0,2,\ldots$, which we give special names, $s$ at $\ell=0$ and $\mathcal J_{S,\ell}$ at $\ell=2,4,\ldots$. The latter are are conserved currents in the free 4d theory. The operator at $\ell=2$ is the stress-tensor and is conserved also in the interacting theory, the other ones are broken.
\item At $\tau=4$, we have operators of the form $\de^\ell\varphi^4_S$ for $\ell=0,2,4,5,6,7,8,\ldots$. At $\ell=0$ and $\ell=5$, these operators are non-degenerate.
\item At $\tau=6,8,10,\ldots$ there are many operators of the form $\square^n\de^\ell\varphi^m_S$ for $m=4,6,\ldots$. At $\tau=6$, $\ell=0$, there are only two operators: $\varphi^6_S$ and $\square\varphi^4_S$.
\end{itemize}

\paragraph{Vector $\boldsymbol V$ representation}
\begin{itemize}
\item At $\tau=1$, we have the operator $\varphi$.
\item At $\tau=3$, we have operators of the form $\de^\ell\varphi^3_V$, for $\ell=1,2,3,4,\ldots$. The operator at $\ell=1$ is non-degenerate.
\item At $\tau=5,7,\ldots$, there are many operators of the form $\square^n\de^\ell\varphi^m_V$, for $m=5,7,\ldots$.
\end{itemize}
\paragraph{Traceless-symmetric $\boldsymbol T$ representation}
\begin{itemize}
\item At $\tau=2$, non-degenerate operators for $\ell=0,2,4,\ldots$, denoted $t$ at $\ell=0$ and $\mathcal J_{T,\ell}$ at $\ell=2,4,\ldots$.
\item At $\tau=4$, operators of the form $\de^\ell\varphi^4_T$ for $\ell=0,2,3,4,5,\ldots$, non-degenerate for $\ell=0$.
\item At $\tau=6,8,10,\ldots$ there are many operators of the form $\square^n\de^\ell\varphi^m_T$ for $m=4,6,\ldots$. At $\tau=6$ and $\ell=0,1$, there is exactly one operator with four fields.
\end{itemize}
\paragraph{Antisymmetric $\boldsymbol A$ representation}
\begin{itemize}
\item At $\tau=2$, non-degenerate operators for $\ell=1,3,5,\ldots$, denoted $\mathcal J_{A,\ell}$. The operator at $\ell=1$ is the global symmetry current and is conserved in the interacting theory.
\item At $\tau=4$, operators of the form $\de^\ell\varphi^4_A$ for $\ell=1,3,5,6,7,8,\ldots$, non-degenerate for $\ell=1$.
\item At $\tau=6,8,10,\ldots$ there are many operators of the form $\square^n\de^\ell\varphi^m_A$ for $m=4,6,\ldots$.
\end{itemize}
\paragraph{$\boldsymbol{T_3}$ and $\boldsymbol{H_3}$ representations}
\begin{itemize}
\item At $\tau=3$, operators of the form $\de^\ell \varphi^3_R$, non-degenerate for $R=T_3$, $\ell=0,2,3,4,5,7$ (absent for $\ell=1$) and for $R=H_3$, $\ell=1,2,3$ (absent for $\ell=0$).
\item
At $\tau=5,7,\ldots$, many operators of the form $\square^n\de^\ell\varphi^m_R$ for $m=5,7,\ldots$.
\end{itemize}
\paragraph{$\boldsymbol{T_4}$, $\boldsymbol{H_4}$ and $\boldsymbol{B_4}$ representations}
\begin{itemize}
\item At $\tau=4$, operators of the form $\de^\ell\varphi^4_R$, non-degenerate for $R=T_4$, $\ell=0,2,3,5$ (absent for $\ell=1$), $R=H_4$, $\ell=1,2$ (absent for $\ell=0$) and $R=B_4$, $\ell=2$ (absent for $\ell=0,1$). \item At $\tau=6,8,10,\ldots$ there are many operators of the form $\square^n\de^\ell\varphi^m_T$ for $m=4,6,\ldots$.
\end{itemize}

If we set $n=1$ in order to specify to the Ising CFT, the counting of operators changes and we find many new cases of non-degenerate operators, \emph{i.e.} cases with only one operator at a given twist and spin. In general, the counting of operators of the form $\square^k\de^\ell\phi^m$ at $n=1$ agrees with that of operators of the form $\square^k\de^\ell\varphi^m_{T_m}$ for general $n$, and in fact the order-$\eps$ anomalous dimensions of these operators are related by a factor $\frac{6}{n+8}$. The exception is the operator $\varphi^3_{T_3}$ and one operator of the form $\de^\ell\varphi^4_{T_4}$, $\ell=3,5,7,\ldots$, where the corresponding operator does not exist in the $n=1$ case.\footnote{More precisely, the corresponding states go into the descendant level of the higher-spin currents $\mathcal J_{\ell+1}=\phi\partial^{\ell+1}\phi$, which in the interacting theory are not conserved for $\ell=3,5,7,\ldots$ \cite{Rychkov:2015naa}.} A complete classification, including order $\eps$ anomalous dimensions, of operators of this type with up to five partial derivatives was given in \cite{Kehrein:1994ff}.

\subsubsection{Multiplet recombination effect}
\label{subsec:multrecomb}

When comparing the spectra of the free theory and the interacting theory, there is an important multiplet recombination effect to take into account \cite{Rychkov:2015naa}; see also \cite{Maldacena:2012sf,Giombi:2016hkj}. In simplified terms, this is due to the fact that the equations of motion in the interacting theory,
\begin{equation}
\square \varphi \propto \varphi^3_V,
\label{eq:EOM}
\end{equation}
are different from those in the free theory, which read $\square\varphi=0$. Equation \eqref{eq:EOM} relates $\square\varphi$ and $\varphi^3_V$ and as a consequence, some operators $\tilde \O$, like $\varphi^3_V$, that are primary operators in the free theory no longer exist in the interacting theory. The corresponding states become descendants of the primary $\O$, in this case $\varphi$. 
Apart from the operator $\varphi$, this effect also happens for the higher-spin currents $\O=\mathcal J_{R,\ell}$, which are conserved in the free theory, $\de_{\mu_1}\mathcal J_{R,\ell}^{\mu_1\mu_2\cdots\mu_\ell}=0$. The corresponding states that in the interacting theory become descendants are those that in the free theory belong to primary operators of the form $\tilde\O=\de^{\ell-1}\varphi^4_{R}$. In summary, the multiplet recombination effect involves the following pairs of operators:
\begin{itemize}
\item $\O=\varphi$, $\tilde \O=\varphi_V^3$.
\item $\O=\mathcal J_{S,\ell}$, $\tilde \O=\de^{\ell-1}\varphi^4_{S,\ell}$ for $\ell=4,6,8,\ldots$. The stress-tensor $\mathcal J_{S,2}$ is conserved also in the interacting theory. In fact, neither the free nor the interacting theory contains a primary operator of the form $\de\varphi^4_S$.
\item $\O=\mathcal J_{T,\ell}$, $\tilde \O=\de^{\ell-1}\varphi^4_{T,\ell}$ for $\ell=2,4,6,\ldots$.
\item $\O=\mathcal J_{A,\ell}$, $\tilde \O=\de^{\ell-1}\varphi^4_{A,\ell}$ for $\ell=3,5,7,\ldots$. The global symmetry current $\mathcal J_{A,1}$ is conserved also in the interacting theory.
\end{itemize}

Multiplet recombination has an interesting effect on the conformal block decomposition. In this paper, we will determine expressions for four-point correlators that expand in a series in small $\eps$,
\begin{equation}
\G(u,v)=\G^{(0)}(u,v)+\eps \G^{(1)}(u,v)+\ldots.
\end{equation}
Here the term $\G^{(0)}(u,v)$ agrees with the correlator for the free theory in $d=4$ dimensions, and can therefore be expanded in conformal blocks as a sum over primary operators in the free theory,
\begin{equation}
\label{eq:freedecomp}
\G^{(0)}(u,v)=\sum_{\O}a_{\O}^{\mathrm{free}}G_{\Delta^{\mathrm{free}}_\O,\ell_\O}(u,v)+\sum_{\tilde \O}a_{\tilde\O}^{\mathrm{free}}G_{\Delta^{\mathrm{free}}_{\tilde\O},\ell_{\tilde\O}}(u,v),
\end{equation}
where we have divided the sum into one sum over operators that remain primaries in the interacting theory, and one of those that correspond to descendants in the interacting theory.
On the other hand, if we were considering the interacting theory at order $\eps^0$, we would expect to find
\begin{equation}
\label{eq:interdecomp}
\G^{(0)}(u,v)=\sum_{\O}a_{\O}G_{\Delta_\O,\ell_\O}(u,v),
\end{equation}
where the operators contributing are the same as those in the first sum in \eqref{eq:freedecomp}.

In fact, both the expansions \eqref{eq:freedecomp} and \eqref{eq:interdecomp} are correct.
This apparent contradiction between the two expressions is reconciled by a special relation of the conformal blocks \cite{Rychkov2013},
\begin{equation}
a^{\mathrm{free}}_{\O} G^{[\Delta_{12}^{\mathrm{free}},\Delta_{34}^{\mathrm{free}}]}_{\Delta^{\mathrm{free}}_\O,\ell_\O}(u,v)+a^{\mathrm{free}}_{\tilde\O} G^{[\Delta_{12}^{\mathrm{free}},\Delta_{34}^{\mathrm{free}}]}_{\Delta^{\mathrm{free}}_{\tilde\O},\ell_{\tilde\O}}(u,v) = a_{\O} G^{[\Delta_{12},\Delta_{34}]}_{\Delta_\O,\ell_\O}(u,v)
\label{eq:reconciliation}
.
\end{equation}
This equation is possible because on the right-hand side, the operator dimensions involved are those in the interacting theory, and therefore include anomalous dimensions. The $\eps$ dependence of these anomalous dimensions is canceled by poles in $\eps$ that appear at descendant level in the interacting conformal block.
We will see an explicit example of such a recombination effect in section~\ref{secc:mixedchanneldatamultrecomb}.\footnote{Notice that the recombination effect involving the broken currents $\mathcal J_{R,\ell}$ and their twist-four descendants of the form $\de^{\ell-1}\varphi^4_R$ is not visible in the correlators of identical external operators. This is because they have spins of opposite parity, and the OPE of operators with identical scaling dimension contains only primaries with spin that are either all even or all odd. Likewise, the operators $\varphi$ and $\varphi^3_V$ can only show up in mixed correlators thanks to the $\mathbb Z_2$ parity of number of fields, so multiplet recombination on the level of conformal block expansions is a feature that is only captured in mixed correlators.}

\subsection{Correlators and conformal data}
\label{sec:kinematicconventions}

The four-point function of general scalar operators $\O_i$ in representations $R_i$, takes the form
\begin{equation}
\label{eq:fourpointsetup}
\left\langle \O^I_{1}(x_1)\O^J_{2}(x_2)\O^K_{3}(x_3)\O^L_{4}(x_4)\right\rangle=\mathbf K_{\O_{1}\O_{2}\O_{3}\O_{4}}(x_i)\sum_R \mathbf P^{R,IJKL}_{R_1R_2;R_3R_4} \G^R_{\O_{1}\O_{2}\O_{3}\O_{4}}(u,v)
,
\end{equation}
where $\mathbf P^{R,IJKL}_{R_1R_2;R_3R_4}$ are projectors, to be described below, and
\begin{equation}\label{eq:Kkinematic}
\mathbf K_{\O_1\O_2\O_3\O_4}(x_1,x_2,x_3,x_4)=\left(\frac{x_{24}^2}{x_{14}^2}\right)^{\frac{\Delta_1-\Delta_2}2}\left(\frac{x_{14}^2}{x_{13}^2}\right)^{\frac{\Delta_3-\Delta_4}2}\frac{1}{(x_{12}^2)^{\frac{\Delta_1+\Delta_2}2}(x_{34}^2)^{\frac{\Delta_3+\Delta_4}2}}
.
\end{equation}
The function $\G^R_{\O_{1}\O_{2}\O_{3}\O_{4}}(u,v)$ depends on the coordinates $x_i$ only through the cross-ratios $u=z\zb$ and $v=(1-z)(1-\zb)$; see \eqref{eq:crossratios}.
We have the conformal block expansion
\begin{equation}
\label{eq:CBintro}
\G^R_{\O_1\O_2\O_3\O_4}(u,v)=\frac1{\mathcal N_R} \sum_\O \lambda_{\O_1\O_2\O}\lambda_{\O_3\O_4\O}G^{[\Delta_{12},\Delta_{34}]}_{\Delta_\O,\ell_O}(u,v)
,
\end{equation}
where $\mathcal N_R$ is a normalization constant.
Here the sum is over primary operators $\O$ that are contained in both of the OPEs $\O_1\times \O_2$ and $\O_3\times \O_4$. For scalar operators, the OPE only contains operators $\O$ transforming in traceless-symmetric representations of the Lorentz subgroup of the conformal group, denoted spin-$\ell$ operators. The functions $G_{\Delta,\ell}(u,v)$ are the conformal blocks, which are fixed functions in terms of $\Delta$, $\ell$, $u$, $v$, spacetime dimension $d$, and the pairwise differences of external scaling dimensions $\Delta_{12}=\Delta_1-\Delta_2$ and $\Delta_{34}=\Delta_3-\Delta_4$. 

\subsubsection{Projectors and normalizations}

Without considering any global symmetry, it is natural to define the OPE coefficients between scalar operators in such a way that they are completely permutation symmetric,
\begin{equation}
\label{eq:threepointsym}
\lambda_{\O_1\O_2\O_3}=\lambda_{\O_2\O_1\O_3}=\lambda_{\O_1\O_3\O_2}.
\end{equation}
In the presence of global symmetry, each operator carries the indices of its irrep, and the three-point functions would need to be promoted to a three-point tensor structure in order to make the permutation symmetry of \eqref{eq:threepointsym} manifest. However, such three-point functions will never appear alone in our work, where we are concerned with four-point functions and therefore only use the projectors for four-point functions.

Inside our four-point function, we would still like the OPE coefficients $\lambda_{\O_1\O_2\O}$ introduced in \eqref{eq:CBintro} to satisfy permutation symmetry. This condition is unimportant when one considers only a single four-point correlator, but becomes increasingly important when larger systems of correlators are considered. Moreover, we would like the contribution from the identity operator in the four-point correlator to be consistent with a product of two-point functions, implying
\begin{equation}
\label{eq:OPEid}
\lambda_{\O\O\1}=1, \qquad \mathbf P^{S,IJKL}_{RR;R'R'}=\delta^{IJ}\delta^{KL}.
\end{equation}

Fortunately, there is a choice of normalizations of the tensor structures $\mathbf P^{R,IJKL}_{R_1R_2;R_3R_4}$ that gives permutation-invariant OPE coefficients, satisfying \eqref{eq:threepointsym} and \eqref{eq:OPEid}, for arbitrarily large sets of external operators \cite{HeFuture}.\footnote{This is summarized in appendix~A.3 of \cite{Henriksson:2022rnm}. We thank Ning Su for useful discussions and for sharing some unpublished results with us.} Using such tensor structures would give $\mathcal N_R=1$ in \eqref{eq:CBintro}. This choice introduces some unpleasant square roots in the crossing equation and is usually not chosen in the literature. Here we will instead introduce a minimal set of normalization constants to give simpler expressions in our formulas. Our choice corresponds to
\begin{align}\label{eq:normVVVV}
\mathcal N_S&=1,
& \mathcal N_T&=\frac{\sqrt{(n+2)(n-1)/2}}{n},& \mathcal N_A&=\frac{\sqrt{n(n-1)/2}}{n},
\end{align}
which for $R=T,A$ agrees with $\mathcal N_R=\frac{\sqrt{\dim R}}{\dim V}$.
Moreover, we will use
\begin{align}
\label{eq:Norm2}
\mathcal N_V&= \mathcal N_{T_3}=\mathcal N_{H_3}=1,
&
\mathcal N_{T_4}&=\frac{\sqrt{(n+6)(n+1)n(n-1)/6}}{(n+2)(n-1)}, \\ \mathcal N_{H_4}&=\frac{\sqrt{(n+4)(n+1)(n-1)(n-2)/2}}{(n+2)(n-1)}, & \mathcal N_{B_4}&=\frac{\sqrt{(n+2)(n+1)n(n-3)/3}}{(n+2)(n-1)},
\label{eq:Norm3}
\end{align}
where for $R=T_4,H_4,B_4$ the formulas agree with $\mathcal N_R=\frac{\sqrt{\dim R}}{\dim T}$.

\subsubsection{Crossing symmetry}
\label{sec:crossingmatrix}

Crossing symmetry is a highly restrictive constraint on CFT data. 
When used in the four-point function \eqref{eq:fourpointsetup}, it gives an equation of the form
\begin{equation}
\G^R_{\O_{1}\O_{2}\O_{3}\O_{4}}(u,v)=
\frac{u^{\frac{\Delta_1+\Delta_2}2}}{v^{\frac{\Delta_2+\Delta_3}2}}
\sum_{R'}
M^{RR'}_{R_1R_2;R_3R_4}
\G_{\O_{3}\O_{2}\O_{1}\O_{4}}^{R'}(v,u)
,
\end{equation}
where the crossing matrices $M^{RR'}$ relate the different global symmetry representations involved in the direct and crossed channel. They can be worked out from the explicit form of the tensor structures. Here we will make use of results found by the authors of \cite{He:2021sto} and we give the explicit expressions in appendix~\ref{app:crossingM}. 

\subsubsection{Conformal block expansion in perturbation theory}

In most of this paper, we will write the conformal block decomposition on the form
\begin{equation}
\label{eq:CBdecomp}
\G^R_{\O_1\O_2\O_3\O_4}(u,v)=\sum_{\tau_0,\ell}a_{R,\tau_0,\ell}G_{\tau_0+\ell+\gamma_{R,\tau_0,\ell},\ell}(u,v)
,
\end{equation}
where
\begin{equation}
a_{R,\tau_0,\ell}=a_{R,\tau_0,\ell}^{1234}=\frac1{\mathcal N_R}\sum_{i}\lambda_{\O_1\O_2\O_i}\lambda_{\O_3\O_4\O_i}
\end{equation}
and the dependence on $\O_j$ will often be omitted when there is no risk of confusion.
The sum over $i$ is over operators that are degenerate for $d=4$, i.e. with dimension $\Delta=\tau_0+\ell+\eps \gamma^{(1)}_{\O_i}+O(\eps^2)$.
This means that when there is degeneracy, the symbol $\gamma_{R,\tau_0,\ell}$ really refers to \emph{weighted averages} of anomalous dimensions, more precisely of the form
\begin{equation}
\gamma_{R,\tau_0,\ell}=\frac1{a_{R,\tau,\ell}\mathcal N_R}\sum_i \eps \lambda_{\O_1\O_2\O_i}\lambda_{\O_3\O_4\O_i} \gamma_{\O_i}^{(1)}+O(\eps^2).
\end{equation}
Note that in our conventions, the $\gamma_{R,\tau_0,\ell}$ contain the factors of $\eps$, whereas any quantity with superscript $^{(k)}$ denotes the $k$th order in a series in $\eps$. In much of what follows, we will often drop the subscript on $\tau_0$, when using it as a label.

Many properties of conformal blocks were found in the works of Dolan and Osborn \cite{Dolan:2000ut,Dolan:2003hv,Dolan:2011dv}. In particular, they determined a Casimir differential equation for the blocks, which could be exactly solved in two and four dimensions. The four-dimensional solution reads
\begin{align}
G^{(4d),[\Delta_{12},\Delta_{34}]}_{\Delta, \ell}(z, \bar z) = \frac{z \bar z}{z - \bar z} \left( k^{[\Delta_{12},\Delta_{34}]}_{\frac{\Delta+\ell}2}(z) k^{[\Delta_{12},\Delta_{34}]}_{\frac{\Delta-\ell}2}(\bar z)
-
k^{[\Delta_{12},\Delta_{34}]}_{\frac{\Delta+\ell}2}(\bar z) k^{[\Delta_{12},\Delta_{34}]}_{\frac{\Delta-\ell}2}(z)\right)
,
\label{eq:4dblock}
\end{align}
which will be of great use because we are working in the vicinity of four dimensions. In this equation,

\begin{equation}\label{eq:khgeneral}
k^{[a,b]}_\beta(x)=x^\beta {}_2 F_1\left(\beta-\frac a2, \beta+\frac b2;2 \beta; x\right),
\end{equation}
which is the same as the one-dimensional conformal blocks, also known as $\mathrm{SL}(2,\R)$ blocks.

Many useful results for conformal blocks are also available in general dimension, albeit often only in series expansions in different limits. Recall that, in Lorentzian signature, the cross-ratios $z,\zb$ are real independent variables $0<z,\zb<1$. The most relevant expansions for this work are the collinear limit and the double-lightcone limit,
\begin{equation}
\text{collinear limit:}\quad z\ll1, \qquad \text{double-lightcone limit:}\quad z\ll(1-\zb)\ll 1.
\end{equation}
In Lorentzian signature, the collinear limit can be approached when the insertion points $x_1$ and $x_2$ become light-like separated, even at finite $x_2^\mu-x_1^\mu$.
In this limit, the blocks reduce to the $\mathrm{SL}(2,\R)$ blocks, times a factor that only depends on the twist $\tau_\ell=\Delta-\ell$,
\begin{align}
\label{eq:blockleading}
G^{(d),[\Delta_{12},\Delta_{34}]}_{\Delta, \ell}(z, \bar z) = z^{\frac{\tau_\ell}{2}} k^{[\Delta_{12},\Delta_{34}]}_{\frac{\tau_\ell}2+\ell}(\bar z) + \mathcal{O}(z^{\frac{\tau_\ell}{2}+ 1}).
\end{align}
Note that this expression is independent of $d$, due to the effectively one-dimensional configuration.
The collinear limit of the correlator captures the contribution from the entire leading-twist family, where the contributions from the different spins are summed up using $\mathrm{SL}(2,\R)$ blocks. It is often convenient to project such a sum to a common factor $z^{\tau/2}$, and trade the contribution from anomalous dimensions $\gamma_{\tau,\ell}=\tau_\ell-\tau$ for factors of $\log z$, 
\begin{equation}
\label{eq:LTcontr}
\G(u,v)=z^{\frac{\tau}2}\sum_\ell a_{\tau,\ell}\left(1+\frac12\gamma_{\tau,\ell}\log z+\ldots\right) k_{\frac{\tau}2+\frac{\gamma_{\tau,\ell}}2+\ell}(\zb)+O(z^{\frac{\tau}2+1})
,
\end{equation}
where we suppressed the dependence on $\Delta_{12}$, $\Delta_{34}$.

Working to leading order in $\eps$, there is a convenient reparametrization of the expression \eqref{eq:LTcontr}, which is valid up to ``regular terms'', which are functions of $\zb$ which have no double-discontinuity. Writing
\begin{equation}
a_{\tau,\ell}=a^{(0)}_{\tau,\ell}(1+\eps\alpha_{\tau,\ell}), \quad \gamma_{\tau,\ell}=\eps\gamma^{(1)}_{\tau,\ell},
\end{equation}
one finds
\begin{equation}
\label{eq:shiftedLTcontr}
\G(u,v) = z^{\frac\tau2}\sum_{\ell}a^{(0)}_{\tau,\ell}\left(1+\frac12\eps\gamma^{(1)}_{\tau,\ell}\log z+\eps\hat\alpha_{\tau,\ell}\right)k_{\hb}(\zb) + O(z^{\frac\tau2+1})+\text{regular terms}
,
\end{equation}
where
\begin{equation}
\hb=\frac{\tau}2+\ell
\end{equation}
and the shifted OPE coefficients have been introduced using\footnote{See \cite{Henriksson:2017eej} for more details on the nature of ``regular terms'' and a derivation of this formula.}
\begin{align}\label{eq:hatalphadef}
		\hat \alpha_{\tau, \ell} = \alpha_{\tau, \ell} - \frac{1}{2 a^{(0)}_{\tau, \ell} } \partial_\ell ( a^{(0)}_{\tau, \ell} \gamma_{\tau, \ell}) \, .
	\end{align}
The upshot of the formula \eqref{eq:shiftedLTcontr} is the following. The whole leading contribution to the correlator has been written as a sum over integer-spaced $\mathrm{SL}(2,\R)$ conformal blocks, up to terms that have vanishing double-discontinuity. By matching singular terms, \emph{i.e.} terms with non-vanishing double-discontinuity in \eqref{eq:shiftedLTcontr}, perturbative CFT data can be computed, converting the insights of \cite{Fitzpatrick:2012yx,Komargodski:2012ek} into a systematic framework dubbed large spin perturbation theory \cite{Alday:2016njk,Alday:2016jfr}.
Below, we will see that the same expressions arise at leading order using the inversion formula.

We conclude this section by noting that the subleading corrections in $z$ in \eqref{eq:blockleading} can be determined order-by-order in general spacetime dimension $d$, by solving the Casimir equation in a series expansion. The solution at each order in $z$ can be written as a sum over $\mathrm{SL}(2,\R)$ blocks with shifted arguments,
\begin{equation}
G^{(d),[\Delta_{12},\Delta_{34}]}_{\Delta, \ell}(z, \bar z) = z^{h}\sum_{k=0}^\infty z^k\sum_{i=-k}^kc_{k,i}(h,\hb) k^{[\Delta_{12},\Delta_{34}]}_{\hb+i}(\zb)
,
\label{eq:subcollinearmain}
\end{equation}
where $h=\frac{\Delta-\ell}2$ and $\hb=\frac{\Delta+\ell}2$.
We defer the details of this expansion to appendix~\ref{app:4-epsblocks}.

\subsection{Lorentzian inversion formula}\label{sec:LIF}

The Lorentzian inversion formula \cite{Caron-Huot:2017vep} is a method for extracting CFT data from the double-discontinuity of the correlator. In its most complete form, it can be written as
\begin{align}
\label{eq:Inversionformulafull}
C^{1234}(\Delta,\ell)&=\frac{\kappa_{\frac{\Delta+\ell}2}^{1234}}4
\int\limits_{[0,1]^2} dz\,d\zb\,\mu^{1234}(z,\zb)G^{(d),1234}_{\ell+d-1,\Delta+1-d}(z,\zb){\dDisc}^{1234}\left[\G_{1234}(z,\zb)\right]
\nonumber
\\&\quad+(-1)^\ell \big(\ 1\leftrightarrow 2\ \big)
,
\end{align}
where
\begin{align}
\kappa_\beta^{1234}&=\frac{\Gamma(\beta-\frac{\Delta_1-\Delta_2}2)\Gamma(\beta+\frac{\Delta_1-\Delta_2}2)\Gamma(\beta-\frac{\Delta_3-\Delta_4}2)\Gamma(\beta+\frac{\Delta_3-\Delta_4}2)}{2\pi^2\Gamma(2\beta)\Gamma(2\beta-1)},
\\
\mu^{1234}(z,\zb)&=\left|\frac{z-\zb}{z\zb}\right|^{d-2}\frac{\left((1-z)(1-\zb)\right)^{\frac{\Delta_2+\Delta_3-\Delta_1-\Delta_4}2}}{(z\zb)^2}.
\end{align}
Note that $\kappa_{\hb}^{1234}=\kappa_{\hb}^{2134}$. Here the labels $1234$ are short for $\O_1\O_2\O_3\O_4$, and will sometimes be suppressed when no confusion occurs.
A central ingredient in \eqref{eq:Inversionformulafull} is the double-discontinuity, which is defined as
\begin{align}
\nonumber
{\dDisc}^{1234}[f(\zb)]&=\cos\left(\pi\tfrac{\Delta_2+\Delta_3-\Delta_1-\Delta_4}2\right)f(\zb)\\&\quad -\frac12e^{i\pi\frac{\Delta_2+\Delta_3-\Delta_1-\Delta_4}2}f^{ \circlearrowleft}(\zb)-\frac12e^{-i\pi\frac{\Delta_2+\Delta_3-\Delta_1-\Delta_4}2}f^{\circlearrowright}(\zb)
\label{eq:dDiscgen}
\end{align}
for a function of $\zb<1$ with a branch cut at $\zb\geqslant1$, and the arrows indicate the analytic continuation around the branch cut.

\subsubsection{The perturbative inversion formula}

The most general inversion formula \eqref{eq:Inversionformulafull} captures the entire spectrum of spinning operators. However it contains a two-dimensional integral against a kernel which contains the conformal block, which for arbitrary spacetime dimensions is not known in closed form.\footnote{In two and four dimensions, where the conformal block is given on exact form, the whole two-dimensional integral can be performed at once in some simple cases; see e.g. \cite{Caron-Huot:2017vep,Alday:2017vkk}.} 
To make progress in the general spacetime dimension, we make use of some simplifications. The first step is to consider the limit of small $z$, in which case the integral over $\zb$ can be performed for each order in $z$, giving an approximate generating function $C(z,\hb)$ \cite{Caron-Huot:2017vep}. The next step is to consider the presence of a small perturbative expansion parameter $\eps$, so that in the limit $\eps\to0$, the spectrum organizes into exact twist families. This gives a \emph{perturbative inversion formula}, first considered in \cite{Alday:2017zzv}, from which the CFT data can be determined exactly order-by-order in $\eps$.

The final product will be a generalization of the $\mathrm{SL}(2,\R)$ inversion formula (see \eqref{eq:SL2Rformula} below), valid also beyond the case of identical external operators. The derivation follows the proof given in \cite{Henriksson:2020jwk}, and the result is that
\begin{equation}
\label{eq:inversionintegralgen}
\mathbb T_{\tau_0}(\log z,\hb)=\kappa^{1234}_{\hb}\int\limits_0^1\frac{d\zb}{\zb^2}k_{\hb}^{1234}(\bar{z})(1-\zb)^{\frac{\Delta_2+\Delta_3-\Delta_1-\Delta_4}2}{\dDisc}^{1234}\left[\G_{1234}(z,\zb)\right]\big|_{z^{\tau_0/2}} + (-1)^\ell\big(\ 1\leftrightarrow 2\ )
,
\end{equation}
where $k_{\hb}^{1234}(\bar{z})$ is defined in \eqref{eq:khgeneral}.
In \eqref{eq:inversionintegralgen}, we collected all powers of $\log z$ into the generating function defined by
\begin{equation}
\mathbb T_{\tau_0}(\log z,\hb)=T^{(0)}_{\hb} + \frac{1}{2} T^{(1)}_{\hb} \log z + \frac{1}{8} T^{(2)}_{\hb} \log^2 z +\ldots
,
\end{equation}
which contains the CFT data of the family of operators of leading twist of the form $\tau_\ell=\tau_0+\gamma_\ell$.
The CFT data may be extracted from the terms in the generating function via
\begin{align}
	a_{\tau_0,\ell} (\gamma_{\tau_0,\ell})^p = T^{(p)}_{\bar h} + \frac{1}{2} \partial_{\bar h} T^{(p+1)}_{\bar h}+ \frac{1}{8} \partial^2_{\bar h} T^{(p+2)}_{\bar h} + \ldots \Big|_{\bar h =\frac{ \tau_0}2+\ell} \, .
	\label{eq:inversion_CFTdata}
\end{align}
To order $\eps$, $T^{(0)}_{\bar h}$ and $T^{(1)}_{\bar h}$ correspond to the terms $a^{(0)}_{\tau_0,\ell}(1+\eps\hat\alpha_{\tau_0,\ell})$ and $a^{(0)}_{\tau_0,\ell}\gamma_{\tau_0,\ell}$ in \eqref{eq:shiftedLTcontr}.

\paragraph{$\boldsymbol{\mathrm{SL}(2,\R)}$ inversion formula} It is sometimes convenient to consider a purely one-dimensional inversion problem, involving only the $\zb$ cross-ratio:
\begin{equation}
\label{eq:SL2Rformula}
\sum_{\hb=h,h+2,\ldots}T(\hb)k_\hb(\zb)=f(\zb)\quad \Leftrightarrow \quad T(\hb)=\kappa_\hb \int\limits_0^1\frac{d\zb}{\zb^2}k_{\hb}\dDisc[f(\zb)]+\text{contr.\ at finite $\hb$}.
\end{equation}
The formula on the right is called the $\mathrm{SL}(2,\R)$, inversion formula. The achievement of the perturbative inversion formula is to reduce the general inversion problem to a collection of $\mathrm{SL}(2,\R)$ inversion problems.

\paragraph{Example: GFF OPE coefficients}

As a warm-up, we will use the perturbative inversion formula \eqref{eq:inversionintegralgen} to find the OPE coefficients of the leading twist operators in the GFF theory. Consider the correlator
\begin{align}
\G_{\O_1\O_2\O_2\O_1}(u,v)&=\frac{u^{\frac{\Delta_1+\Delta_2}2}}{v^{\Delta_2}} =\sum_{n,\ell}a_{n,\ell}^{\mathrm{GFF}} G_{\frac{\Delta_1+\Delta_2}2+2n,\ell}^{1221}(u,v)
,
\end{align}
where the squared OPE coefficients $a_{n,\ell}^{\mathrm{GFF}} $ are known in the literature to take the form \cite{Fitzpatrick:2011dm}
\begin{align}
&
a_{n,\ell}^{\mathrm{GFF}}
\\&=\frac{(\Delta_1+1-\frac d2)_n(\Delta_2+1-\frac d2)_n(\Delta_1)_{n+\ell}(\Delta_2)_{n+\ell}}{\ell !\, n!\, (\ell+\frac d2)_n(\Delta_1+\Delta_2+n+1-d)_n(\Delta_1+\Delta_2+\ell+n-\frac d2)_n(\Delta_1+\Delta_2+2n+\ell-1)_\ell}
,
\label{eq:aGFFgen}
\end{align}
where $(a)_n$ denotes the Pocchammer symbol. If $\O_1=\O_2$, there is an additional contribution, identical up to a factor $(-1)^\ell$.

Our goal is to verify the formula \eqref{eq:aGFFgen} at leading twist ($n=0$) by inverting the contribution from the identity operator:
\begin{equation}
\G_{\O_2\O_2\O_1\O_1}(u,v)|_{\1}=1
.
\end{equation}
Multiplying with the crossing factor and computing the $\dDisc$, we get
\begin{equation}
{\dDisc}^{1221}\left[\frac{(z\zb)^{\frac{\Delta_1+\Delta_2}2}}{((1-z)(1-\zb))^{\Delta_2}}\right]\Big|_{z^{\frac{\Delta_1+\Delta_2}2}} = 2\sin(\pi\Delta_1)\sin(\pi\Delta_2)\frac{\zb^{\frac{\Delta_1+\Delta_2}2}}{(1-\zb)^{\Delta_2}}
.
\end{equation}
Next, we multiply by the factor $(1-\zb)^{\Delta_2-\Delta_1}\kappa_\hb^{1221}$, and integrate against the kernel $\zb^{-2}k_{\hb}^{1221}(\zb)$. It turns out that with the help of an integral representation\footnote{Specifically $_2F_1(a,b;c;x)=\frac{\Gamma(c)}{\Gamma(b)\Gamma(c-b)}\int_0^1 dt\frac{t^{b-1}(1-t)^{c-b-1}}{(1-tx)^a}$.
} for the hypergeometric function, the integral can be worked out explicitly, and we get
\begin{equation}\label{eq:inversionGFF}
\mathbb T_{\frac{\Delta_1+\Delta_2}2}(\log z,\hb)=\frac{\Gamma(\hb+\frac{\Delta_1-\Delta_2}2)\Gamma(\hb-\frac{\Delta_1-\Delta_2}2)\Gamma(\hb+\frac{\Delta_1+\Delta_2}2-1)}{\Gamma(\Delta_1)\Gamma(\Delta_2)\Gamma(\hb-\frac{\Delta_1+\Delta_2}2+1)\Gamma(2\hb-1)}
,
\end{equation}
where the absence of any dependence on $\log z$ shows that there are no anomalous dimensions.
Upon putting $\hb=\frac{\Delta_1+\Delta_2}2+\ell$, this agrees exactly with the GFF OPE coefficients \eqref{eq:aGFFgen} for $n=0$.

\subsubsection{Projections onto subleading twists}\label{sec:projsubleadtwist}

The formula \eqref{eq:inversionintegralgen} extracts the CFT data for leading-twist operators with twist $\tau_0$,\footnote{In general, it can be used whenever there is no twist family with twist $\tau_0-2k+O(\eps)$ for integer $k$.} however in the following we will need to find data also for subleading twist families. 

To find the CFT data for the leading twist family and a collection of subleading twists, we consider  the following quantity,
\begin{equation}
\label{eq:boldTtower}
\mathbb T(z,\hb)=\sum_{\tau_0}z^{\frac{\tau_0}2}\mathbb T_{\tau_0}(\log z,\hb)\,.
\end{equation}
It is defined as the sum of the result of the inversion integral at various powers $z^{\tau_0/2}$; see \eqref{eq:inversionintegralgen}.
In general, the term in \eqref{eq:boldTtower} at a given power $z^{\tau_0/2}$ will correspond to the CFT data from primaries at twist $\tau_0$, and from descendants from primaries of twists $\tau_0-2,\ \tau_0-4,\ \ldots$. In order to determine the CFT data for primaries, we therefore need to find a way to project away the descendant contribution.

The method to do this in arbitrary spacetime dimensions was briefly described in section~3.2.4.2 of \cite{Henriksson:2020jwk} and makes use of the subleading terms in the collinear expansion of the conformal blocks. Here we will give an explicit description of how to perform these projections.

For concreteness, let us consider the case where $\tau_0=2,4,\ldots$, which matches what we use in this paper.
Then the subtracted generating functions $\mathbb S_{\tau_0}$, which yields the CFT data of primaries, can be found by the following subtractions,
\begin{align}
\mathbb{S}_2(\log z, \bar h) \ & = \ \mathbb{T}_2(\log z, \bar h) , \\
\mathbb{S}_4(\log z, \bar h) \ & = \ \mathbb{T}_4(\log z, \bar h) - \sum_{m = -1}^1 c_{1,m}(1 + \tfrac{1}{2} \gamma_{2,\ell} \eps , \bar h - m) \mathbb{S}_2(\log z, \bar h - m) , \\
\mathbb{S}_6(\log z, \bar h) \ &= \ \mathbb{T}_6(\log z, \bar h) - \sum_{m = -1}^1 c_{1,m}(2 + \tfrac{1}{2} \gamma_{4, \ell} \eps , \bar h - m) \mathbb{S}_4(\log z, \bar h - m) \nonumber\\
& \quad - \sum_{m = -2}^2 c_{2,m}(1 + \tfrac{1}{2} \gamma_{2, \ell} \eps , \bar h - m) \mathbb{S}_2(\log z, \bar h - m) ,
\label{eq:projectionTtoS}
\end{align}
where the $c_{k, m}(h,\hb)$ are the coefficients that appear in the subleading corrections to the collinear conformal blocks, \eqref{eq:subcollinearmain} (see appendix \ref{app:4-epsblocks} for more details). An important detail slightly unclear from this formula is that the $\gamma_{\tau_0, \ell}$ implicitly depend on $\bar h$ through the relation $\ell = \bar h - \tau_0 / 2$. The $\ell$ dependence should be converted to $\bar h$ before the replacement $\bar h \to \bar h + m$ is made. This formula also reflects the fact that no projection is needed for twist two. It is easy to see how to generalize for twist $\tau_0 = 8$, $10$, etc.

\paragraph{Example: twist-four data in the free theory}
Let us see how this works by considering the $\langle \varphi_S^2\varphi_S^2\varphi_S^2\varphi_S^2  \rangle$ correlator in the free theory in $4-\eps$ dimensions,
\begin{equation}
\G_{\varphi_S^2\varphi_S^2\varphi_S^2\varphi_S^2}(u,v)=1+\frac{u^{2\Delta_\varphi}}{v^{2\Delta_\varphi}}+u^{2\Delta_\varphi} +\frac4n\left(\frac{u^{\Delta_\varphi}}{v^{\Delta_\varphi}}+\frac{u^{2\Delta_\varphi}}{v^{\Delta_\varphi}}+u^{\Delta_\varphi}\right),
\end{equation}
where $\Delta_\varphi=1-\frac\eps2$. This can easily be computed from Wick contractions. To compute the twist-four data, we must consider both the $\O(z)$ and $\O(z^2)$ expansions of the $\dDisc$, and then we isolate the terms which are non-regular near $\bar z = 1$. We find
\begin{align}
 \text{dDisc}\left[ \mathcal{G}(u, v)\right]\Big|_{z} \ & = \ \text{dDisc}\left[ \frac2n \left( (3 - \eps \log z) \left(\frac{\bar z}{1-\bar z} \right) - \left(\frac{\bar z}{1-\bar z} \right)^{1+\eps} \right) \right] 
 ,
 \label{eq:toinvz1}
 \\
    \text{dDisc}\left[ \mathcal{G}(u, v)\right]\Big|_{z^2} \ &= \ \text{dDisc} \Bigg[ -\left(\frac{\bar z}{1-\bar z} \right)^{2+\eps} - \frac4n \left(\frac{\bar z}{1-\bar z} \right)^{1+\eps}  + (2 - \eps \log z) \left(\frac{\bar z}{1-\bar z} \right)^2
    \nonumber \\
 \label{eq:toinvz2}
& \qquad \qquad \qquad  -\frac2n(-6 + \eps + 3 \eps \log z)\left( \frac{\bar z}{1-\bar z} \right) \Bigg] .
\end{align}

Here we have written the double-discontinuity as a sum of different terms, each of which is a pure power of $\frac\zb{1-\zb}$. The inversion integral of such pure powers can easily be computed using\footnote{We simplified the expression using the standard identity $\sin (\pi  \alpha)= \frac\pi{\Gamma (\alpha) \Gamma (1-\alpha)}$.}
\begin{equation}
\kappa_{\hb}\int_0^1\frac{d\zb}{\zb^2}k_\hb(\zb)\dDisc\left[ \left(\frac{\bar z}{1-\bar z} \right)^{\alpha}\right]=
\frac{2\Gamma(\hb)^2\Gamma(\hb-1+\alpha)}{\Gamma(2\hb-1)\Gamma(\alpha)^2\Gamma(\hb+1-\alpha)},
\end{equation}
found by setting $\Delta_1=\Delta_2=\alpha$ in the example above, \eqref{eq:inversionGFF}. 

The inversion problem therefore amounts to making the replacement
\begin{equation}
\left(\frac{\bar z}{1-\bar z} \right)^{ \alpha} \to
\frac{2\Gamma(\hb)^2\Gamma(\hb-1+\alpha)}{\Gamma(2\hb-1)\Gamma(\alpha)^2\Gamma(\hb+1-\alpha)}
\end{equation}
in \eqref{eq:toinvz1}--\eqref{eq:toinvz2}. This leads to
\begin{align}
	\mathbb{T}_{2}  &=  \frac{ \Gamma(\bar h)^2}{\Gamma(2 \bar h -1)} \frac4n \left(2 - \varepsilon (\log z + 2 S_1(\bar h-1) \right),
	\\
	\mathbb{T}_{4}  &=  \frac{ \Gamma(\bar h)^2}{\Gamma(2 \bar h -1)} \bigg[ \frac{16}n + 2 \bar h (\bar h-1)
	\nonumber \\
	& \quad +\eps \left( 2 -\frac4n + 4 \bar h(\bar h - 1) - \left( \frac{12}n + 2 \bar h(\bar h - 1)\right)  \log z - \left( \frac{16}n + 4 \bar h (\bar h - 1) \right) S_1(\bar h -1) \right) \bigg].
\end{align}

These expressions may be used to compute $\mathbb{S}_4$ from~\eqref{eq:projectionTtoS}. Then the CFT data may be extracted from~\eqref{eq:inversion_CFTdata}, yielding
\begin{align}
    a^{(0)}_{4,\ell}  &=  \frac{2\Gamma(\ell + 2)^2}{\Gamma ( 2\ell + 3)}  \left(\frac4n + \ell^2 + 3 \ell + 2 \right),
    \\
    \gamma_{4,\ell}  &= -2 \eps
    ,
    \\
    \hat{\alpha}_{4,\ell}  &=   \frac{n(2 \ell^2 + 6 \ell + 5)}{4 +n(\ell^2 + 3\ell + 2)} - \frac{4 +n (2 \ell^2 + 6 \ell + 4)}{4 +n( \ell^2 + 3\ell + 2)} S_1(\ell + 1).
\end{align}

Continuing to higher twists, more data can be found. At all higher twist we find $\gamma_{\tau,\ell}=-2\eps$, which, recalling that $\Delta_{\varphi}^{\mathrm{free}}=1-\frac\eps2$, shows that the operators in the $s\times s$ OPE at these twists must be constructed out of four fields.
In summary, we have the following useful observation:
\begin{description}
\item[Twist 2] Operator $s$ at spin zero, and operators $\mathcal J_{S,\ell}$ at $\ell=2,4,6,\ldots$. 
\item[Twist 4,\,6,\,\ldots] Operators of the form $\partial^\ell\square^n\varphi^4_S$, with twist $4+2n-2\eps$. In the interacting theory, they will acquire anomalous dimensions at order $\eps$.
\end{description}
In particular, any operator with six or more fields does not appear in the free theory correlator. In the interacting theory, such operators can appear, but with OPE coefficients suppressed by at least one power of $\eps$. Since the operators appear with squared OPE coefficients, they will not appear in our considerations at order $\eps$. This means that if there is only one operator of ``$\varphi^4$ type'' at a given twist and spin, for instance $\square\varphi^4_S$ at $\tau_0=6,\ell=0$, it will effectively be non-degenerate, even if there is another operator at that twist and spin (at $\tau_0=6,\ell=0$ we also have $\varphi^6_S$).

We finish this section by giving the closed-form of the leading-order OPE coefficients in the free theory,
\begin{align}
	a^{(0)}_{2,\ell} =&\frac{8\Gamma(\ell+1)^2}{n\Gamma(2\ell+1)},
	\label{eq:ssssa02ell}
	\\
	a^{(0)}_{\tau,\ell} =&\frac{2\left((-1)^{\tau/2}c +(\tau+\ell-2)(\ell+1) \right)\Gamma\left( \frac \tau2 +\ell \right)^2\Gamma\left( \frac \tau2 -1\right)^2}{\Gamma(\tau+2\ell-1 )\Gamma( \tau- 3)}, \quad \tau=4,6,\dots,
	\label{eq:ssssa0tauell}
\end{align}
where $c=\frac4n$. 
They can be verified by computing the free theory four-point function and decompositing using the conformal blocks \eqref{eq:4dblock}.

\section{Mixed $\boldsymbol \varphi$ $\boldsymbol s$ system}
\label{sec:phi-s-system}

In this section we will consider the correlators involving the operators $\varphi$ and $s$. Apart from the $\langle\varphi\varphi\varphi\varphi\rangle$ correlator, this system is simple since only one $O(n)$ irrep is exchanged in each OPE channel.
This means that we can focus the discussion to the inversion problem, which we will describe in detail.
We begin by reviewing some results for $\langle\varphi\varphi\varphi\varphi\rangle$, before moving to $\langle\varphi\varphi ss\rangle$ and $\langle ssss \rangle$.

\subsection[Input: results from the $\langle\varphi\varphi\varphi\varphi\rangle$ correlator]{Input: results from the $\boldsymbol{\langle\varphi\varphi\varphi\varphi\rangle}$ correlator}
\label{sec:phiphiphiphi}

We now review the results from the large spin perturbation theory analysis of the $\langle\varphi\varphi\varphi\varphi\rangle$. This correlator was considered to order $\eps^4$ for $n=1$ in \cite{Alday:2017zzv}, which was generalized to any $n$ in \cite{Henriksson:2018myn}. Here we will only concern ourselves with computation and results to order $\eps^2$.
The assumptions that went into the mentioned works are the following:
\begin{itemize}
\item The $O(n)$ CFT is a perturbation from the free theory in $d=4-\eps$ dimensions, and CFT data admits an expansion in integer powers of $\eps$.
\item The leading twist operators are non-degenerate and the stress-tensor and global symmetry current are conserved.
\item The operator dimensions for leading-twist operators can be analytically continued to spin zero.
\item The only undetermined constants are the order-$\eps$ corrections to the OPE coefficients
\begin{equation}\label{eq:undetconsts}
	\lambda^2_{\varphi\varphi s}=\frac 2n+a_s^{(1)}\eps+O(\eps^2), \qquad \lambda^2_{\varphi\varphi t}=\mathcal N_T\left(2+a_t^{(1)}\eps+O(\eps^2)\right).
\end{equation}
\end{itemize}
In practice, to order $\eps^3$, the whole double-discontinuity is given by the crossed-chanel operators $\1$, $s=\varphi^2_S$ and $t=\varphi^2_T$. After computing the inversion problem at order $\eps^2$, the CFT data depends on the second-order anomalous dimension of $\varphi$ -- denoted $\gamma_\varphi^{(2)}$ -- and the first-order anomalous dimensions $\gamma_s$ and $\gamma_t$. Conservation of the stress-tensor and analytic continuation to spin zero provide three equations that fix the these leading constants:
\begin{equation}
\gamma_\varphi^{(2)}=0,\ \gamma_s=0,\ \gamma_t=0, \qquad \text{or}\qquad \gamma_\varphi^{(2)}=\frac{n+2}{4(n+8)^2},\ \gamma_s=\frac{n+2}{n+8},\
\gamma_t=\frac2{n+8}
.
\label{eqgsgtsol}
\end{equation}

With these expressions, and by supplementing the ansatz for the undetermined OPE coefficients in \eqref{eq:undetconsts}, we can compute the sum of conformal blocks at order $\eps$. This gives correlators of the form
\begin{align}
\G^S_{\varphi\varphi\varphi\varphi}(u,v)&=1+\frac1n\left(u+\frac uv\right)+\eps\left(-\frac{(n+2)u}{n(n+8)}\Phi(u,v)-\frac{u(1+v)}{2nv}\log u+\frac{u}{2nv}\log v\right)
\nonumber\\ & \quad +
\eps\left(a^{(1)}_s+\frac{2(n+2)}{n(n+8)}\right)G^{(4d)}_{2,0}(u,v)+O(\eps^2),
\label{eq:GSffff}
\\
\G^T_{\varphi\varphi\varphi\varphi}(u,v)&=u+\frac uv+\eps\left(-\frac{2u}{n+8}\Phi(u,v)-\frac{u(1+v)}{2v}\log u+\frac{u}{2v}\log v\right)
\nonumber\\ & \quad +
\eps\left(a^{(1)}_t+\frac{4}{n+8}\right)G^{(4d)}_{2,0}(u,v)+O(\eps^2),
\label{eq:GTffff}
\\
\G^A_{\varphi\varphi\varphi\varphi}(u,v)&=u-\frac uv+\eps\left(\frac{u(1-v)}{2v}\log u-\frac{u}{2v}\log v\right)+O(\eps^2),
\label{eq:GAffff}
\end{align}
where $\Phi(u,v)$ is the standard box integral, also known as $\bar D_{1111}$, given by \cite{Usyukina:1992jd}
\begin{equation}\label{eq:boxfunction}
\Phi(u,v)=\frac{2\left( \mathrm{Li}_2(z)- \mathrm{Li}_2(\zb)\right)+\log \left(z\zb\right)\log (\frac{1-z}{1-\zb}) }{z-\zb}
,
\end{equation}
where $\mathrm{Li}_k(x)$ denotes the polylogarithm.

A direct analysis of crossing for \eqref{eq:GSffff}--\eqref{eq:GAffff} shows that
\begin{equation}\label{eq:ffffas1at1}
a^{(1)}_s=-\frac{2(n+2)}{n(n+8)}, \qquad a^{(1)}_t=-\frac{4}{n+8}.
\end{equation}
To compute the next order in the correlator we would need the order $\eps^2$ corrections to these OPE coefficients. 
In fact, they were determined by the higher-order considerations in \cite{Henriksson:2018myn},\footnote{See also \cite{Dey:2016mcs}, which determined them using bootstrap in Mellin space.} and we compute the next order in the correlators in appendix~\ref{app:fffforder2}.

Some results that will be needed in the sections below are the OPE coefficients for the twist-two operators, extracted from the correlators above. They take the forms
\begin{align}
a_{S,2,\ell}&=\frac{2\Gamma(\ell+1)^2}{n\Gamma(2\ell+1)}\left(1+\left(S_1(2\ell)-2S_1(\ell)-\frac{n+2}{n+8}\delta_{\ell,0}\right)\eps\right)+O(\eps^2),
\label{eq:ffS2l}
\\
a_{T,2,\ell}&=\frac{2\Gamma(\ell+1)^2}{\Gamma(2\ell+1)}\left(1+\left(S_1(2\ell)-2S_1(\ell)-\frac{2}{n+8}\delta_{\ell,0}\right)\eps\right)+O(\eps^2),
\label{eq:ffT2l}
\\
a_{A,2,\ell}&=\frac{2\Gamma(\ell+1)^2}{\Gamma(2\ell+1)}\left(1+\big(S_1(2\ell)-2S_1(\ell)\big)\eps\right)+O(\eps^2),
\label{eq:ffA2l}
\end{align}
where we used the values \eqref{eq:ffffas1at1}.

\subsection[$\expv{\varphi \varphi ss}$ correlator]{$\boldsymbol{\expv{\varphi \varphi ss}}$ correlator}
\label{sec:ffsssection}
In this section we will compute the correlator $\G_{\varphi\varphi s s}(u,v)$ at order $\eps$ for general $n$. 
We begin by inverting into the pure channel, \emph{i.e.} we consider the pairwise OPEs $\varphi \times \varphi$ and $s \times s$.
The relevant inversion formula \eqref{eq:inversionintegralgen} reads
\begin{equation}\label{eq:Tffssgen}
	\mathbb T_{\varphi\varphi ss}( z,\hb)=\left( 1+(-1)^\ell \right)\kappa_{\hb}^{\varphi\varphi s s} \int_0^1\frac{d\zb}{\zb^2}k_{\hb}^{\varphi\varphi s s}(\zb)\dDisc\left[\G_{\varphi\varphi ss}(u,v)\right]
	,
\end{equation}
where we used $\Delta_1=\Delta_2=\Delta_\varphi$, $\Delta_3=\Delta_4=\Delta_s$.
As explained in the introduction, we will use the crossing equation to determine the double-discontinuity in terms of crossed-channel operators. In the case at hand, this amounts to operators in the $V$ irrep.
To order $\eps$, the only operator we have to consider in the double-discontinuity is $\varphi$. We will then compute the OPE coefficients and anomalous dimensions at order $\eps$ for a number of twists and conclude by resumming the correlator.

\subsubsection{Inversion into the pure channel}\label{ffssinversionsection}

The dimensions of the external fields are
\begin{align}
	\Delta_\varphi &=1-\frac{\eps}{2} +O(\eps^2),
	\\
	\Delta_s &=2-\eps +\gamma_s \eps +O(\eps^2),
	\label{eq:deltafdeltas}
\end{align}
where $\gamma_s =\frac{n+2}{n+8}$ is the anomalous dimension of $s$. The inversion formula then reads
\begin{equation}\label{eq:Tffss}
	\mathbb T_{\varphi\varphi ss}( z,\hb)=\left( 1+(-1)^\ell \right)\kappa_{\hb}^{\varphi\varphi s s} \int_0^1\frac{d\zb}{\zb^2}k_{\hb}^{\varphi\varphi s s}(\zb)\dDisc\left[\frac{u^{1-\frac{\eps}2}}{v^{(3-\frac{3}2 \eps +\gamma_s \eps)/2}}\G_{s\varphi\varphi s}(v,u)\right].
\end{equation}
Note that the double-discontinuity is determined in terms of the mixed-channel correlator only. This is a simplification compared to the inversion problem for the mixed channel, which involves finding double-discontinuities originating from both the pure channel and the mixed channel; see sections~\ref{subsec:auxiliarymixedsstt} and~\ref{subsec:inversionintomixed} below.

To perform the integration in \eqref{eq:Tffss} we need to write $\G_{s\varphi\varphi s}(v,u)$ as a sum of mixed-channel conformal blocks, however as already mentioned, only the operator $\varphi$ contributes to order $\eps$. Let us motivate why this is the case. All of the other operators are of the form $\square^n \partial^\ell \varphi^m_V$, where $m=3,5,\ldots$. In the double-lightcone $u,v \rightarrow 0$ limit the conformal block (before crossing) associated to such operator is proportional to $z^{\tau'/2}$, where $\tau'$ is the twist. It can be written as
\begin{equation}
\tau'=\Delta_\varphi +\Delta_s +(m-3)+2n +\hat\gamma_{n,m,\ell}
,
\end{equation}
where we have defined an anomalous dimension $\hat\gamma_{n,m,\ell}$, which is of order $\eps$. Ignoring regular terms, the double-discontinuity in \eqref{eq:Tffss} is
\begin{equation}
	\dDisc\left[ \frac{1}{(1-\bar{z})^{(\Delta_\varphi +\Delta_s)/2}} (1-\bar{z})^{\frac12 (\Delta_\varphi +\Delta_s +2n +\hat\gamma_{n,m,\ell})} \right] \sim \left( \hat\gamma_{n,m,\ell} \right)^2.
\end{equation}
Since $\hat\gamma_{n,m,\ell} \sim \eps$, this operator contributes to the $\dDisc$ at earliest at order $\eps^2$.

In conclusion, the only contribution at order $\eps$ is therefore given by the exchange of $\varphi$, which has twist $\tau = 1+O(\eps)$ and spin 0. Hence
\begin{equation}
	\G_{s\varphi\varphi s}(v,u) = \lambda^2_{\varphi s \varphi} G^{\Delta_s,\Delta_\varphi,\Delta_\varphi,\Delta_s}_{\Delta\varphi,\ell=0}(v,u) +\text{terms with no dDisc at this order}.
\end{equation}
Moreover, since the OPE coefficients are permutation invariant, $\lambda_{\varphi s\varphi}=\lambda_{\varphi\varphi s}$, we can use the value of this OPE coefficient as determined by the $\langle\varphi\varphi\varphi\varphi\rangle$ correlator,
\begin{equation}
		\lambda^2_{\varphi s\varphi}=\frac 2n+a_s^{(1)}\eps+O(\eps^2),
\end{equation}
where $a_s^{(1)}$ is given in \eqref{eq:ffffas1at1}.

We then need to evaluate the conformal block. The general formula for the scalar conformal block with non-identical external scalars of dimensions $\tilde{\Delta}_i$ is given in \cite{Dolan:2000ut} in terms of the dimension $\Delta_\varphi$ of the exchanged scalar and the spacetime dimension $d$:
\begin{align}
	G^{\tilde{\Delta}_1,\tilde{\Delta}_2,\tilde{\Delta}_3,\tilde{\Delta}_4}_{\Delta_\varphi,\ell=0}(u,v) &=u^{\frac{\Delta_\varphi}2} \sum_{m,p=0}^{\infty}\frac{\left(\frac{\Delta_\varphi+\tilde{\Delta}_1-\tilde{\Delta}_2}2 \right)_m \left(\frac{\Delta_\varphi-\tilde{\Delta}_3+\tilde{\Delta}_4}2 \right)_{m}}{\left(\Delta_\varphi+1-\frac{d}2 \right)_m}
	\nonumber \\
	&\quad \times \frac{\left(\frac{\Delta_\varphi-\tilde{\Delta}_1+\tilde{\Delta}_2}2 \right)_{m+p} \left(\frac{\Delta_\varphi+\tilde{\Delta}_3-\tilde{\Delta}_4}2 \right)_{m+p}}{(\Delta_\varphi)_{2m+p}}\frac{u^m (1-v)^p}{m!p!},
	\label{eq:scalarblockmixed}
\end{align}
In our problem we will exchange $u$ and $v$ and plug in $d=4-\eps$, $\tilde{\Delta}_1 =\tilde{\Delta}_4 =\Delta_s$ and $\tilde{\Delta}_2 =\tilde{\Delta}_3 =\Delta_\varphi$.

When evaluating the double-discontinuity deriving from the scalar conformal block, we can truncate the sum at $m=0$. The discarded terms at $m>0$ correspond to descendant states at higher twist, and their contribution is suppressed by the same mechanism that suppresses the contribution from higher-twist primary operators. Thus, keeping only $m=0$, we perform the sum over $p$ and expand in $\eps$ to obtain
\begin{align}
	\mathbb T_{\varphi\varphi ss}(z,\hb) &=\left( 1+(-1)^\ell \right)\kappa_{\hb}^{\varphi\varphi s s} \int_0^1\frac{d\zb}{\zb^2}k_{\hb}^{\varphi\varphi s s}(\zb)
	\nonumber \\
	& \quad \times \dDisc\left[\frac{2u}{n v} +\frac{\left(-u\log u +u\log v -\gamma_s u\log v \right)}{n v}\eps + \frac{a_s^{(1)} u}{v}\eps \right] +O(\eps^2).
	\label{eq:Tffsstoinvert}
\end{align}

\subsubsection{Extracting the CFT data in the pure channel}
To get the CFT data in the pure channel we expand \eqref{eq:Tffsstoinvert} in powers of $u$ and then compute the integral by means of the inversion dictionary; see appendix \ref{app:inversions}.
Focusing first on the leading twist contribution, $\tau_0=2$, we find
\begin{align}
	T^{(0)}_2 &=\frac{4\Gamma(\bar{h})^2}{n\Gamma(2\bar{h}-1)}\left( 1 -\eps\frac{6}{n+8}S_1(\bar{h}-1) +\eps\frac{a^{(1)}_s n}2\right) +O(\eps^2),
	\\
	T^{(1)}_2 &=-\eps\frac{4\Gamma(\bar{h})^2}{n\Gamma(2\bar{h}-1)} +O(\eps^2).
\end{align}
Extracting the data using \eqref{eq:inversion_CFTdata}, we get
\begin{align}
\label{eq:ffSSaahat}
	a_{2,\ell}&=\frac{4\Gamma(\ell+1)^2}{n\Gamma(2\ell+1)}\left( 1 -\eps\frac{(14+n)S_1(\ell)}{8+n} +\eps S_1(2\ell) -\eps \frac{n+2}{n+8} \right) +O(\eps^2),
	\\
\label{eq:ffSSanomdim}
	\gamma_{2,\ell}&=-\eps +O(\eps^2).
\end{align}

To proceed to subleading twists, we use the method explained in section~\ref{sec:projsubleadtwist}. For the first few twists, we find
\begin{align}\label{eq:S4formula}
	S_4^{(0)} &=\eps\frac{2(n+2)\Gamma(\bar{h})^2}{n(n+8)\Gamma(2\bar{h}-1)} +O(\eps^2),
	\\
	S^{(0)}_{\tau_0=6,8,\ldots} &=0+O(\eps^2),
\end{align}
and all $S^{(k)}_{\tau_0}=0$ for $k\geqslant1$.
This is in agreement with the result from \cite{Alday:2017zzv}, that the operators at twist 6 and higher have OPE coefficients that are of order $\eps^2$. For twist four, \eqref{eq:S4formula} straightforwardly gives
\begin{align}
	a_{4,\ell}&= \eps \frac{2(2+n)\Gamma(\ell+2)^2}{n(n+8)\Gamma(2\ell+3)} +O(\eps^2).
	\label{eq:ffssOPEcoeff}
\end{align}


\subsubsection{Resummation of the correlator}
We proceed with the resummation of the correlator $\G_{\varphi\varphi s s}(u,v)$ at order $\eps$. We use the subcollinear expansion \eqref{eq:blocksgenerald} for the conformal blocks as discussed in appendix~\ref{app:4-epsblocks}. The only operators that contribute to this order are the identity and the families of operators at twist 2 and 4, with data given by equations \eqref{eq:ffSSaahat}, \eqref{eq:ffSSanomdim} and \eqref{eq:ffssOPEcoeff}.

Since the inversion formula is not guaranteed to hold at spin zero, we are forced to introduce an additional contribution with support at spin zero only. From the free theory decomposition, we know that this does not affect the leading-order OPE coefficients, $a_{2,0}=\frac4n+O(\eps)$.
For the dimension and the OPE coefficient corrections, we supplement finite support contributions:
\begin{align}
	a_{2,\ell}&=\frac{4\Gamma(\ell+1)^2}{n\Gamma(2\ell+1)}\left( 1 -\eps\frac{(n+14)S_1(\ell)}{n+8} +\eps S_1(2\ell) -\eps \frac{n+2}{n+8} +\eps \tilde\alpha \delta_{\ell,0} \right) +O(\eps^2),
	\label{eq:ffssOPEfinal}
	\\
	\gamma_{2,\ell}&=-\eps + \eps \frac{n+2}{n+8} \delta_{\ell,0} +O(\eps^2).
	\label{eq:ffssCFTdatafinal}
\end{align}
Here we have used the fact that the dimension of the twist-2, spin-0 operator $s=\varphi^2_S$ is known from the $\langle\varphi\varphi\varphi\varphi\rangle$ correlator, the non-trivial solution in \eqref{eqgsgtsol}, whereas for the OPE coefficients we have introduced an unknown finite-support solution. At twist 4, we assume that the OPE coefficient formula extends to spin zero -- we will comment more on this assumption in section~\ref{sec:finitespincontributions}.

In summary, the correlator at order $\eps$ is given by the infinite sum
\begin{equation}
	\G_{\varphi\varphi s s}(z,\bar{z}) =1 +\sum_{\ell=0,2,\dots} \left( a_{2,\ell} G_{2+\gamma_{2,\ell}, \ell}(z, \bar z) + a_{4,\ell} G_{4, \ell}(z, \bar z) \right),
\end{equation}
where 1 represents the contribution of the identity operator and the other contributions are given by \eqref{eq:ffssOPEfinal}, \eqref{eq:ffssCFTdatafinal} at twist 2 and \eqref{eq:ffssOPEcoeff} at twist 4.
The result takes the following form,
\begin{align}
	\mathcal{G}_{\varphi \varphi ss} (u,v)&=1+\frac2n\left(u+\frac uv\right)+\eps\left[-\frac{u(1+v)}{nv}\log u+\frac{u(6+(n+2)v)}{n(n+8)v}\log v \right.
	\nonumber \\
	&\quad-\left. \frac{2(n+2)u}{n(n+8)}\Phi(u,v)-\frac{2(n+2)u(1+v)}{n(n+8)v}
	\right]
	\nonumber \\
	&\quad+4\eps \left(\tilde\alpha+\frac{n+2}{n+8}\right)G^{(d=4)}_{2,0}(u,v)+O(\eps^2),
	\label{eq:GffssAns}
\end{align}
where $\Phi(u,v)$ is the box function defined in \eqref{eq:boxfunction}.\footnote{Since the contribution proportional to $\tilde \alpha$ is already at order $\eps$, we have written the finite-spin contribution at the last line using the four-dimensional block, evaluated at dimension $\Delta=2$ and spin $\ell=0$. Explicitly $G^{(d=4)}_{2,0}(u,v)=\frac{z\zb}{z-\zb}\log(\frac{1-\zb}{1-z})$.} At this point, we will not be able to determine the value of $\tilde \alpha$. But in the next step, when we have written the result for the correlator $\expv{ssss}$, we will find that consistency with crossing will fix its value to
\begin{equation}
	\tilde{\alpha} =-\frac{n+2}{n+8}.
\end{equation}

\subsubsection{Decomposition in the mixed-channel}
\label{subsec:sffs}
Having found the correlator \eqref{eq:GffssAns}, we can use crossing to obtain
\begin{equation}
\G_{s\varphi\varphi s}(u,v)=\frac{u^{\frac{\Delta_s+\Delta_\varphi}2}}{v^{\Delta_\varphi}}\G_{\varphi\varphi s s}(v,u).
\end{equation}
We can then use the mixed-channel conformal blocks to find the CFT data to order $\eps$. We find that only the operator $\varphi$ at twist $1$, and operators at twist $\tau=3$ contribute.
For the twist-3 operators we find
\begin{align}
\label{eq:OPEasffsT3}
a^{(0)}_{3,\ell}&=\frac{2\Gamma(\ell+2)^2(2(-1)^\ell+n(\ell+1))}{n\Gamma(2\ell+3)}
,\\
\nonumber
\hat\alpha_{3,\ell}&=\frac1{2(n+8)(2(-1)^\ell+n(\ell+1))}\bigg(
12n\ell-n(n+20)(\ell+1)S_1(\ell)
\\&\quad-(-1)^\ell\left(24S_1(\ell)+\frac{4\ell(n+2)+3(n+8)}{\ell+1}\right)
\bigg),
\label{eq:alphaasffsT3}
\\
\gamma_{3,\ell}&=\frac1{(n+8)(2(-1)^\ell+n(\ell+1))}\left((n-16)(-1)^\ell+\frac{4(n+2)}{\ell+1}-\frac{n(n+20)(\ell+1)}2\right)\eps+O(\eps^2),
\label{eq:gammaasffsT3}
\end{align}
and the full OPE coefficients are given by \eqref{eq:hatalphadef}, where the partial derivative is not acting on the $\ell$ dependence in $(-1)^\ell$. At spin $\ell=1$, there is no degeneracy, and our formula reproduces the known anomalous dimension, and gives a new result for the OPE coefficient:
\begin{equation}
\gamma_{3,1}=-\frac32\eps+\frac{n+2}{n+8}\eps+O(\eps^2), \qquad \lambda^2_{\varphi s \partial\varphi^3_V}=\frac{2(n-1)}{3n} \left(1 - \frac{n+11}{3(n+8)} \eps\right)+O(\eps^2).
\end{equation}

\subsection[$\expv{ssss}$ correlator]{$\boldsymbol{\expv{ssss}}$ correlator}\label{sec:ssss}
In this section we compute the correlator $\G_{ssss}(u,v)$ at order $\eps$ for general $n$. 
Contrary to the previous section, an infinite family of operators now contributes to the double-discontinuity, namely the family of operators of approximate twist 2.
The data for all higher-twist operators can be determined from twist-two data alone, using the inversion formula. The twist-two data, in turn, is fixed by our previous considerations. 
For an alternative approach, taken in  \cite{Henriksson:2017eej}, the required twist-two data could be specified with a finite ansatz that depends only on a few undetermined parameters.\footnote{\cite{Henriksson:2017eej} considered the general problem of determining the four-point function of identical operators of dimension $2+O(g)$ in $d=4$. The relation to the $O(n)$ model in the $\eps$-expansion was discussed in \cite{Henriksson:2020jwk}.} 

In this section, we define
\begin{equation}
	a_{\tau,\ell} =a^{(0)}_{\tau,\ell}(1+\eps \alpha_{\tau,\ell}) +O(\eps^2),
\end{equation}
and we will work with the leading order $a^{(0)}_{\tau,\ell}$ and order-$\eps$ correction $\alpha_{\tau,\ell}$. For the anomalous dimension we use the same notation as above.

\subsubsection{Twist-two CFT data and inversion}

The only irrep appearing in this correlator is $S$. The inversion formula reads
\begin{equation}\label{eq:Tssss}
	\mathbb T_{ssss}(z,\hb) =\left( 1+(-1)^\ell \right)\kappa_{\hb}^{ssss} \int_0^1\frac{d\zb}{\zb^2}k_{\hb}^{ssss}(\zb)\dDisc\left[\left(\frac uv \right)^{2-\eps +\gamma_s\eps}\G_{ssss}(v,u)\right].
\end{equation}
The correlator that appears in the inversion integral is equal to the one we are trying to compute. The strategy is to write it as a sum of conformal blocks and then isolate the terms contributing to the double-discontinuity. The latter is completely determined by its contribution from the identity, and from operators at twist 2.\footnote{This is easy to see by simply counting powers of $z$, which become $1 -\bar{z}$ upon crossing. The $u/v$ crossing factor gives (slightly less than) two inverse powers of $v \sim (1 -\bar{z})$, so anything that is proportional to $z^2$ or higher will be regular after crossing. In the collinear limit the blocks behave as $G(z,\bar{z})=z^{\tau/2} k_{\bar{h}} (\bar{z}) +O(z^{\tau/2+1})$, and therefore it is clear that only twist 2 or lower contribute.}

To proceed, we need to know the $d=4-\eps$ OPE coefficients and anomalous dimensions of the exchanged operators at twist 2. For this purpose, we will use the results of the previous sections for $\expv{\varphi\varphi\varphi\varphi}$ and $\expv{\varphi\varphi ss}$. 
In particular, since at twist 2 all the operators -- $s$ and $\mathcal{J}_{S,\ell}$ -- are non-degenerate, the OPE coefficients \eqref{eq:ffssOPEfinal} can be expressed as
\begin{align}
	a^{\varphi\varphi ss}_{2,\ell} =\sqrt{a^{\varphi\varphi\varphi\varphi}_{2,\ell} a^{ssss}_{2,\ell}},
\end{align}
which implies
\begin{align} a^{ssss}_{2,\ell}=\frac{\left( a^{\varphi\varphi ss}_{2,\ell} \right)^2}{a^{\varphi\varphi\varphi\varphi}_{2,\ell}},
	\label{}
\end{align}
where $a^{\varphi\varphi\varphi\varphi}_{2,\ell}$ are the OPE coefficients at twist 2 \eqref{eq:ffS2l} associated to the $S$ irrep. Expanding at order $\eps$ we obtain
\begin{align}
	a^{(0)}_{2,\ell}&=\frac{8\Gamma(\ell+1)^2}{n\Gamma(2\ell+1)},
	\label{eq:sssstwist2inputa0}
	\\
	\alpha_{2,\ell} &=-\frac{2(n+2)}{n+8} -\frac{12}{n+8}S_1(\ell) +S_1(2\ell) + \left(\frac{n+2}{n+8} +2\tilde{\alpha} \right)\delta_{\ell,0},
	\label{eq:sssstwist2inputalpha}
	\\
	\gamma_{2,\ell} &=-\eps +\eps \frac{n+2}{n+8} \delta_{\ell,0} +O(\eps^2),
	\label{eq:sssstwist2inputgamma}
\end{align}
where as expected $a^{(0)}_{2,\ell}$ agrees with \eqref{eq:ssssa02ell} and $\tilde{\alpha}$ is the constant of \eqref{eq:ffssOPEfinal}.

Now we need to determine the form of the twist-2 contribution to the correlator.
We use the $d=4-\eps$ conformal block of appendix \ref{app:4-epsblocks}, which for $\tau=2+O(\eps)$ at order $z$ reduces to the collinear block $z^{\tau/2}k_{\tau/2+\ell}(\bar{z})$. Thus we define
\begin{equation}\label{eq:defFzlogz}
	\sum_\ell a^{(0)}_{2, \ell}(1 + \eps \alpha_{2, \ell}) z^{1+\gamma_{2,\ell}/2} k_{1 +\ell +\gamma_{2,\ell}/2} (\bar{z}) +O(z^2)= z F(\bar{z}, \log z) +O(z^2).
\end{equation}
Together with the contribution from the identity operator, it accounts for the entire double-discontinuity to order $\eps$.
Note that $F(\bar{z}, \log z)$ is the coefficient on the leading power in $z$, so it depends on $\log z$ but not on $z$ directly. We will break $F(\bar{z}, \log z)$ into its leading order and order-$\eps$ pieces,
\begin{equation}
	F(\bar{z}, \log z) =F^{(0)}(\bar{z}, \log z) +\eps F^{(1)}(\bar{z}, \log z).
\end{equation}
Using the data \eqref{eq:sssstwist2inputa0}--\eqref{eq:sssstwist2inputgamma}, we can compute these pieces exactly. The result is
\begin{align}
	F^{(0)}(\bar{z}, \log z) &= \frac{4 \bar{z} (\bar{z} - 2)}{n(\bar{z}-1)},
	\nonumber \\
	F^{(1)}(\bar{z}, \log z) &=-\frac{16}{n}\tilde{\alpha}\log (1-\bar{z}) +\frac{8(\bar{z}^2 -6\bar{z} +4)+2n(2\bar{z}^2 -9\bar{z}+8)}{n(n+8)(\bar{z}-1)}\log(1-\bar{z})
	\nonumber \\
	&\quad +\left( -\frac{2\bar{z}(\bar{z}-2)}{n(\bar{z}-1)} -\frac{4(n+2)}{n(n+8)}\log (1-\bar{z}) \right)(\log z +\log \bar{z})
	\nonumber \\
	&\quad -\frac{8(n+2)\bar{z}(\bar{z}-2)}{n(n+8)(\bar{z}-1)} -\frac{8(n+2)}{n(n+8)} \mathrm{Li}_2 (\bar{z}).
	\label{}
\end{align}

Next we would like to compute the double-discontinuity in \eqref{eq:Tssss}. It is given by the identity operator, and the twist-2 contribution $F(\bar{z}, \log z)$ with cross-ratios exchanged under crossing:
\begin{align}
	\mathbb T_{ssss}(z,\hb) &=\left( 1+(-1)^\ell \right)\kappa_{\hb}^{ssss} \int_0^1\frac{d\zb}{\zb^2}k_{\hb}^{ssss}(\zb)
	\nonumber \\
	&\quad \times \dDisc\left[\left(\frac{z \bar{z}}{(1-z)(1-\bar{z})} \right)^{2-\eps +\gamma_s\eps} \Big( 1+(1-\bar{z})F\big(1-z,\log(1-\bar{z})\big) \Big) \right].
	\label{eq:TsssswithF}
\end{align}
The argument of the double-discontinuity can be expanded in $\eps$ and, in principle, at arbitrary order in powers of $z$ to compute the inversion at any twist using the method of section~\ref{sec:projsubleadtwist}. At twist 2 and spin greater than zero we correctly recover the input \eqref{eq:sssstwist2inputa0}, \eqref{eq:sssstwist2inputalpha}, \eqref{eq:sssstwist2inputgamma}. As an example, we present the $d=4-\eps$ CFT data at twist 4:
\begin{align}
	a^{(0)}_{4,\ell} &=\frac{2\big( n(\ell+2)(\ell+1)+4 \big)\Gamma(\ell+2)^2 }{n\Gamma(2\ell+3)},
	 \\
	\alpha_{4,\ell} &=\frac{4}{(n+8)\big(n(\ell+2)(\ell+1)+4\big)}\Big(n(3\ell^2+6\ell+2) -2
	\nonumber \\
	& \quad -\big(n(6\ell^2+18\ell+17)+28 \big)S_1(\ell+1) +3\big( n(\ell^2+3\ell+4)+8 \big)S_1(2\ell+2)
	\nonumber \\
	& \quad +4(n+2)S_1(\ell+1)^2 -4(n+2)S_1(\ell+1) S_1(2\ell+2) -2(n+2)S_2(\ell+1)
	\nonumber \\
	& \quad -8\tilde\alpha (n+8)\big( S_1(\ell+1) -S_1(2\ell+2) \big) \Big),
	\label{eq:alpha4ellssss}
	 \\
	\gamma_{4,\ell} &=\frac{4\big(-24-3n( \ell(\ell+3) +4)-8(n+8)\tilde{\alpha} +4(n+2)S_1(\ell+1) \big)}{(n+8)\big( n(\ell+2)(\ell+1)+4 \big)} \eps +O(\eps^2),
	\label{eq:ssss4-epstwist4data}
\end{align}
where $S_2(x)$ is the second harmonic number.
The strategy would now be to keep going for twist 6, etc. However, we find that this quickly becomes technically difficult because it involves the coefficients $c_{k,m}$ that determine the subcollinear expansion of the conformal blocks, which grow very rapidly in size. Fortunately, since we are working in the vicinity of four dimensions, we are able to circumvent this problem by making use of the four-dimensional conformal blocks, which take the simple compact form \eqref{eq:4dblock}.

\subsubsection{Shortcut: inversion and decomposition using 4d blocks}\label{sec:shortcutto4d}
To solve the aforementioned problem, we will proceed with the computation working with four-dimensional conformal blocks. The advantage is that the projections to subleading twists in the inversion formula takes a very simple form. Since we are working only at order $\eps$, the dimensional corrections induced from this procedure can be completely disentangled -- in practice they will show up as an additional contribution to the OPE coefficients at order $\eps$ -- while the contribution from anomalous dimensions, already at order $\eps$, will be unaffected. To this end, we induce a shift
\begin{equation}
\label{eq:deltaalphaMain}
\alpha^{(d=4-\eps)}_{\tau,\ell} =\alpha^{(d=4)}_{\tau,\ell} +\Delta \alpha_{\tau,\ell},
\end{equation}
where $\alpha_{\tau,\ell}^{(d=4-\eps)}$ represents the physical (true) OPE coefficient corrections, and $\alpha^{(d=4)}_{\tau,\ell}$ represents the apparent OPE coefficient corrections when using four-dimensional blocks.
As we explain in appendix \ref{app:dimensionalshift}, the shifts $\Delta\alpha_{\tau,\ell}$ can be found in an exact form. In fact, since we are working at leading order in $\eps$, they can be determined by decomposing the free theory correlator in two different ways: using 4d blocks and using the true, ($4-\eps$)d blocks.

Consider the conformal block decomposition in the presence of a small parameter $\eps$, working to linear order. To capture the contribution of the reference twist $\tau_0$ to the power $z^{\tau_0/2}$, we define three functions (compare with \eqref{eq:blocksgenerald})
\begin{equation}
	\sum_\ell a^{(0)}_{\tau_0,\ell}\left( 1+\eps \hat{\alpha}_{\tau_0,\ell} +\gamma_{\tau_0,\ell} \frac{\log z}{2} \right)k_{\frac{\tau_0}{2}+\ell}(\bar{z}) =F_{\tau_0}(\bar{z}) +\eps H_{\tau_0}(\bar{z}) +\eps G_{\tau_0}(\bar{z}) \frac{\log z}{2},
\end{equation}
where we factored out $z^{\tau_0/2}$ and the functions $F_{\tau_0}(\bar{z})$, $H_{\tau_0}(\bar{z})$ and $G_{\tau_0}(\bar{z})$ are defined up to regular (\emph{i.e.} vanishing $\dDisc$) terms. The shifted OPE coefficients $\hat{\alpha}_{\tau_0,\ell}$ are defined in \eqref{eq:hatalphadef}.

As discussed in section~\ref{sec:projsubleadtwist}, for a given reference twist $\tau_0$, smaller twists will in general contribute to the same power $z^{\tau_0/2}$. In order to isolate the contribution of smaller twists we have to understand the contribution of a twist $\tau_0$ to the power $z^{k+\tau_0/2}$, with $k=0,1,2,\dots$. This is simple when using the four-dimensional blocks. Specifically, using
\begin{equation}\label{eq:4dblocktauell}
	G^{(d=4)}_{\tau,\ell}(z,\bar{z}) =\frac{z \bar z}{z - \bar z} \left( k_{\frac{\tau}{2} + \ell}(z) k_{\frac{\tau}{2} - 1}(\bar z) - k_{\frac{\tau}{2} + \ell}(\bar z) k_{\frac{\tau}{2} -1}(z)\right)
\end{equation}
we see that
\begin{align}
	&\left(\sum_\ell a^{(0)}_{\tau_0,\ell}\left( 1+\eps \hat{\alpha}^{(d=4)}_{\tau_0,\ell} +\gamma_{\tau_0,\ell} \partial_\tau \right)\left[ \frac{z \bar{z}}{\bar{z}-z}k_{\frac{\tau}2 -1}(z) \right]\bigg|_{\tau=\tau_0} k_{\frac{\tau_0}{2}+\ell}(\bar{z})\right)\Bigg|_{z^{k+\frac{\tau_0}2}}
	\nonumber \\
	&\quad=\left( \big( F_{\tau_0}(\bar{z}) +\eps H_{\tau_0}(\bar{z}) +\eps G_{\tau_0}(\bar{z}) \partial_\tau \big) \left[ \frac{z \bar{z}}{\bar{z}-z}k_{\frac{\tau}2 -1}(z) \right]\bigg|_{\tau=\tau_0} \right)\Bigg|_{z^{k+\frac{\tau_0}2}},
	\label{}
\end{align}
where $F_{\tau_0}(\bar{z})$, $H_{\tau_0}(\bar{z})$ and $G_{\tau_0}(\bar{z})$ are the same as above.

The strategy is to use $d=4$ CFT data and conformal blocks to compute the correlator (or rather the non-zero $\dDisc$ part of it) at each power $z^{T/2}$ from all twists $\tau_0=2,4,\dots,T$, namely
\begin{equation}
	\G(z,\bar{z}) \big|_{z^{\frac T2}} =\sum_{\tau_0 \leqslant T} \left( \big( F_{\tau_0}(\bar{z}) +\eps H_{\tau_0}(\bar{z}) +\eps G_{\tau_0}(\bar{z}) \partial_\tau \big) \left[ \frac{z \bar{z}}{\bar{z}-z}k_{\frac{\tau}2 -1}(z) \right]\bigg|_{\tau=\tau_0} \right)\Bigg|_{z^{\frac T2}}
	.
\end{equation}
From this we derive
\begin{align}
	&F_T(\bar{z}) +\eps H_T(\bar{z}) +\eps G_T(\bar{z}) \frac{\log z}{2}
	\nonumber \\
	&=\G(z,\bar{z}) \big|_{z^{\frac T2}} -\sum_{\tau_0 < T} \left( \big( F_{\tau_0}(\bar{z}) +\eps H_{\tau_0}(\bar{z}) +\eps G_{\tau_0}(\bar{z}) \partial_\tau \big) \left[ \frac{z \bar{z}}{\bar{z}-z}k_{\frac{\tau}2 -1}(z) \right]\bigg|_{\tau=\tau_0} \right)\Bigg|_{z^{\frac T2}},
	\label{}
\end{align}
and $a^{(0)}_{T,\ell}$, $a^{(0)}_{T,\ell}\hat{\alpha}^{(d=4)}_{T,\ell}$, $a^{(0)}_{T,\ell}\gamma_{T,\ell}$ can be found from $F_T(\bar{z})$, $H_T(\bar{z})$, $G_T(\bar{z})$ using the $\mathrm{SL}(2,\R)$ inversion formula \eqref{eq:SL2Rformula}.

Instead of the correlator $\G_{ssss}(z,\bar{z})$ we will use the argument of the double-discontinuity in (\ref{eq:TsssswithF}): for each twist $T$, we expand that argument to order $z^{T/2}$ and subtract the contribution of lower twists to obtain the three functions $F_T(\bar{z})$, $G_T(\bar{z})$ and $H_T(\bar{z})$. To perform the $\mathrm{SL}(2,\R)$ inversions we need a number of inversion results for specific terms, which we collect in the inversion dictionary of appendix \ref{app:inversions}. Since the projection to higher twists reduces to expanding the function $\frac{z \bar{z}}{\bar{z}-z}k_{\frac{\tau}2 -1}(z)$ and its derivatives, the method can be implemented for arbitrarily high twists.

The results for $a^{(0)}_{\tau,\ell}$ agree with \eqref{eq:ssssa02ell}--\eqref{eq:ssssa0tauell}, and the anomalous dimensions take the form
\begin{align}
	\gamma_{\tau,\ell} &=\frac{8 (n+2) \left(2 S_1(\ell+\frac{\tau}{2}-1)-2 S_1(\frac\tau{2}-2)+(-1)^{\tau/2}\right)}{(n+8) \left((\ell+1) n (\ell+\tau-2)+4 (-1)^{\tau/2}\right)}-\frac{12}{n+8}
	\nonumber \\
	&-\frac{32}{\left( 4(-1)^{\tau/2} +n(\tau-2+(\tau-1)\ell +\ell^2) \right)}\left(\tilde\alpha+\frac{n+2}{n+8}\right)
	\label{}
\end{align}
for $\tau=4,6,\ldots$.
For the corrections to the OPE coefficients, the most compact way to present the results of the inversion is to give $\hat\alpha^{(4d)}_{\tau,\ell}$, which for $\tau=4,6,\ldots$ takes the form
\begin{align}
\nonumber
\hat\alpha^{(4d)}_{\tau,\ell} &= \frac{12}{n+8}\bigg(1-2S_1(\tfrac\tau2 -2)-S_1(\tfrac\tau2+\ell-1)+S_1(\tau-4)\bigg)
\\
&\quad+\frac{1}{\frac4n(-1)^{\tau/2}+(\ell+1)(\ell+\tau-2)}\bigg[
\frac{6(\tau-3)}{n+8}-\frac8n\frac{n+2}{n+8}\bigg(2 S_1(\tfrac\tau2+\ell-1) S_1(\tau -4)
\nonumber \\
&\quad
-2 S_1(\tfrac{\tau }{2}-2) (S_1(\tfrac\tau2+\ell-1)+S_1(\tau -4))+2 S_1(\tfrac{\tau }{2}-2)^2-S_2(\tfrac{\tau }{2}-2)+\zeta_2\bigg)
\nonumber \\
&\quad +\frac{(-1)^{\tau/2}}{n(n+8)}\bigg(
3 n \delta _{4,\tau }+4 (n+8) S_1(\tfrac\tau2+\ell-1)
-8 n S_1(\tau -4)+12 (n+4) S_1(\tfrac{\tau }{2}-2)
\nonumber \\
&\quad
-8 (n+8)-16 S_1(\tau -4)
\bigg)
\bigg]
\nonumber
\\&\quad-\frac{32}{n(\frac4n(-1)^{\tau/2}+(\ell+1)(\tau+\ell-2))}\bigg(S_1(\tfrac\tau2-2)-S_1(\tau-4)\bigg)\left(\tilde\alpha+\frac{n+2}{n+8}\right)
,
\end{align}
where $\zeta_2=\frac{\pi^2}6$.
Ultimately, the physical OPE coefficients can be found by using first \eqref{eq:hatalphadef} within the $d=4$ data, and then apply the shift \eqref{eq:deltaalphaMain}.

Finally, let us emphasize that the use of four-dimensional blocks in this section is purely a method to simplify the computations. We have checked for the first few twists that the outcome of using the 4d block and the shift $\Delta\alpha_{\tau,\ell}$ agrees with the general method using ($4-\eps$)d blocks.

\subsubsection{Resummation of the correlator}
Now we proceed with the determination of a closed-form expression for the correlator. We can use four-dimensional conformal blocks and four-dimensional data, because the difference between resumming with 4d and ($4-\eps$)d CFT data and conformal blocks is zero at linear order in $\eps$.
Thus the correlator is given by the infinite sum
\begin{equation}
	\G_{ssss}(z,\bar{z}) =1 +\sum_{\substack{\tau=2,4,\dots \\ \ell=0,2,\dots}} a^{(0)}_{\tau,\ell}\left(1+\eps \alpha^{(d=4)}_{\tau, \ell}\right) G^{(d=4)}_{\tau+\gamma_{\tau,\ell}, \ell}(z, \bar z) +O(\eps^2),
\end{equation}
where we have isolated once again the contribution of the identity operator. The result is
\begin{align}
	\mathcal{G}_{ssss} (u,v)&=1+u^2+\frac{u^2}{v^2}+\frac 4n\left( u+\frac uv +\frac{u^2}v \right) +\eps\left[-\frac{8(n+2)u(u+v+1)}{n(n+8)v} \right.
	\nonumber \\
	&\quad -\frac{2u\big( nv(v+1)+8v(u+v+1)+nu(3v^2 -2v+3) \big)}{n(n+8)v^2}\log u
	\nonumber \\
	&\quad -\frac{2u\big(-4v(u+v+1)+n(-3u-2v^2+v+uv) \big)}{n(n+8)v^2}\log v
	\nonumber \\
	&\quad \left. -\frac{4(n+2)u(u+v+uv)}{n(n+8)v}\Phi(u,v) \right] +\eps \tilde\G(u,v) +O(\eps^2),
	\label{eq:Gssssfinal}
\end{align}
where
\begin{align}
	\tilde\G(z,\bar{z})&=-\left( \tilde\alpha +\frac{n+2}{n+8} \right)\left(\frac{32 u^3\Phi(u,v)}{n\big( (u-v+1)^2 -4u \big)}
	+\frac{16 u^2\big( (1-v)^2-u(v+1) \big)}{nv\big( (u-v+1)^2 -4u \big)}\log u \right.
	\nonumber \\
	&\quad \left. +\frac{16 u^2 (u-v+1)}{n\big( (u-v+1)^2 -4u \big)}\log v +\frac{16z \bar{z}}{n(z-\zb)}\log \Bigl( \frac{1-z}{1-\zb}\Bigr) \right).
\end{align}
As we anticipated, the term $\tilde\G(z,\bar{z})$ that we have separated out is not crossing-invariant. In fact, by explicit computations, we find that $\mathcal{G}_{ssss} (u,v)-\left( \frac uv \right)^{2-\eps+\gamma_s \eps}\mathcal{G}_{ssss} (v,u)$ is proportional to $\big(\tilde\alpha+\frac{n+2}{n+8}\big)$.
This leads to the conclusion
\begin{equation}
	\tilde\alpha =-\frac{n+2}{n+8},
\end{equation}
\emph{i.e.} $ \tilde\G(u,v) =0$. Therefore the correlator is given by \eqref{eq:Gssssfinal} without the $\tilde\G(u,v)$ term.

\subsubsection{Finite-spin contributions}
\label{sec:finitespincontributions}

In the computation above, we allowed for a finite-spin contribution at twist 2, parametrized by the constant $\tilde\alpha$. Since $\tilde\alpha$ appears in the twist-2 data it contributes to the double-discontinuity, and indeed we see that the CFT data of higher-twist operators depends on $\tilde\alpha$.

We did not introduce any similar finite-spin contribution at higher twists, and we shall now comment on why. It is clear that adding a finite-spin contribution to the higher twist data would never contribute to the double-discontinuity. It must also be a perturbation to the correlator that does not destroy crossing invariance.
In \cite{Heemskerk:2009pn}, such crossing-compatible finite-spin contributions were considered that are perturbations to the correlator of a generalized free field of dimension $\Delta$. Allowing only for spin 0 and working in $d=4$,\footnote{The corresponding expression for general $d$ was found in \cite{Fitzpatrick:2010zm}.} any such solution has anomalous dimensions of the form
\begin{equation}
\label{eq:HPPSsoln}
\gamma_{p,\ell}=\mathrm{constant}\times\frac{(2 \Delta -1) (p+1) (\Delta +p-1) (2 \Delta +p-3)}{(\Delta -1) (2 \Delta +2 p-3) (2 \Delta +2 p-1)}\delta_{\ell,0}
,
\end{equation}
where $\gamma_{p,\ell}$ denotes the anomalous dimension at twist $2\Delta+2p$ and spin $\ell$. Moreover, in the same solution, $a_{p,\ell}^{(0)}\alpha_{p,\ell}=\frac12\partial_p(a_{p,\ell}^{(0)}\gamma_{p,\ell})$. The conclusion is that any such finite-spin solution must be proportional to the anomalous dimension at twist $2\Delta$. But we have found that the inversion formula gives the correct result at twist 4 and spin 0, and therefore conclude that we should not add any such finite-spin contribution.\footnote{Of course, our correlator is not identical to the GFF correlator, but the structure is very similar (and becomes the GFF correlator in the limit $n\to\infty$). It is reasonable to believe that also in the our case, any finite-spin solution, if it exists, must have a contribution to anomalous dimensions and not just OPE coefficients.}

Later on we shall consider other correlators, where also odd spin operators are exchanged. In principle, in these correlators there could be finite-spin contributions appearing at spin $\ell=1$. However, the Lorentzian inversion formula holds down to $\ell>\ell_*$, where $\ell_*$ is the Regge intercept, which in the $\eps$-expansion has been found to be below 1 \cite{Liu:2020tpf,Caron-Huot:2020ouj,CaronHuot2022}.

\section{Mixed $\boldsymbol\varphi$ $\boldsymbol s$ $\boldsymbol t$ system}
\label{sec:phi-s-t-system}

In this section we turn to the full system of correlators involving the operators $\varphi$, $s$ and $t$. 
The inversion integrals encountered in this section will be similar to those used in the previous section.

\subsection[$\expv{\varphi\varphi tt}$ correlator]{$\boldsymbol{\expv{\varphi\varphi tt}}$ correlator}
We begin by working out the computation of $\G_{\varphi\varphi tt}(u,v)$ at order $\eps$ for general $n$. The logic is equivalent to that of section \ref{sec:ffsssection}. We will again invert into the pure channel, exploiting the known results for $\langle\varphi\varphi\varphi\varphi\rangle$. The novelty of this computation is the appearance of more than one irrep in the pure channel. For each of these we perform the inversion to get the CFT data. We conclude by resumming to obtain the corresponding correlators, which together fix $\G_{\varphi\varphi tt}(u,v)$.

\subsubsection{Inversion into the pure channel}
In the pure channel we have to consider irreps in both of the tensor products
\begin{align}
	V \otimes V &= S \oplus T \oplus A,
	 \\
	T \otimes T &= S \oplus T \oplus A \oplus T_4 \oplus H_4 \oplus B_4
	.
\end{align}
As anticipated, we now have to deal with three irreps: $S$, $T$ and $A$. Thus the correlator will have a $S$-, $T$- and $A$-term, each with its own CFT data. Under the exchange $1\leftrightarrow2$, the $A$ irrep is odd, and therefore contains odd-spin operators, whereas the other two irreps are even and contain even-spin operators.

We now use a generalization of the Lorentzian inversion formula for a correlator of four externals in the irreps $R_i$. Here we will be completely generic and keep all the indices (we shall simplify this formula in a moment):
\begin{align}
	\mathbb T^{R;R_1,R_2;R_3,R_4}_{1234}( z,\hb)&=\kappa^{1234}_{\hb}\int_0^1\frac{d\zb}{\zb^2}k_{\hb}^{1234}(\zb)\dDisc\left[\frac{u^{\frac{\Delta_1+\Delta_2}2}}{v^{\frac{\Delta_2+\Delta_3}2}} \sum_{R'}
	M^{RR'}_{R_1R_2;R_3R_4} \G^{R'}_{3214}(v,u)\right]
	\nonumber \\
	&\quad + (-1)^\ell\kappa^{2134}_{\hb}\int_0^1\frac{d\zb}{\zb^2}k_{\hb}^{2134}(\zb)\dDisc\left[\frac{u^{\frac{\Delta_1+\Delta_2}2}}{v^{\frac{\Delta_1+\Delta_3}2}} \sum_{R'} M^{RR'}_{R_1R_2;R_3R_4}
	 \G^{R'}_{3124}(v,u)\right]
	 .
	\label{eq:T1234generalwithR}
\end{align}
Here $M^{RR'}_{R_1R_2;R_3R_4}$ denotes the crossing matrix introduced in section \ref{sec:crossingmatrix}.

Let us specialize to the case at hand, \emph{i.e.} $R_1=R_2 =V$, $R_3=R_4=T$, $R=S,T,A$. In the mixed channel after crossing we have
\begin{equation}
	T \otimes V = V \otimes T = V \oplus H_3 \oplus T_3,
\end{equation}
which implies that the sums in \eqref{eq:T1234generalwithR} are over $R'=V,H_3,T_3$. Moreover, $\Delta_1 =\Delta_2 =\Delta_\varphi$, $\Delta_3 =\Delta_4 =\Delta_t$, where
\begin{equation}
	\Delta_t = 2 -\eps +\gamma_t \eps +O(\eps^2),
\end{equation}
with $\gamma_t =\frac{2}{n+8}$. 

A word of caution is needed here. Before crossing, the second term of the inversion formula \eqref{eq:inversionintegralgen} contains $\G_{2134}(z,\bar{z})$. In the present case, when $R=A$, exchanging 1 and 2 produces a minus sign. Thus \eqref{eq:T1234generalwithR} simplifies to
\begin{align}
	\mathbb T^{R;V,V;T,T}_{\varphi\varphi tt}(z,\hb) &=\left( 1\pm(-1)^\ell \right)\kappa^{\varphi\varphi tt}_{\hb}\int_0^1\frac{d\zb}{\zb^2}k_{\hb}^{\varphi\varphi tt}(\zb)
	\nonumber \\
	&\quad \times \dDisc\left[\frac{u^{1-\frac{\eps}2}}{v^{\left( 3 -\frac32\eps +\gamma_t \eps \right)/2}} \sum_{R'=V,H_3,T_3} M^{RR'}_{VV;TT} \G^{R'}_{t\varphi\varphi t}(v,u)\right], \quad R=S,T,A.
	\label{eq:TRVVTT}
\end{align}
Here the plus sign holds for $S,T$, whereas the minus sign is for $A$. The $3\times 3$ crossing matrix $M^{R,R'}_{VV;TT}$ is given in \eqref{ffttcrossmatrix}.

We now need to determine the double-discontinuity, which we will find by writing $\G^{R'}_{t\varphi\varphi t}(v,u)$ as a sum over crossed-channel conformal blocks. A discussion similar to that in section \ref{ffssinversionsection} shows that, at order $\eps$, the contributions to the double-discontinuity of $H_3$ and $T_3$ are suppressed, since they only contain operators at twist 3 and higher. Thus only $R'=V$ contributes to the sum. Furthermore, among the operators in the $V$ irrep, only $\varphi$ contributes to order $\eps$. We write
\begin{equation}
	\G^V_{t\varphi\varphi t}(v,u) =\lambda_{t\varphi\varphi}^2G^{\Delta_t,\Delta_\varphi,\Delta_\varphi,\Delta_t}_{\Delta_\varphi,\ell=0}(v,u)+\text{terms with no dDisc at this order},
\end{equation}
where the block is given in \eqref{eq:scalarblockmixed} with $d=4-\eps$, $\tilde{\Delta}_1 =\tilde{\Delta}_4 =\Delta_t$ and $\tilde{\Delta}_2 =\tilde{\Delta}_3 =\Delta_\varphi$ and the OPE coefficient at order $\eps$ is (see \eqref{eq:undetconsts})
\begin{equation}
	\lambda^2_{t\varphi\varphi }=\lambda^2_{\varphi\varphi t}=\mathcal{N}_T \left(2+a_t^{(1)}\eps+O(\eps^2)\right),
\end{equation}
where $a_t^{(1)}$ is given in \eqref{eq:ffffas1at1} and $\mathcal{N}_T$ in \eqref{eq:normVVVV}.

Eventually \eqref{eq:TRVVTT} becomes
\begin{align}
	\mathbb T^{R;V,V;T,T}_{\varphi\varphi tt}(z,\hb) &=\left( 1\pm(-1)^\ell \right)\kappa^{\varphi\varphi tt}_{\hb}\int_0^1\frac{d\zb}{\zb^2}k_{\hb}^{\varphi\varphi tt}(\zb) M^{RV}_{VV;TT}\,\mathcal{N}_T
	\nonumber \\
	&\quad\times \dDisc\left[ \frac{2 u}v +\frac{ ( -u\log u +u\log v -u\gamma_t \log v )}{v}\eps +\frac{ a^{(1)}_t u}{v}\eps \right]
	 +O(\eps^2),
	\label{eq:TRfftt}
\end{align}
where $R=S,T,A$.
We see that the inversions for the three irreps are identical up to a factor $M^{RV}_{VV;TT}$, which means that we only will need to perform the integral once.

\subsubsection{Extracting the CFT data in the pure channel}
We proceed exactly as in section \ref{sec:ffsssection}, projecting to subleading twists.. Plugging in the values for the matrix elements we obtain the $d=4-\eps$ dimensional OPE coefficients at order $\eps$. As in $\expv{\varphi \varphi ss}$, at twist 6 and higher the OPE coefficients are of order $\eps^2$, \emph{i.e.} suppressed. Explicitly we obtain
\begin{align}
	a_{S,2,\ell}&=\frac{4\Gamma(\ell+1)^2}{n\Gamma(2\ell+1)}\left( 1 -\eps\frac{(14+2n)S_1(\ell)}{8+n} +\eps S_1(2\ell) -\eps\frac{2}{n+8} \right) +O(\eps^2),
		 \label{eq:ffttaS2ell}
	 \\
	a_{T,2,\ell}&=\frac{2\sqrt{2(n^2+2n-8)}\Gamma(\ell+1)^2}{\sqrt{n^2+n-2}\Gamma(2\ell+1)} \left( 1 -\eps\frac{(14+2n)S_1(\ell)}{8+n} +\eps S_1(2\ell) -\eps\frac{2}{n+8} \right) +O(\eps^2),		 \label{eq:ffttaT2ell}
	 \\
	a_{A,2,\ell}&=-\frac{2\sqrt{2n}\Gamma(\ell+1)^2}{\sqrt{n-1}\Gamma(2\ell+1)} \left( 1 -\eps\frac{(14+2n)S_1(\ell)}{8+n} +\eps S_1(2\ell) -\eps\frac{2}{n+8} \right) +O(\eps^2),		 \label{eq:ffttaA2ell}
	\\
	a_{S,4,\ell}&=\eps \frac{4\Gamma(\ell+2)^2}{n(n+8)\Gamma(2\ell+3)} +O(\eps^2),
	\label{eq:ffttaS4ell}
	 \\
	a_{T,4,\ell}&=\eps\frac{2\sqrt{2(n^2+2n-8)}\Gamma(\ell+2)^2}{(n+8)\sqrt{n^2+n-2}\Gamma(2\ell+3)} +O(\eps^2),	\label{eq:ffttaT4ell}
	 \\
	a_{A,4,\ell}&=-\eps\frac{2\sqrt{2n}\Gamma(\ell+2)^2}{(n+8)\sqrt{n-1}\Gamma(2\ell+3)} +O(\eps^2),	\label{eq:ffttaA4ell}
	 \\
	a_{R,k,\ell}&=O(\eps^2), \quad k=6,8,\dots.
\end{align}
It is important to recall that, due to the $\pm$ sign in \eqref{eq:TRfftt}, the OPE coefficients for $S$, $T$ have even integer $\ell$, while $A$ has odd integer $\ell$.

For the anomalous dimension at twist 2 we obtain
\begin{equation}\label{}
	\gamma_{R,2,\ell}=-\eps +O(\eps^2).
\end{equation}
Again, this data will be supplemented with its value at spin zero, introducing two new unknowns in the problem -- for the OPE coefficient at twist 2 in $S$ and $T$.

\subsubsection{Resummation of the correlator}
In this section we perform the resummation of the correlator for $S$, $T$ and $A$, using the conformal blocks \eqref{eq:blocksgenerald}. As usual, the $S$ irrep will contain the identity; and at twist 2, spin 0, the irreps $S$ and $T$ will have the operators $s$ and $t$ respectively.

\paragraph{$\boldsymbol{\G^S_{\varphi\varphi tt}(u,v)}$ correlator} We start from the OPE coefficients \eqref{eq:ffttaS2ell} at twist 2, and \eqref{eq:ffttaS4ell} at twist four, but supplement the twist-2 values with a compact support contribution at spin zero
\begin{align}
	a_{S,2,\ell} \, \to \, a_{S,2,\ell}+\frac4n\eps\tilde\alpha_S \delta_{\ell,0}
	.
	\end{align}
	For the anomalous dimensions, we use the expression \eqref{eq:ffssCFTdatafinal}.
Having obtained all the necessary contributions, we compute the infinite sum
\begin{equation}
	\G^S_{\varphi\varphi tt}(z,\bar{z}) =1 +\sum_{\ell=0,2,\dots} \left( a_{S,2,\ell} G_{2+\gamma^S_{2,\ell}, \ell}(z, \bar z) + a_{S,4,\ell} G_{4, \ell}(z, \bar z) \right)
\end{equation}
to obtain
\begin{align}
	\G^S_{\varphi\varphi tt}(u,v) &=1+\frac{2u(v+1)}{nv}+\eps\left( -\frac{4u(v+1)}{n(n+8)v} -\frac{u(v+1)}{nv}\log u \right.
	\nonumber \\
	&\quad \left. +\frac{u(2v+n+6)}{n(n+8)v}\log v -\frac{2(n+2)u}{n(n+8)}\Phi(u,v) \right)
	\nonumber \\
	&\quad +\eps \frac4n \left( \tilde\alpha_S +\frac{n+2}{n+8} \right)G^{(d=4)}_{2,0}(u,v) +O(\eps^2).
	\label{eq:GffttSfinal}
\end{align}

\paragraph{$\boldsymbol{\G^T_{\varphi\varphi tt}(u,v)}$ correlator} We start from the data OPE coefficients \eqref{eq:ffttaT2ell} and \eqref{eq:ffttaT4ell}, and supplement a finite-support contribution,
\begin{equation}
	a_{T,2,\ell} \, \to \, a_{T,2,\ell}+\frac{2\sqrt{2(n+4)(n-2)}}{\sqrt{(n+2)(n-1)}}\eps\tilde\alpha_T \delta_{\ell,0}
	.
	\end{equation}
For the anomalous dimensions we use
\begin{equation}
\gamma_{T,2,\ell}=-\eps+\eps\left( 1-\frac{n+6}{n+8}\right) \delta_{\ell,0} +O(\eps^2).
\end{equation}
We then find
\begin{align}
	\G^T_{\varphi\varphi tt}(u,v) &=\sqrt{\frac{2(n+4)(n-2)}{(n+2)(n-1)}}\frac{u(v+1)}{v}+\eps\sqrt{\frac{(n+4)(n-2)}{2(n+2)(n-1)}}\left( -\frac{4u(v+1)}{(n+8)v} \right.
	\nonumber \\
	&\quad \left. -\frac{u(v+1)}{v}\log u +\frac{u(2v+n+6)}{(n+8)v}\log v -\Phi(u,v)\frac{4u}{n+8} \right)
	\nonumber \\
	&\quad +\eps \sqrt{\frac{8(n+4)(n-2)}{(n+2)(n-1)}} \left( \tilde\alpha_T +\frac{2}{n+8} \right)G^{(d=4)}_{2,0}(u,v) +O(\eps^2).
	\label{eq:GffttTfinal}
\end{align}

\paragraph{$\boldsymbol{\G^A_{\varphi\varphi tt}(u,v)}$ correlator} Here there is no finite support contribution, and we get
\begin{align}
	\G^A_{\varphi\varphi tt}(u,v) &=\frac{\sqrt{2n}}{\sqrt{n-1}}\frac{u(v-1)}{v}+\eps\frac{\sqrt{n}}{\sqrt{2(n-1)}}\left( -\frac{4u(v-1)}{(n+8)v} \right.
	\nonumber \\
	&\quad \left. -\frac{u(v-1)}{v}\log u +\frac{u(2v-n-6)}{(n+8)v}\log v \right)+O(\eps^2).
	\label{eq:GffttAfinal}
\end{align}

In the next section we will be able to fix the constants in \eqref{eq:GffttSfinal} and \eqref{eq:GffttTfinal} through a consistency check; they are
\begin{align}
	\tilde\alpha_S =-\frac{n+2}{n+8}
,\qquad
	\tilde\alpha_T =-\frac{2}{n+8},
	\label{eq:ffttconstants}
\end{align}
and hence the terms proportional to the factor $G^{(d=4)}_{2,0}(u,v)$ vanish.

\subsection[{$\expv{tttt}$ correlator}]{$\boldsymbol{\expv{tttt}}$ correlator}

In this section we compute the correlator $\G_{tttt}(u,v)$ at order $\eps$ for general $n$. The logic is similar to that of $\expv{ssss}$, but we have more than one exchanged representation as in $\expv{\varphi\varphi tt}$. Since many of the technical aspects have been covered in these two sections, here we will sketch the computations and give the main results.

In this section, we use the notation $a_{\tau,\ell} =a^{(0)}_{\tau,\ell}(1+\eps \alpha_{\tau,\ell}) +O(\eps^2)$ for the OPE coefficients, and the definition of the anomalous dimension does not change compared to the previous section.

\subsubsection{Twist-two CFT data and inversion}
The relevant tensor product is
\begin{equation}\label{eq:TotimesT}
	T \otimes T =S \oplus T \oplus A \oplus T_4 \oplus H_4 \oplus B_4,
\end{equation}
so the correlator will be the sum of six pieces. The crossing matrix entering the Lorentzian inversion formula can be extracted from \cite{Reehorst:2020phk} and is given by \eqref{eq:Mtttt} in appendix~\ref{app:crossingM}. The spin $\ell$ is odd for $A$ and $H_4$ and even otherwise. 

For the three irreps $S$, $T$ and $A$ we can find the OPE coefficients at twist 2 (where no mixing occurs) from the previous correlators. We have
\begin{equation}
	a^{tttt}_{R,2,\ell} =\frac{\left( a^{\varphi\varphi tt}_{R,2,\ell} \right)^2}{a^{\varphi\varphi\varphi\varphi}_{R,2,\ell}}, \qquad R=S,T,A,
\end{equation}
where the denominator is given by \eqref{eq:ffS2l}, \eqref{eq:ffT2l} and \eqref{eq:ffA2l}.
From this we obtain
\begin{align}
	a^{(0)}_{S,2,\ell} &=\frac{8\Gamma(\ell+1)^2}{n\Gamma(2\ell+1)},
	\label{eq:tttta0S} \\
	\alpha_{S,2,\ell} &=-\frac{4}{n+8} -\frac{2(n+6)}{n+8}S_1(\ell) +S_1(2\ell) +\left( \frac{n+2}{n+8} +2\tilde{\alpha}_S \right) \delta_{\ell,0},
	\label{eq:ttttalphaS} \\
	\gamma_{S,2,\ell} &=-\eps +\eps\frac{n+2}{n+8} \delta_{\ell,0} +O(\eps^2),
	\\
	a^{(0)}_{T,2,\ell} &=\frac{4(n+4)(n-2)\Gamma(\ell+1)^2}{(n+2)(n-1)\Gamma(2\ell+1)},
	\label{eq:tttta0T} \\
	\alpha_{T,2,\ell} &=-\frac{4}{n+8} -\frac{2(n+6)}{n+8} S_1(\ell) +S_1(2\ell) +\left( \frac{2}{n+8} +2\tilde{\alpha}_T \right) \delta_{\ell,0},
	\label{eq:ttttalphaT} \\
	\gamma_{T,2,\ell} &=-\eps+\eps\frac{2}{n+8} \delta_{\ell,0} +O(\eps^2),
	\\
	a^{(0)}_{A,2,\ell} &=-\frac{4n\Gamma(\ell+1)^2}{(n-1)\Gamma(2\ell+1)},
	\label{eq:tttta0A} \\
	\alpha_{A,2,\ell} &=-\frac{4}{n+8} -\frac{2(n+6)}{n+8} S_1(\ell) +S_1(2\ell),
	\label{eq:ttttalphaA} \\
	\gamma_{A,2,\ell} &=-\eps +O(\eps^2).
\end{align}
We use this CFT data and the $d=4-\eps$ conformal block at order $z$ to define the three non-zero twist-2 contributions to the correlator,
\begin{equation}
	\sum_\ell a^{(0)}_{R,2, \ell}(1 + \eps \alpha_{R,2, \ell}) z^{1+\gamma_{R,2,\ell}/2}k_{1+\ell+\gamma_{R,2,\ell}/2}(\bar{z}) +O(z^2) = z F^R(\bar{z}, \log z) + O(z^2), \quad R=S,T,A.
\end{equation}
They are
\begin{align}
	F^S(\bar{z}, \log z) &=\frac{4\bar{z}(\bar{z}-2)}{n(\bar{z}-1)} +\eps \frac2n\left[ \left( -\frac{\bar{z}(\bar{z}-2)}{\bar{z}-1} -\frac{2(n+2)}{n+8}\log(1-\bar{z}) \right)\log (z\bar{z}) \right.
	\nonumber \\
	&\quad + \frac{-8\bar{z}(\bar{z}-2)+\big( n(8-9\bar{z})+4(\bar{z}^2 -6\bar{z} +4) \big)\log(1-\bar{z}) -4(n+2)(\bar{z}-1)\mathrm{Li}_2(\bar{z})}{(n+8)(\bar{z}-1)}
	\nonumber \\
	&\quad -\left. 8\tilde\alpha_S\log(1-\bar{z}) \right],
	\\
	F^T(\bar{z}, \log z) &=\frac{2(n+4)(n-2)\bar{z}(\bar{z}-2)}{(n+2)(n-1)(\bar{z}-1)} \nonumber \\
	&\quad+\eps \frac{(n+4)(n-2)}{(n+2)(n-1)}\left[ \left( -\frac{\bar{z}(\bar{z}-2)}{\bar{z}-1} \right. \right.
	-\left.\frac{4\log(1-\bar{z})}{n+8} \right)\log (z\bar{z})  -8\tilde\alpha_T\log(1-\bar{z})
	\nonumber \\
	&\quad -\left.\frac{8\bar{z}(\bar{z}-2)-\big( 4\bar{z}^2-(n+24)\bar{z}+16 \big)\log(1-\bar{z}) +8(\bar{z}-1)\mathrm{Li}_2(\bar{z})}{(n+8)(\bar{z}-1)} \right],
	\\
	F^A(\bar{z}, \log z) &=\frac{2n\bar{z}^2}{(n-1)(\bar{z}-1)}+\eps\frac{n}{n-1}\left[ -\frac{\bar{z}^2}{\bar{z}-1}\log (z\zb) -\frac{ 8\bar{z}^2 -\bar{z}(n+4\bar{z})\log(1-\bar{z}) }{(n+8)(\bar{z}-1)} \right].
	\label{}
\end{align}

The Lorentzian inversion formula can be deduced from \eqref{eq:TsssswithF} and \eqref{eq:TRVVTT}; it takes the form
\begin{align}
	\mathbb T^{R;T,T;T,T}_{tttt}(z,\hb)
	&=\left( 1\pm(-1)^\ell \right)\kappa^{tttt}_{\hb}\int_0^1\frac{d\zb}{\zb^2}k_{\hb}^{tttt}(\zb) \dDisc\left[\left(\frac{z \bar{z}}{(1-z)(1-\bar{z})} \right)^{2-\eps+\gamma_t\eps} \right.
	\nonumber \\
	&\quad \times \left. \sum_{R'} M^{RR'}_{TT;TT} \Big( \delta_{R,S}+(1-\bar{z})F^{R'} \big(1-z,\log(1-\bar{z})\big) \Big) \right],
	\label{eq:TRtttt}
\end{align}
where $ R=S,T,A,T_4,H_4,B_4$.
The $+$ sign corresponds to even ($S$, $T$, $T_4$, $B_4$) and the $-$ sign to odd ($A$, $H_4$) spins. The term with the Kronecker delta $\delta_{R,S}$ corresponds to the identity operator which is present only in the $S$ irrep.

Using the methods from the previous section, the CFT-data in all six representations can be computed case by case in $\tau$, giving results for OPE coefficients and anomalous dimensions. These lengthy expressions are not particularly illuminating so we omit them, proceeding instead to the determination of the correlators.\footnote{Once the correlators have been determined, the corresponding CFT-data can be extracted by performing a conformal block decomposition.}
Before doing this however, we would like to fix the free parameters $\tilde\alpha_S$ and $\tilde\alpha_T$. This can be done exploiting the knowledge coming from the free theory: there, we have that the OPE coefficients $a_{T_4,6,0}$ and $a_{B_4,4,0}$ -- which start at order $\eps$ for every $\ell$ -- vanish.\footnote{Actually, in the interacting theory, $T_4$ has an operator at twist 6, spin 0, but its OPE coefficient is suppressed at order $\eps$. For $B_4$, instead, it is impossible to construct any operator at all.} Imposing this constraint we obtain two equations for the two constants. From \eqref{eq:TRtttt} we find
\begin{align}
	a_{T_4,6,0} &=\eps\frac{104}{9(n+8)} +\eps\frac{52}{9n}\tilde\alpha_S +\eps\frac{26(n-2)}{9n}\tilde\alpha_T,
	\\
	a_{B_4,4,0} &=\eps\frac{8}{n+8} +\eps\frac{16}{n}\tilde\alpha_S -\eps\frac{4(n+4)}{n}\tilde\alpha_T,
	\label{}
\end{align}
which can be set simultaneously to zero to get\footnote{Note that the same result could have been achieved -- as in section \ref{sec:ssss} -- by keeping these constants until the end and noting that they are fixed by crossing symmetry.}
\begin{align}
	\tilde\alpha_S &=-\frac{n+2}{n+8},
	\qquad
	\tilde\alpha_T =-\frac{2}{n+8}.
\end{align}

\subsubsection{Resummation of the correlator}
As for $\expv{ssss}$, it is hard to compute the $d=4-\eps$ CFT data at high twist. We therefore expand \eqref{eq:TRtttt} in powers of $z$ and use the technique introduced in \ref{sec:shortcutto4d} to extract the four-dimensional CFT data to arbitrary twist. Again, it is possible to obtain the physical ($4-\eps$)d OPE coefficients by means of a formula similar to \eqref{eq:4dto4minusepsshift}.

Our final result is given by resumming using four-dimensional blocks to obtain six correlators, one for each irrep:
\begin{equation}
	\G^R_{tttt}(z,\bar{z}) =\delta_{R,S} +\sum_{\tau,\ell} a^{(0)}_{R,\tau,\ell}\left(1+\eps \alpha^{(d=4)}_{R,\tau, \ell}\right) G^{(d=4)}_{\tau+\gamma_{R,\tau,\ell}, \ell}(z, \bar z) +O(\eps^2), \quad R=S,T,A,T_4,H_4,B_4.
\end{equation}
Note that the range of $\tau$ depends on $R$ -- the sum starts at 2 for $S$, $T$, $A$ and 4 otherwise --, and that the values of $\ell$ is odd for $A$ and $H_4$, and even otherwise.

The result is
\begin{align}
	\G^S_{tttt}(u,v) &=\frac{n^3 v^2-8uv(u+v+1) +n^2 v(4uv+4u+v) +2n\big( u^2(v+1)^2 -v^2 +2uv(v+1) \big)}{n(n+2)(n-1)v^2}
	\nonumber \\
	&\quad +\eps\left( -\frac{4u\big( (n^2+4n-4)u(v+1)+(n+2)^2(n-1)v \big)}{n(n+2)(n-1)(n+8)v}\Phi(u,v) \right.
	\nonumber \\
	&\quad-\tfrac{2u\big(nu(6v^2+4v+6) -16v(u+v+1)+n^2(u+9v+uv)(v+1)+n(n^2+6)v(v+1) \big)}{n(n+2)(n-1)(n+8)v^2}\log u
	\nonumber \\
	&\quad+\tfrac{2u\big(n(n+6)u+(n+4)(n+2)(n-1)v+(n+4)(n-2)uv+4(n+2)(n-1)v^2 \big)}{n(n+2)(n-1)(n+8)v^2}\log v
	\nonumber \\
	&\quad \left. -\frac{16u\big((n-2)(u+v+1)+ n^2(v+1) \big)}{n(n+2)(n-1)(n+8)v} \right) +O(\eps^2),
	\label{eq:GttttS}
	\\
	\G^T_{tttt}(u,v) &=\frac{2u\big(n^2 v(v+1)+n(v+1)(u+2v+uv)-8v(u+v+1) \big)}{(n+2)(n-1)v^2}
	\nonumber \\
	&\quad+\eps\left( \frac{2u\big( -2(n+4)(n-2)v-(n^2+8n-16)u(v+1) \big)}{(n+2)(n-1)(n+8)v}\Phi(u,v) \right.
	\nonumber \\
	&\quad-\tfrac{u\big(2n^2(u+5v+uv)(v+1) +12nu(v^2+1)-64v(u+v+1)+ n(n^2+8)v(v+1) \big)}{(n+2)(n-1)(n+8)v^2}\log u
	\nonumber \\
	&\quad-\frac{u\big( 4n(3u+2v^2) +2n^2(2v^2+u+3v+uv) -32v(u+v+1) +n^3 v \big)}{(n+2)(n-1)(n+8)v^2}\log v
	\nonumber \\
	&\quad \left. -\frac{8u\big( 2(n-4)(u+v+1) +n^2(v+1) \big)}{(n+2)(n-1)(n+8)v}\right) +O(\eps^2),
	\label{eq:GttttT}
	\\
	\G^A_{tttt}(u,v) &=\frac{2nu(v-1)\big( u(v+1)+(n+2)v \big)}{(n+2)(n-1)v^2}
	\nonumber \\
	&\quad+\eps\left( -\frac{8nu(v-1)}{(n-1)(n+8)v} -\frac{2n(n+6)u^2(v-1)}{(n+2)(n-1)(n+8)v}\Phi(u,v)  \right.
	\nonumber \\
	&\quad-\frac{nu(v-1)\big( 2(n+6)u(v+1)+(n+2)(n+8)v \big)}{(n+2)(n-1)(n+8)v^2}\log u
	\nonumber \\
	&\quad \left. -\frac{nu\big( 2(n+6)u +(n+2)v(-4v+n+4) \big)}{(n+2)(n-1)(n+8)v^2}\log v \right) +O(\eps^2),
	\label{eq:GttttA}
	\\
	\G^{T_4}_{tttt}(u,v) &=\frac{u^2(v^2+4v+1)}{v^2} +\eps\bigg( -\frac{16u^2}{(n+8)v} 
	 -\frac{u^2\big(n+6+ (n+6)v^2+4(n+4)v \big)}{(n+8)v^2}\log u 
	 \nonumber \\
	&\quad
	 +\frac{u^2\big( n+6+2(n+4)v \big)}{(n+8)v^2}\log v -\frac{8u^2(v+1)}{(n+8)v}\Phi(u,v)\bigg) +O(\eps^2),
	\label{eq:GttttT4}
	\\
	\G^{H_4}_{tttt}(u,v) &=\frac{u^2(v^2-1)}{v^2} +\eps\bigg( -\frac{4u^2(v-1)}{(n+8)v}\Phi(u,v)
	\nonumber \\
	&\quad -\frac{(n+6)u^2(v+1)(v-1)}{(n+8)v^2}\log u -\frac{(n+6)u^2}{(n+8)v^2}\log v \bigg) +O(\eps^2),
	\label{eq:GttttH4}
	\\
	\G^{B_4}_{tttt}(u,v) &=\frac{u^2(v-1)^2}{v^2} +\eps\bigg( \frac{8u^2}{(n+8)v} -\frac{2u^2(v+1)}{(n+8)v}\Phi(u,v)
	\nonumber \\
	&\quad -\frac{u^2\big( n+6+(n+6)v^2-2(n+4)v \big)}{(n+8)v^2}\log u +\frac{u^2\big( n+6-(n+4)v \big)}{(n+8)v^2}\log v \bigg) +O(\eps^2).
	\label{eq:GttttB4}
\end{align}

\subsection[$\expv{sstt}$ correlator]{$\boldsymbol{\expv{sstt}}$ correlator}

In this section we compute the correlator $\G_{sstt}(u,v)$ at order $\eps$ for general $n$. As usual, we would like to compute the double-discontinuity and inversion integral for the pure channel. However, the contribution to this double-discontinuity is from mixed-channel twist-2 operators, whose OPE coefficients we have not yet determined. We will therefore first determine the mixed-channel data at twist 2 only, by solving the inversion problem for the mixed channel. Then we will turn to the pure channel and determine the data at all twists, and ultimately the correlator.
In this section, we again use the notation $a_{\tau,\ell} =a^{(0)}_{\tau,\ell}(1+\eps \alpha_{\tau,\ell}) +O(\eps^2)$ for the OPE coefficients.

\subsubsection{Inversion into the mixed channel}
\label{subsec:auxiliarymixedsstt}
Let us begin by explaining why we do not want to invert directly into the pure channel. As in $\expv{\varphi\varphi ss}$, in the pure channel only the $S$ irrep appears. The inversion formula greatly simplifies to
\begin{equation}\label{eq:ssttLIFpure}
	\mathbb T_{sstt}(z,\hb) =\left(1+(-1)^\ell\right)\kappa^{sstt}_{\hb}\int_0^1\frac{d\zb}{\zb^2}k_{\hb}^{sstt}(\zb)\dDisc\left[\frac{u^{\Delta_s}}{v^{\frac{\Delta_s+\Delta_t}2}}\G_{tsst}(v,u)\right].
\end{equation}
In the mixed channel $\G_{tsst}$, only the $T$ irrep appears on the right hand side. We would like to proceed as usual by replacing $\G_{tsst}(v,u)$ in $\dDisc$ by its conformal block decomposition at twist 2. The problem is that we do not know the associated OPE coefficients $\lambda^2_{ts\mathcal{J}^T_\ell}$ at order $\eps$. 
The first goal will therefore be to find these extract those OPE coefficients by inverting into the mixed channel.

When considering the mixed channel, the inversion formula \eqref{eq:inversionintegralgen} reads
\begin{align}
	\mathbb T_{tsst}(z,\hb) &=\kappa^{tsst}_{\hb}\int_0^1\frac{d\zb}{\zb^2}k_{\hb}^{tsst}(\zb)(1-\bar{z})^{\Delta_s-\Delta_t}{\dDisc}^{tsst}\left[\frac{u^{\frac{\Delta_t+\Delta_s}2}}{v^{\Delta_s}}\G_{sstt}(v,u)\right]
	\nonumber \\
	&\quad + (-1)^\ell\kappa^{stst}_{\hb}\int_0^1\frac{d\zb}{\zb^2}k_{\hb}^{stst}(\zb){\dDisc}^{stst}\left[\left( \frac uv \right)^{\frac{\Delta_t+\Delta_s}2}\G_{stst}(v,u)\right].
	\label{eq:ssttstrategy}
\end{align}
Now the first line of \eqref{eq:ssttstrategy} is in principle known, because we can write the part of $\G_{sstt}(v,u)$ that has non-vanishing double-discontinuity as the contribution from the identity plus the twist 2, (even) spin $\ell$ operators, with OPE coefficients given by our previous results $\G_{ssss}$ and $\G^S_{tttt}$. We find
\begin{equation}\label{eq:Gsstttwist2}
	\G_{sstt}(u,v)\big|_{\tau=2,\ell} =\sum_\ell \lambda_{ss\mathcal{J}_{S,\ell}}\lambda_{tt\mathcal{J}_{S,\ell}} G^{sstt}_{2+\gamma_{S,2,\ell}}(u,v) =zF_{\mathrm{pure}}(\bar{z},\log z) +O(z^2).
\end{equation}
In the second line of \eqref{eq:ssttstrategy} we have the same problem we had in the pure channel \eqref{eq:ssttLIFpure}. However, we know that
that in the mixed channel at twist 2 only even spins appear. This can be seen either from the considerations about the spectrum reviewed in section~\ref{sec:ONCFT} -- it is known that primary operators of the form $[\varphi,\varphi]_{T,0,\ell}$ only exist for even spin -- or by decomposing the free theory correlator
\begin{equation}
	\G^{(d=4)}_{tsst,free}(u,v) =\left(\frac uv\right)^2 \left( 1+\frac 4n \left( u+ \frac uv +\frac{u^2}v \right) \right).
\end{equation} We stress that this is no longer true at higher twists. Therefore we obtain the equation
\begin{align}
	\mathbb T^{\text{odd }\ell}_{\tau_0=2,tsst}(\log z,\hb)=0 &=\kappa^{tsst}_{\hb}\int_0^1\frac{d\zb}{\zb^2}k_{\hb}^{tsst}(\zb)(1-\bar{z})^{\Delta_s-\Delta_t}{\dDisc}^{tsst}\left[\frac{u^{\frac{\Delta_t+\Delta_s}2}}{v^{\Delta_s}}\G_{sstt}(v,u)\right]\bigg|_z
	\nonumber \\
	&\quad -\kappa^{stst}_{\hb}\int_0^1\frac{d\zb}{\zb^2}k_{\hb}^{stst}(\zb){\dDisc}^{stst}\left[\left( \frac uv \right)^{\frac{\Delta_t+\Delta_s}2}\G_{stst}(v,u)\right]\bigg|_z,
	\label{}
\end{align}
which tells us that the second line must be equal to the first. Analyticity in spin allows us to generalize this result to even spins; namely, at twist 2 and for even spin, the two lines of \eqref{eq:ssttstrategy} are again equal. Neglecting the contribution of the identity (which is of order $z^2$), we can replace $\G_{sstt}(v,u)$ with \eqref{eq:Gsstttwist2} to obtain
\begin{align}
	&\mathbb T^{\text{even }\ell}_{\tau_0=2,tsst}(\log z,\hb)
	\nonumber \\
	&=2\kappa^{tsst}_{\hb}\int_0^1\frac{d\zb}{\zb^2}k_{\hb}^{tsst}(\zb)(1-\bar{z})^{\Delta_s-\Delta_t}{\dDisc}^{tsst}\left[\frac{u^{\frac{\Delta_t+\Delta_s}2}}{v^{\Delta_s}}(1-\bar{z})F_{\mathrm{pure}}(1-z,\log(1-\bar{z}))\right]\bigg|_z.
	\label{eq:ssttTtwist2mixed}
\end{align}

Due to the new kernel -- that now depends on $\eps$ through the difference $\Delta_s-\Delta_t$ -- it is hard to proceed directly with the integration. Consider the generic situation in which one has two operators $\O_1$, $\O_2$ of dimensions $\Delta_1=2+\eps\gamma_1$ and $\Delta_2=2+\eps\gamma_2$ respectively. It turns out that we are able to compute the following integral:
\begin{align}
	\kappa^{1221}_{\hb}\int_0^1\frac{d\zb}{\zb^2}k_{\hb}^{1221}(\zb)(1-\bar{z})^{\Delta_2-\Delta_1}{\dDisc}^{1221}\left[\frac{\bar{z}^{2+\eps\frac{\gamma_1+\gamma_2}2}}{(1-\bar{z})^{1+\eps\gamma_2}}\right] =\frac{\Gamma(\bar{h})^2\big( 1+\eps(\gamma_1+\gamma_2)S_1(\bar{h}-1) \big)}{\Gamma(2\bar{h}-1)}.
	\label{eq:mixedintegral}
\end{align}
The idea is to write $F_{\mathrm{pure}}(\bar{z},\log z)$ in such a way that the argument of the double-discontinuity in \eqref{eq:ssttTtwist2mixed} is given by a term proportional to $\bar{z}^{2+\eps\frac{-2+\gamma_t+\gamma_s}2}(1-\bar{z})^{-1+\eps-\eps\gamma_s}$ plus order-$\eps$ corrections; the latter can be then integrated using the old $\eps$-independent kernel.

We use $F_{\mathrm{pure}}(\bar{z},\log z) =\hat{F}_{\mathrm{pure}}(\bar{z},\log z) +\eps \Delta F_{\mathrm{pure}}(\bar{z},\log z)$, where
\begin{equation}
	\hat{F}_{\mathrm{pure}}(\bar{z},\log z) =-\frac{4\bar{z}(\bar{z}-2)}{n(1-\bar{z})^{1+\eps\frac{-2+\gamma_s+\gamma_t}2}}\bigg|_\eps
\end{equation}
is such that it captures the $\eps^0$ part of $F_{\mathrm{pure}}(\bar{z},\log z)$ and $\Delta F_{\mathrm{pure}}(\bar{z},\log z)$ is just the difference $F_{\mathrm{pure}}(\bar{z},\log z) -\hat{F}_{\mathrm{pure}}(\bar{z},\log z)$. Plugging these into \eqref{eq:ssttTtwist2mixed} we obtain
\begin{align}
	\mathbb T^{\text{even }\ell}_{\tau_0=2,tsst}(\log z,\hb) &=2\kappa^{tsst}_{\hb}\int_0^1\frac{d\zb}{\zb^2}k_{\hb}^{tsst}(\zb)(1-\bar{z})^{\Delta_s-\Delta_t}{\dDisc}^{tsst}\left[ \frac 4n \frac{\bar{z}^{2+\eps\frac{-2+\gamma_t+\gamma_s}2}}{(1-\bar{z})^{1+\eps(-1+\gamma_s)}} \right]
	\nonumber \\
	&\quad +2\kappa_{\hb}\int_0^1\frac{d\zb}{\zb^2}k_{\hb}(\zb){\dDisc}\left[\left(\frac uv \right)^2(1-\bar{z})\eps \Delta F_{\mathrm{pure}}(1-z,\log(1-\bar{z}))\right]\bigg|_z,
	\label{}
\end{align}
where the first line is exactly of the form \eqref{eq:mixedintegral} and the second line can be worked out with the help of the usual dictionary. Note also that in the first line we already took the coefficient of the power $z$. In the second line, the inversion is at leading order in $\eps$, so we could replace the double-discontinuity with that for identical external operators. We finally obtain the OPE coefficients $\lambda^2_{ts\mathcal{J}^T_\ell}$ and anomalous dimensions in the mixed channel at twist 2:
\begin{align}
	a^{(0),\mathrm{mixed}}_{2,\ell} &=\frac{8\Gamma(\ell+1)^2}{n\Gamma(2\ell+1)},
	\nonumber \\
	\alpha^{\mathrm{mixed}}_{2,\ell} &=-\frac{n+4}{n+8} -\frac{n+12}{n+8}S_1(\ell) +S_1(2\ell),
	\nonumber \\
	\gamma^{\mathrm{mixed}}_{2,\ell} &=-\eps +O(\eps^2),
	\label{eq:ssttmixedCFTdata}
\end{align}
valid for $\ell>0$.

\subsubsection{Pure channel data and resummation}
The strategy is now to go back to \eqref{eq:ssttLIFpure} and use the twist-2 mixed CFT data to compute $\G_{tsst}(u,v)$ at order $z$. Before we proceed, we have to consider the spin-0 contribution to \eqref{eq:ssttmixedCFTdata}, \emph{i.e.} $\lambda^2_{tst}$. But this is known from out previous computation by permutation symmetry of OPE coefficients
\begin{equation}
	\lambda^2_{tst} =\lambda^2_{tts},
\end{equation}
where $\lambda^2_{tts}$ is the $\expv{tttt}$ OPE coefficient in the irrep S at spin 0, which is given by \eqref{eq:tttta0S}, \eqref{eq:ttttalphaS} and \eqref{eq:ffttconstants}. Considering also the dimension of $t$, \eqref{eq:ssttmixedCFTdata} gets corrected to 
\begin{align}
	a^{(0),\mathrm{mixed}}_{2,\ell} &=\frac{8\Gamma(\ell+1)^2}{n\Gamma(2\ell+1)},
	\\
	\alpha^{\mathrm{mixed}}_{2,\ell} &=-\frac{n+4}{n+8} -\frac{n+12}{n+8}S_1(\ell) +S_1(2\ell) -\frac{2}{n+8}\delta_{\ell,0},
	\label{eq:lambdast20}\\
	\gamma^{\mathrm{mixed}}_{2,\ell} &=-\eps +\eps\frac{2}{n+8} \delta_{\ell,0} +O(\eps^2).
	\label{}
\end{align}
From these we compute the twist-2 contribution
\begin{equation}\label{eq:Gtssttwist2}
	\G_{tsst}(u,v)\big|_{\tau=2,\ell} =\sum_\ell a^{(0),\mathrm{mixed}}_{2,\ell}(1+\eps \alpha^{\mathrm{mixed}}_{2,\ell}) G^{tsst}_{2+\gamma^{\mathrm{mixed}}_{2,\ell}}(u,v) =zF_{\mathrm{mixed}}(\bar{z},\log z) +O(z^2).
\end{equation}
We are now ready to compute the CFT data in the pure channel for any twist. With the help of \eqref{eq:Gtssttwist2}, \eqref{eq:ssttLIFpure} becomes
\begin{equation}\label{}
	\mathbb T_{sstt}(z,\hb) =\left(1+(-1)^\ell\right)\kappa^{sstt}_{\hb}\int_0^1\frac{d\zb}{\zb^2}k_{\hb}^{sstt}(\zb)\dDisc\left[\frac{u^{\Delta_s}}{v^{\frac{\Delta_s+\Delta_t}2}}(1-\bar{z})F_{\mathrm{mixed}}(1-z,\log(1-\bar{z})) \right].
\end{equation}
The argument of the $\dDisc$ can now be expanded in $\eps$ and powers of $z$, and the integral is calculated with the usual dictionary. As an example, we display the $d=4-\eps$ CFT data at twist 4:
\begin{align}
	a^{(0)}_{4,\ell} &= \frac{8\Gamma(\ell+2)^2}{n\Gamma(2\ell+3)},
	\nonumber \\
	\alpha_{4,\ell} &=-\frac{1}{2(n+8)}\left( n+4 -16 S_1(\ell+1)^2 +24 S_1(\ell+1) -16S_1(2\ell+2) \right.
	\nonumber \\
	&\quad +\left. 16S_1(\ell+1)S_1(2\ell+2) +8S_2(\ell+1) \right),
	\nonumber \\
	\gamma_{4,\ell} &=-\eps\frac{8}{n+8}\left( 1-S_1(\ell+1) \right) +O(\eps^2).
	\label{}
\end{align}

To find the expression for the correlator we again proceed with the four-dimensional CFT data, \emph{i.e.} $a^{(0)}_{\tau,\ell}$, $\alpha^{(d=4)}_{\tau,\ell}$, $\gamma_{\tau,\ell}$. Then we conclude by resumming the correlator
\begin{equation}
	\G_{sstt}(z,\bar{z}) =1 +\sum_{\tau,\ell} a^{(0)}_{\tau,\ell}\left(1+\eps \alpha^{(d=4)}_{\tau, \ell}\right) G^{(d=4)}_{\tau+\gamma_{\tau,\ell}, \ell}(z, \bar z) +O(\eps^2),
\end{equation}
obtaining
\begin{align}
	\G_{sstt}(u,v) &=1 +\frac{4u(u+v+1)}{nv} +\eps\bigg( -\frac{4(n+4)u(u+v+1)}{n(n+8)v} -\frac{2u\big( 8u+(n+8)(1+v) \big)}{n(n+8)v}\log u \nonumber \\
	&\quad+\frac{2u\big( 4u+(n+4)v +4 \big)}{n(n+8)v}\log v -\frac{4u\big( 2u(v+1)+(n+2)v \big)}{n(n+8)v}\Phi(u,v) \bigg) +O(\eps^2).
	\label{eq:Gssttans}
\end{align}

\subsection{Mixed-channel data and multiplet recombination}
\label{secc:mixedchanneldatamultrecomb}

With the correlator \eqref{eq:Gssttans} at hand, we can perform the conformal block decomposition in the mixed channel, using
\begin{equation}
\G_{tsst}(u,v)=\frac{u^{\frac{\Delta_t+\Delta_s}2}}{v^{\Delta_s}}\G_{sstt}(v,u).
\end{equation}
When considering the mixed channel, the multiplet recombination effect discussed in section~\ref{subsec:multrecomb} is now manifest.
In the interacting theory, the twist-2 currents $\mathcal J_{T,\ell}$ are no longer conserved, and their multiplets recombine with an operator of the form $\partial^{\ell-1}\varphi_T^4$. Consider the operator content at twist $\tau=4$: the number of operators is given by the generating functions
\begin{align}
\text{Free theory:} &\quad \sum_\ell d_\ell q^\ell =1+q+3q^2+3q^3+6q^4+6q^5+\ldots,
\\
\text{Interacting theory:} &\quad \sum_\ell d_\ell q^\ell= 1+3q^2+2q^3+6q^4+5q^5+\ldots,
\end{align}
where $d_\ell$ is the number of twist-4 operators at spin $\ell$, \emph{i.e.} the degeneracy.
In particular, there is no operator at twist 4, spin 1 in the interacting theory.

The multiplet recombination affects the conformal blocks, which we evaluate using the subcollinear expansion explained in appendix~\ref{app:4-epsblocks}. Here we have to consider the case $\Delta_{34}=-\Delta_{12}$.
There is an order of limits problem in the coefficient $c_{1,0}$ of \eqref{eq:c10}.
We are interested in the case
\begin{equation}
\Delta_{\mathcal J_{T,\ell}}=2+\ell+\eps \gamma_{2,\ell}^{(1)}+\eps^2\gamma_{2,\ell}^{(2)}, \quad \Delta_{12}=\eps (\gamma_t-\gamma_s),
\end{equation}
with $\gamma_{2,\ell}^{(1)}=-1$.
Taking first $\Delta_{12}\to0$, one finds $c_{1,0}=\frac12$. Taking instead first $\gamma_{2,\ell}^{(1)}\to-1$, one finds
\begin{equation}
c_{1,0}=\frac{\Delta_{12}^2+2\eps^2\gamma_{2,\ell}^{(2)}}{4\eps^2\gamma_{2,\ell}^{(2)}}+O(\eps).
\end{equation}
So we find that in the interacting case, when $\Delta_{12}\neq0$, the coefficient $c_{1,0}$ gets a constant shift at order $\eps^0$, which takes the form $\Delta_{12}^2/(4\eps^2\gamma_{2,\ell}^{(2)})$. The constants $c_{1,\pm1}$ are not affected by this problem, and take the values
\begin{equation}
c_{1,-1}=1+O(\eps), \qquad c_{1,1}=\frac{(\ell+1)^2}{4(2\ell+1)(2\ell+3)} +O(\eps).
\end{equation}

We can now use these constants to decompose the correlator to leading order in $\eps$.\\
Free theory:
\begin{equation}
a^{(0)}_{4,\ell}=\frac{n+4}4,\ 1,\ \frac{3n+2}{5n},\ \frac72,\ \frac{15n+4}{126n} \ldots,\quad \ell=0,1,2,3,4,\ldots.
\end{equation}
Interacting theory:
\begin{equation}
a^{(0)}_{4,\ell}=\frac{n+4}4,\ 0,\ \frac{3n+2}{5n},\ \frac{2n+4}{9n+14},\ \frac{15n+4}{126n} \ldots, \quad\ell=0,1,2,3,4,\ldots,
\end{equation}
where as anticipated there is no operator at spin 1 in the interacting case.
By performing the entire decomposition, we find for the non-degenerate operators
\begin{align}
a_{4,0}&=\frac{n+4}n\left(1-\frac{10}{n+8}\eps\right)+O(\eps^2),& \gamma_{4,0}&=-\eps\frac{n}{n+8}+O(\eps^2),
\\
a_{6,0}&=\frac{n-2}{3n}\left(1-\frac{n+40}{12(n+8)}\eps\right)+O(\eps^2),&\gamma_{6,0}&=-\eps\frac{n+12}{n+8}+O(\eps^2),
\\
a_{6,1}&=\frac14\left(1-\frac{27n+608}{288(n+8)}\eps\right)+O(\eps^2), & \gamma_{6,1}&=-\eps\frac{3n+32}{3(n+8)}+O(\eps^2).
\end{align}
The anomalous dimensions are in complete agreement with those reported in \cite{Henriksson:2022rnm}.

\section{Correlator $\boldsymbol{\langle \phi\phi\phi^2\phi^2\rangle}$ at $\boldsymbol{n=1}$ to order $\boldsymbol{\varepsilon^2}$}
\label{sec:fseps2}

In this section we consider the correlator
$
\langle\phi\phi\phi^2\phi^2\rangle
$
to order $\eps^2$ at $n=1$. We will consider both the pure channel $\G_{\phi\phi\phi^2\phi^2}(u,v)$ and the mixed channel $\G_{\phi^2\phi\phi\phi^2}(u,v)$ decompositions. In general, there is an obstacle to applying large spin perturbation theory at subleading orders due to a mixing problem. Here we will outline the status of this problem for the correlator at hand, and explain how it can be circumvented in the special case of $n=1$. We will then find order-$\eps^2$ conformal data in both channels.

Let us begin by considering the crossed-channel operators $\O'$ that contribute to the double-discontinuity at each order. The orders at which they appear are as follows:

\paragraph{$\boldsymbol{\O'}$ appearing in $\boldsymbol{\G_{\phi^2\phi\phi\phi^2}(v,u)}$} \begin{itemize}
\item Order $\eps^0$: $\O'=\phi$, contribution proportional to $\lambda_{\phi^2\phi\phi}^2$.
\item Order $\eps^2$: $\O'=\de^\ell\phi^3$, contribution proportional to $\lambda_{\phi^2\phi\O'}^2\gamma_{\O'}^2$.
\end{itemize}
\paragraph{$\boldsymbol{\O'}$ appearing in $\boldsymbol{\G_{\phi\phi\phi^2\phi^2}(v,u)}$}
\begin{itemize}
\item Order $\eps^0$: $\O'=\1$, contribution $1$.
\item Order $\eps^1$: $\O'=\phi^2$, contribution proportional to $\lambda_{\phi\phi\phi^2}\lambda_{\phi^2\phi^2\phi^2}\gamma_{\phi^2}$.
\item Order $\eps^2$: $\O'=\mathcal J_\ell$, contribution proportional to $\lambda_{\phi\phi\mathcal J_\ell}\lambda_{\phi^2\phi^2\mathcal J_\ell}\gamma_{\mathcal J_\ell}$, where $\gamma_{\mathcal J_\ell}\sim\eps^2$.\footnote{The fact that it is proportional to a single factor of the anomalous dimension, and not a squared factor, will be discussed in section~\ref{subsec:inversionintomixed}.}
\end{itemize}
If all of these operators were non-degenerate, then the all relevant double-discontinuities could be readily computed. Unfortunately, this is not always the case.
Recall that when there are degenerate operators, the symbols $a_{\tau,\ell}$ and $a_{\tau,\ell}\gamma_{\tau,\ell}$ etc. do not denote CFT data of individual operators, but sums over all spin-$\ell$ operators with 4d twist equal to $\tau$,
\begin{equation}
a_{\tau,\ell}\gamma_{\tau,\ell}^k = \sum_i \lambda_{\O_1\O_2\O_{\tau,\ell,i}}\lambda_{\O_3\O_4\O_{\tau,\ell,i}}\gamma_{\O_{\tau,\ell,i}}^k,
\end{equation}
where $i$ ranges over the degenerate operators. Therefore, in general, the expressions $a_{\tau,\ell}\gamma_{\tau,\ell}^2$ that enter the double-discontinuity cannot be found from $a_{\tau,\ell}$ and $\gamma_{\tau,\ell}$ alone. To determine the contribution from the individual operators is referred to as resolving a mixing problem, and is in general a task that requires additional input.

In the $n=1$ theory it turns out that we have exactly the information needed to resolve the mixing, \emph{if one assumes the knowledge of the leading-order anomalous dimensions}. Thus, in order to make progress, we will put the bootstrap philosophy aside and use as much input from the literature that is needed to proceed. Thus:
\begin{description}
\item[Until this section:] only use conformal field theory principles, and information about the spectrum of the free 4d theory.
\item[In this section:] input information about order-$\eps$ anomalous dimensions.
\end{description}
The anomalous dimensions needed were computed in the 1990's by Kehrein, Wegner and Pis'mak \cite{Kehrein:1992fn}, and constitute a simple example of the diagonalization of the one-loop dilatation operator; see also \cite{Kehrein:1994ff}.

\subsection{Unmixing of twist-three operators}

Let us consider operators of the form $\O_{\ell,i}\sim \de^\ell\phi^3$ for $n=1$, where $i$ is an extra index labeling the operators that are degenerate in the 4d theory. The range of $i$ is the number $d_\ell$ of primary operators at each spin $\ell$, which is counted by the generating function
\cite{Kehrein:1994ff,Roumpedakis:2016qcg}
\begin{equation}
\sum d_\ell q^\ell=-1+\frac{1}{(1-q^2)(1-q^3)}=q^2+q^3+q^4+q^5+2q^6+q^7+2q^8+\ldots
.
\end{equation}
In the correlator $\G_{\phi^2\phi\phi\phi^2}(v,u)$, the operators $\O_{\ell,i}$ appear as double-twist operators, and we define their anomalous dimensions by
\begin{equation}
\Delta_{\O_{\ell,i}}=3-\frac32\eps+\ell+\gamma_{\ell,i}, \qquad \gamma_{\ell,i}=\gamma^{(1)}_{\ell,i}\eps+O(\eps^2)
.
\end{equation}
Kehrein, Wegner and Pis'mak \cite{Kehrein:1992fn} found that exactly one operator, which we will call $i=1$, at each $\ell=2,3,4,5,6\ldots$ has a non-vanishing anomalous dimension, 
\begin{equation}
\gamma^{(1)}_{\ell,i}=\begin{cases}
\gamma^{\neq0}_\ell, & i=1,
\\
0 & i>1,
\end{cases} \qquad \gamma_\ell^{\neq0}=\frac{2(-1)^\ell}{3(\ell+1)}+\frac13.
\label{eq:gammanonzero}
\end{equation}
The fact that there are only two different values for the anomalous dimension means that we can resolve the mixing problem. First, let us recall the results from section~\ref{subsec:sffs}, which for $n=1$ yields
\begin{align}
\label{eq:a3lN1}
a_{3,\ell}&=\frac{2\Gamma(\ell+2)^2(1+\ell+2(-1)^\ell)}{\Gamma(2\ell+3)}+O(\eps)
,
\\
a_{3,\ell}\gamma_{3,\ell}&=\frac{2\Gamma(\ell+2)^2(1+\ell+2(-1)^\ell)}{\Gamma(2\ell+3)}\left(-\frac76+\frac{2(-1)^\ell}{3(\ell+1)}\right)\eps+O(\eps^2)
.
\end{align}
Here $\gamma_{3,\ell}$ is defined by $\Delta=3+\ell+\gamma_{3,\ell}$, and differs from $\gamma_{\O_{\ell,i}}$ by the constant shift $-\frac32\eps+O(\eps^2)$. We have to solve two equations,
\begin{equation}
a_\ell^{\neq0}+a_\ell^{=0}=a_{3,\ell}, \qquad a_\ell^{\neq0}\left(\gamma_\ell^{\neq0}-\frac32\right)\eps+a_\ell^{=0}\left(0-\frac32\right)\eps=a_{3,\ell}\gamma_{3,\ell}
,
\end{equation}
for two unknowns, $a_\ell^{\neq0}$, $a^{=0}_\ell$. The unique solution to these equations is that $a_{\ell}^{\neq0}=a_{3,\ell}$, $a^{=0}_\ell=0$, which implies\footnote{Given that $a_{\ell}^{=0}$ represents a squared OPE coefficient, the vanishing of the leading OPE coefficient implies that the square is of order $\eps^2$.}
\begin{equation}
\lambda^2_{\phi^2\phi \O_{\ell,i}}=\begin{cases}
\text{$a_{3,\ell}$ in \eqref{eq:a3lN1} } & i=1 \text{ (for $\gamma_{\ell,i}=\gamma^{\neq0}$)}
\\
O(\eps^2) & i>1\text{ (for $\gamma_{\ell,i}=0$)}
\end{cases}
.
\label{eq:surprisingobservation}
\end{equation}
This is a surprising result: the operators with zero anomalous dimensions must also have suppressed OPE coefficients in the $\phi^2\times \phi$ OPE!

\subsection{Pure channel inversion and conformal data}

We have
\begin{align}
	\mathbb T_{\phi\phi\phi^2\phi^2}( z,\hb)&=\big(1+(-1)^\ell\big)\kappa^{\phi\phi\phi^2\phi^2}_{\hb}\int_0^1\frac{d\zb}{\zb^2}k^{\phi\phi\phi^2\phi^2}_{\hb}(\zb)\dDisc\left[\frac{u^{\Delta_\phi}}{v^{\frac{\Delta_\phi+\Delta_{\phi^2}}2}}\G_{\phi^2\phi\phi\phi^2}(v,u)\right].
\end{align}

There are now two contributions to the double-discontinuity,
\begin{enumerate}
\item Contribution from $\phi$: $F_\phi(1-\zb,1-z)=\lambda_{\phi^2\phi\phi}^2 G_{\Delta_\phi,0}(v,u)$.
\item Contribution from twist-3 operators
\begin{equation}
\label{eq:sumtwist3}
F_{\tau=3}(1-\zb,1-z)=\sum_{\ell=2,3,4,\ldots}a_{3,\ell}\left(\gamma_\ell\right)^2 G_{3+\ell,\ell}(v,u).
\end{equation}
where we now define $\gamma_\ell=\Delta_{\O_{\ell,1}}-(\Delta_\phi+\Delta_{\phi^2}+\ell)=\frac23\frac{(-1)^\ell}{\ell+1}$.
\end{enumerate}
These combine to give 
\begin{equation}
\G_{\phi^2\phi\phi\phi^2}(v,u) = F_\phi(1-z, 1-\bar z) + F_{\tau = 3}(1-z, 1-\bar z)+\ldots,
\end{equation}
where the terms in the ellipses do not contribute to the double-discontinuity to this order.
 For the $\phi$ contribution, we use the formula \eqref{eq:scalarblockmixed} evaluated to the next order in $\eps$. Since we are now generating contributions to the double-discontinuity proportional to $\gamma^2$, we can no longer truncate the sum at $m=0$. But the contribution from $m\geqslant1$ is canceled by extending the sum in \eqref{eq:sumtwist3} to include spin $\ell=0$ (the contribution at $\ell=1$ vanishes).\footnote{By the multiplet recombination effect discussed in section~\ref{subsec:multrecomb}, the descendant states that contribute to the $m>0 $ part of the sum defining $G_{\Delta_\phi,0}(v,u)$ correspond to the states of the $\phi^3$ multiplet in the free theory. This holds at least for the leading order OPE coefficients. The leading order anomalous dimension, $\Delta_0=3(1-\frac\eps2)+\eps +\ldots = \Delta_\phi+2 +O(\eps^2)$, which allows our redefinition \eqref{eq:Ftilde1}, \eqref{eq:Ftilde2}, up to regular terms.} Our strategy, then, will be to define 
\begin{align}
\label{eq:Ftilde1}
 \widetilde{F}_\phi(z, \bar z)   \ &=  \ F_\phi(z, \bar z)  - \text{descendants at twist $\geqslant3$}, \\
\label{eq:Ftilde2}
 \widetilde{F}_{\tau = 3}(z, \bar z) \ & = \ F_{\tau = 3}(z, \bar z) + a_{3,\ell=0}\left(\gamma_{\ell=0}\right)^2 G_{3,\ell = 0}(v,u).
\end{align}
Then because the descendants in the first line cancel the spin zero part in the second line, we have $  \widetilde{F}_\phi(1-z, 1-\bar z)  +  \widetilde{F}_{\tau = 3}(1-z,1- \bar z) = {F}_\phi(1-z, 1-\bar z)  +  {F}_{\tau = 3}(1-z,1- \bar z) $, which is true inside the double-discontinuity. In summary, $\widetilde F_\phi$ only gets contributions from the $m = 0$ part of the sum defining $\G_{\phi^2\phi\phi\phi^2}(v,u)$ in \eqref{eq:scalarblockmixed}. It takes the form
\begin{align}
\widetilde F_\phi(z,\zb)&=\lambda_{\phi^2\phi\phi}^2 \sqrt{z \bar z} \Big(1 - \frac{1}{4} \log (z \bar z) \eps \nonumber \\ 
&  \quad + \frac{1}{864} \big(4 \log (z \bar z) + 27 \log (z \bar z)^2 +  24\, \mathrm{Li}_2(z + \bar z + z \bar z) \big) \eps^2 \Big).
\end{align}
For the OPE coefficient, we use the result \cite{Gopakumar:2016cpb}
\begin{equation}
\lambda_{\phi\phi\phi^2}^2=2\left(1-\frac\eps3-\frac{17}{81}\eps^2\right)+O(\eps^3).
\end{equation}

For twist-3 operators, we compute the sum \eqref{eq:sumtwist3}, starting from $\ell=0$. Since we are already working at order $\eps^2$, this can be found using the four-dimensional conformal blocks, and gives
\begin{equation}
\widetilde F_{\tau=3}(z,\zb)=(z\zb)^{3/2}\log^2z\frac{2\,\mathrm{Li}_2(z)-2\,\mathrm{Li}_2(\zb)+\log(1-z)-\log(1-\zb)}{18(z-\zb)}.
\end{equation}

Hence, we arrive at the inversion problem
\begin{align}
	\mathbb T_{\phi\phi\phi^2\phi^2}( z,\hb) &=\big( 1+(-1)^\ell \big)\kappa^{\phi\phi\phi^2\phi^2}_{\hb}\int_0^1\frac{d\zb}{\zb^2}k^{\phi\phi\phi^2\phi^2}_\hb(\zb)
	\\
	&\quad\times
	\dDisc\left[\frac{u^{\Delta_{\phi}}}{v^{\frac{\Delta_\phi+\Delta_{\phi^2}}2}}\left(\widetilde F_\phi (1-\zb,1-z)+ \widetilde F_{\tau=3}(1-\zb,1-z) \right)\right].
\end{align}
Computing the inversion integral, we find the CFT data order-by-order in twist.
The results for the first three twists are
\begin{align}
\Delta_{2, \ell} &= 2 +\ell - \eps + \frac{1}{54} \left( 1 - \frac{6}{\ell(\ell+1)} \right) \eps^2 + \mathcal{O}(\eps^3) , \\
\Delta_{4, \ell} &= 4 + \ell - \frac{2 \ell (\ell+3)}{3(\ell+1)(\ell+2)} \eps+ \mathcal{O}(\eps^2), \\
\Delta_{6, \ell} &= 6 + \ell + \mathcal{O}(\eps) ,
\end{align}
for the scaling dimensions, and
\begin{align}
a_{2,\ell}&=\frac{4\Gamma(\ell+1)^2}{\Gamma(2\ell+1)}\bigg(1-\frac\eps3\big[1+5S_1(\ell)-3S_1(2\ell)\big]+\eps^2\bigg[\frac{9-17\ell(\ell+1)^2}{81\ell(\ell+1)^2}
\nonumber
\\&\quad +\left(\frac{56\ell^2+56\ell-9}{81\ell(\ell+1)}-\frac53S_1(2\ell)\right)S_1(\ell)
+\frac{25}{18}S_1(\ell)^2-\frac{19\ell^2+19\ell-6}{54\ell(\ell+1)}S_1(2\ell)
\nonumber\\
&\quad
+\frac12S_1(2\ell)^2-\frac7{12}S_2(\ell)+\frac12S_2(2\ell)-\frac2{9\ell(\ell+1)}S_{-2}(\ell)
\bigg]\bigg)+O(\eps^3),
\label{eq:opeT2ffss}
\\
a_{4, \ell} &= \frac{2 \Gamma(\ell+2)^2}{\Gamma(2\ell+3)} \Bigg( \eps + \eps^2 \Bigg[ \frac{-924-2683 \ell -2290 \ell^2 - 641 \ell^3 -38 \ell^4}{108 (\ell +1)(2\ell+ 3)} \\
& \quad - \frac{4}{3} (\ell^2 + 3 \ell + 1) S_1(\ell) + \frac{2}{3} \ell (\ell + 3) S_1({2 \ell + 4}) \Bigg] \Bigg)\nonumber+O(\eps^3),
\label{eq:opeT4ffss}
\\
a_{6, \ell} &= \frac{\ell (\ell + 5) \Gamma(\ell + 3)^2}{36 (\ell^2 + 5 \ell + 6) \Gamma(2 \ell + 5)} \eps^2+O(\eps^3) ,
\end{align}
for the OPE coefficients. The vanishing of the last expression at $\ell=0$ is consistent with the fact that $\lambda_{\phi\phi\phi^6}=O(\eps^2)$ and $\lambda_{\phi^2\phi^2\phi^6}=O(\eps)$.

The operators $\mathcal J_\ell$ at twist $\tau_0=2$ are non-degenerate, and using the data in \eqref{eq:opeT2ffss} for $\lambda_{\phi\phi\mathcal J_\ell}\lambda_{\phi^2\phi^2\mathcal J_\ell}$ together with the known results for the OPE coefficients $\lambda_{\phi\phi\mathcal J_\ell}^2$, computed in
\cite{Gopakumar:2016cpb},  we are able to extract the order-$\eps^2$ part of $\lambda_{\phi^2\phi^2\mathcal J_\ell}$. This is a new result and takes the form
\begin{align}
\lambda^2_{\phi^2\phi^2\mathcal J_\ell}&=
\frac{(\lambda_{\phi\phi\mathcal J_\ell}\lambda_{\phi^2\phi^2\mathcal J_\ell})^2}{\lambda_{\phi\phi\mathcal J_\ell}^2}
\nonumber \\
&= \frac{8\Gamma(\ell+1)^2}{\Gamma(2\ell+1)}\bigg(
1-\frac\eps3\big(2+4S_1(\ell)-3S_1(2\ell)\big)+\eps^2\bigg[\frac{9-25\ell(\ell+1)^2}{81\ell(\ell+1)^2}
\nonumber\\
&\quad
+\left(\frac{91\ell^2+91\ell-9}{81\ell(\ell+1)}-\frac43S_1(2\ell)\right)S_1(\ell)+\frac89S_1(\ell)^2-\frac{37\ell^2+37\ell-6}{54\ell(\ell+1)}S_1(2\ell)
\nonumber\\
&\quad+
\frac12S_1(2\ell)^2-\frac5{12}S_2(\ell)+\frac12S_2(2\ell)-\frac{4}{9\ell(\ell+1)}S_{-2}(\ell)
\bigg]
\bigg)+O(\eps^3).
\label{eq:OPEcoefsssJ}
\end{align}
For the stress-tensor at spin $\ell=2$, the conformal Ward identity holds as expected.

Likewise, for the non-degenerate operators at twist 4, spin $\ell=0$ and $\ell=2$, we can combine our results \eqref{eq:alpha4ellssss} (for $\tilde\alpha=-\frac{n+2}{n+8}$ and $n=1$) and \eqref{eq:opeT4ffss} to find $\lambda_{\phi\phi\O}^2$:
\begin{equation}
\lambda_{\phi\phi\phi^4}^2=\frac{\eps^2}{54}-\frac{47}{1458}\eps^3+O(\eps^4), \qquad \lambda_{\phi\phi\de^2\phi^4}^2=\frac{\eps^2}{1440}-\frac{\eps^3}{1440}+O(\eps^4).
\label{eq:opetwist4res}
\end{equation}
Both of these agree with \cite{Carmi:2020ekr}, where the second value was only given numerically.\footnote{In fact, they can also be computed from the approach of \cite{Gopakumar:2016cpb} and we thank Aninda Sinha for providing us with unpublished results that precisely match \eqref{eq:opetwist4res}. Moreover, we have computed them within the approach of \cite{Alday:2017zzv} and found full agreement.}

\subsection{Mixed channel inversion and conformal data}
\label{subsec:inversionintomixed}
We now have
\begin{align}
	\mathbb T_{\phi^2\phi\phi\phi^2}( z,\hb)&=\kappa^{\phi^2\phi\phi\phi^2}_{\hb}\int_0^1\frac{d\zb}{\zb^2}k_{\hb}^{\phi^2\phi\phi\phi^2}(\zb){\dDisc}^{\phi^2\phi\phi\phi^2}\left[\frac{u^{\frac{\Delta_{\phi^2}+\Delta_\phi}2}}{v^{\Delta_\phi}}\G_{\phi\phi\phi^2\phi^2}(v,u)\right]
\nonumber
\\&\quad + (-1)^\ell\kappa^{\phi\phi^2\phi\phi^2}_{\hb}\int_0^1\frac{d\zb}{\zb^2}k_{\hb}^{\phi\phi^2\phi\phi^2}(\zb){\dDisc}^{\phi\phi^2\phi\phi^2}\left[\left(\frac uv\right)^{\frac{\Delta_{\phi^2}+\Delta_\phi}2}\G_{\phi\phi^2\phi\phi^2}(v,u)\right].
\label{eq:inversionintomixed}
\end{align}

The contributions from the second line in \eqref{eq:inversionintomixed} are the same as above; however, they must now be summed with different conformal blocks, since we have replaced $\Delta_{12}\to-\Delta_{12}$. We find
\begin{align}
\widetilde F_{\phi}(z,\bar{z})&=\lambda_{\phi\phi^2\phi}^2\sqrt{z\zb}\bigg(1+\frac{\eps}{12}(2\log(1-\zb)-3\log(z\zb))+\frac{\eps^2}{2592}\big(81\log^2(z\zb)+12\log(z\zb)
\nonumber \\
&\quad
-108\log (z\zb)\log(1-\zb)+128\log(1-\zb)-72\mathrm{Li}_2(\zb)\big)
\bigg)
,
\\
\widetilde F_{\tau=3}(z,\bar{z})&=\frac{(z\zb)^{\frac32}\eps^2\log^2z}{18(z-\zb)}\bigg(\log^2(1-z)-\log^2(1-\zb)-\log\left(\tfrac{1-z}{1-\zb}\right)-2\log z\log(1-z)
\nonumber \\
&\quad
+2\log\zb\log(1-\zb)-2\mathrm{Li}_2(1-z)+2\mathrm{Li}_2(1-\zb)
\bigg).
\end{align}

For the contribution from the first line in \eqref{eq:inversionintomixed}, we would need to consider the identity operator, and the contribution from twist-2 operators. As usual, we are interested in the limit
\begin{equation}
\frac1{v^{\Delta_\phi}}G_{\tau+\ell,\ell}(v,u)\sim 1+\frac12\gamma\log(1-\zb)+\frac18\gamma^2\log^2(1-\zb).
\end{equation}
In the case of identical external operators, the double-discontinuity is zero until second order in $\gamma$, since $\dDisc[\log(1-\zb)]=0$ while $\dDisc[\log^2(1-\zb)]=4\pi^2$. However, this is no longer the case with unequal external operators! Instead,
\begin{equation}
{\dDisc}^{2112}[\log(1-\zb)]=2\pi\sin(\pi\Delta_{12}),
\end{equation}
where $\Delta_{12}=\Delta_1-\Delta_2$.
Likewise,
\begin{equation}
{\dDisc}^{2112}[\log^2(1-\zb)]=4\pi\sin(\pi\Delta_{12})\log(1-\zb)+4\pi\cos(\pi\Delta_{12}).
\end{equation}

Taking these considerations into account, we can write down the contributions from the pure channel. We find
\begin{equation}
{\dDisc}^{\phi^2\phi\phi\phi^2}\bigg[\frac{u^{\frac{\Delta_{\phi^2}+\Delta_\phi}2}}{v^{\Delta_\phi}}\G_{\phi\phi\phi^2\phi^2}(v,u)\bigg]={\dDisc}^{\phi^2\phi\phi\phi^2}\bigg[\frac{u^{\frac{\Delta_{\phi^2}+\Delta_\phi}2}}{v^{\Delta_\phi}}F(1-\zb,1-z)\bigg],
\end{equation}
where
\begin{equation}
F(z,\zb)=1+F_{\phi^2}(z,\zb)+F_2(z,\zb)
\end{equation}and the three terms represent the contributions from $\1$, $\phi^2$ and $\mathcal J_\ell$, $\ell=2,4,\ldots$ respectively. The last part is the sum over the spinning twist-2 operators times their anomalous dimension $\Delta_{\mathcal J_{\ell}}-(2\Delta_\phi+\ell)=-\frac{\eps^2}{9\ell(\ell+1)}$; see \eqref{eq:deltaJellIntro}. This gives
\begin{align}
F_{2}(z,\zb)&=\frac{z\zb\log z}{18(z-\zb)}\left(4\,\mathrm{Li}_2(\zb)-4\,\mathrm{Li}_2(z)+4\log\left(\frac{1-\zb}{1-z}\right)+\log^2(1-\zb)-\log^2(1-z)\right)
\eps^2.
\end{align}
The contribution from $\phi^2$ is most conveniently expressed as
\begin{align}
\frac{F_{\phi^2}(z,\zb)}{(z\zb)^{\Delta_\phi}}&=\frac{2\log z\log(\frac{1-z}{1-\zb})}{3(1-z)}\eps
+\eps^2\bigg(
\frac{\log^2z\log(\frac{1-z}{1-\zb})}{18(1-z)}
\nonumber
\\
&\quad +\frac{\log z}{81(1-z)}\left((27\log(1-z)-56)\log\left(\frac{1-\zb}{1-z}\right)+18\mathrm{Li}_2(z)+36\zeta_2\right)\!\!
\bigg)
.
\end{align}
To find it, we used the formula \eqref{eq:scalarblockmixed} for the conformal block and the values $\Delta{\phi^2}-2\Delta_\phi=\frac\eps3+\frac{8\eps^2}{81}$ and $\lambda_{\phi\phi\phi^2}\lambda_{\phi^2\phi^2\phi^2}=4-\frac{8\eps}3$ for the involved conformal data. Moreover, we only included the leading term in the limit $\zb\to1$, since we are only interested in computing the conformal data of mixed-channel twist-3 operators.

Finally, we would also expect to get a contribution from twist-4 operators, which also have a single factor of $\log(1-\zb)$. However, it turns out that this contribution vanishes, since it is also proportional to an extra factor of $1-\zb$, and the inversion of $(1-\zb)\log(1-\zb)$ vanishes.

The inversions in the first line of \eqref{eq:inversionintomixed} can be computed using
\begin{align}
\kappa^{2112}_{\hb}\int\frac{d \zb}{\zb^2}k_\hb^{[\Delta_{21},\Delta_{12}]} &{\dDisc}^{2112}\bigg[\frac{\zb^{\frac{\Delta_1+\Delta_2}2}}{(1-\zb)^{\Delta_1}}\left(\frac{1-\zb}\zb\right)^\alpha \bigg]
\nonumber
\\&=
\frac{\Gamma(\hb+\frac{\Delta_{12}}2)\Gamma(\hb-\frac{\Delta_{12}}2)\Gamma(\hb+\frac{\Delta_1+\Delta_2}2-\alpha-1)}{\Gamma(2\hb-1)\Gamma(\hb-\frac{\Delta_1+\Delta_2}2+\alpha+1)},
\end{align}
which can be found from the definition using an integral representation for $k_\hb^{[\Delta_{21},\Delta_{12}]}(\zb)$. We will use it for $\Delta_1=\Delta_\phi$ and $\Delta_2=\Delta_{\phi^2}$.
Likewise, the inversions in the second line of \eqref{eq:inversionintomixed} can be computed using\footnote{Seemingly, there is a mismatch between the negative power $v^{-\frac{\Delta_\phi+\Delta_{\phi^2}}2}$ in \eqref{eq:inversionintomixed}, and the power $v^{-\Delta_1}$ chosen here. However, all the operators that enter the double-discontinuity in \eqref{eq:inversionintomixed} have at least an extra power of $v^{\Delta_\phi/2}$, meaning that all terms are of the form $v^{-\Delta_1+k}(\text{const.}+O(\eps))$ for integer $k$, where $\Delta_1=\Delta_\phi$.}
\begin{align}
\kappa_\hb^{1212}\int\frac{d\zb}{\zb^2}k_{\hb}^{[\Delta_{12},\Delta_{12}]} & {\dDisc}^{1212}\bigg[\frac{\zb^{\frac{\Delta_1+\Delta_2}2}}{(1-\zb)^{\Delta_1}}\left(\frac{1-\zb}\zb\right)^\alpha\bigg]
\nonumber
\\&= \frac{(\Gamma(\hb+\frac{\Delta_{12}}2)^2\Gamma(\hb+\frac{\Delta_1+\Delta_2}2-\alpha-1)}{\Gamma(2\hb-1)\Gamma(\hb-\frac{\Delta_1+\Delta_2}2+\alpha+1)\Gamma(\Delta_1-\alpha)^2}.
\end{align}

Putting all the pieces together, and computing the inversion integrals, we find the following anomalous dimensions for the twist-3 operators:
\begin{align}
\label{eq:gamma3ellres}
\gamma_{3,\ell}^{\neq0}&=\frac{2(-1)^\ell}{3(\ell+1)}\eps-\frac{4\eps^2}{9(\ell+3)(\ell-1)}\left(S_1(\ell)-\frac{\ell^3+2\ell^2+2\ell+3}{(\ell+1)^3}\right)
\nonumber \\
&\quad+(-1)^\ell\frac{9(\ell+1)(3\ell^2+6\ell-1)S_1(\ell)-38\ell^3-69\ell^2+38\ell-75}{81(\ell+3)(\ell+1)^2(\ell-1)}\eps^2+O(\eps^3).
\end{align}
The complete dimensions of the operators are given by adding $\Delta_\phi+\Delta_{\phi^2}+\ell$.
We also extracted the OPE coefficients to order $\eps$, and found that they agree with \eqref{eq:OPEasffsT3} upon putting $n=1$. To this order, in the $n=1$ case they represent OPE coefficients of the degenerate operators $\O_{\ell,1}$.
We have not determined the order-$\eps^2$ part of the OPE coefficients. They would contain an admixture of OPE coefficients of the operators $\O_{\ell,i}$ with $i>1$, \emph{i.e.} those with $\gamma^{(1)}_\ell=0$.

The first few cases read
\begin{align}
\Delta_{\partial^2\phi^3}&=\big[5-\frac32\eps\big]+\frac59\eps+\frac{161}{972}\eps^2 +O(\eps^3),
&
\Delta_{\partial^3\phi^3}&=\big[6-\frac32\eps\big]+\frac16\eps+\frac{17}{2592}\eps^2 +O(\eps^3),
\\
\Delta_{\partial^4\phi^3}&=\big[7-\frac32\eps\big]+\frac7{15}\eps+\frac{3559}{20250}\eps^2 +O(\eps^3),
&
\Delta_{\partial^5\phi^3}&=\big[8-\frac32\eps\big]+\frac29\eps+\frac{61}{1620}\eps^2 +O(\eps^3),
\end{align}
and so on. We also note that putting $\ell=0$, we find that
\begin{equation}
\Delta_{\O_{\ell,1}}\big|_{\ell=0}=3-\frac\eps2-\frac{\eps^2}{108}+O(\eps^3)=d-\Delta_\phi ,
\end{equation}
which is a shadow relation of a similar type as that noticed for $\phi^3$ theory in $6-\eps$ dimensions, where $\Delta_{\partial^\ell\phi^2}|_{\ell=0}+\Delta_\phi=d$ \cite{Goncalves:2018nlv}. We do not understand the exact meaning of such shadow relations beyond leading order.

\subsection{Comparison with numerics}
\begin{figure}
\centering
\includegraphics[width=\textwidth]{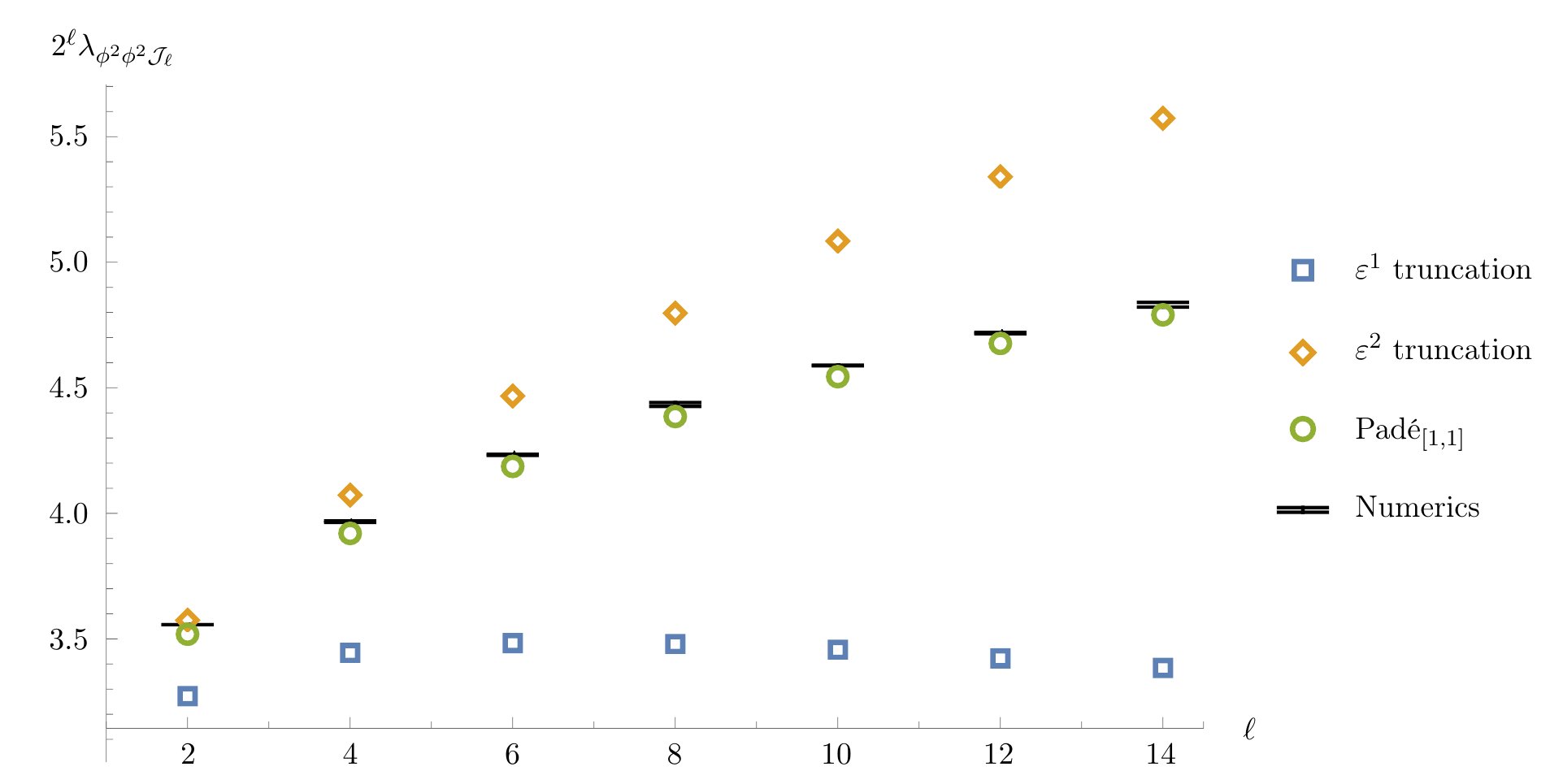}
\caption{Estimates for the OPE coefficients $\lambda_{\phi^2\phi^2\mathcal J_\ell}$ in three dimension, and comparison with the numerical values of \cite{Simmons-Duffin:2016wlq}.}\label{fig:opecoefs-ssss}
\end{figure}
With some new order-$\eps^2$ data at hand, we can make comparison with the numerical bootstrap data of \cite{Simmons-Duffin:2016wlq}.
This data was found by applying the extremal functional method to a collection of points on the boundary of the allowed region for the 3d Ising island.
It comprises a list of operators together with the estimates of their scaling dimensions and their OPE coefficients in the relevant OPEs involving $\phi$ and $\phi^2$, in that paper denoted $\sigma$ and $\epsilon$. We will compare the following sets of data:
\begin{figure}
\centering
\includegraphics[width=\textwidth]{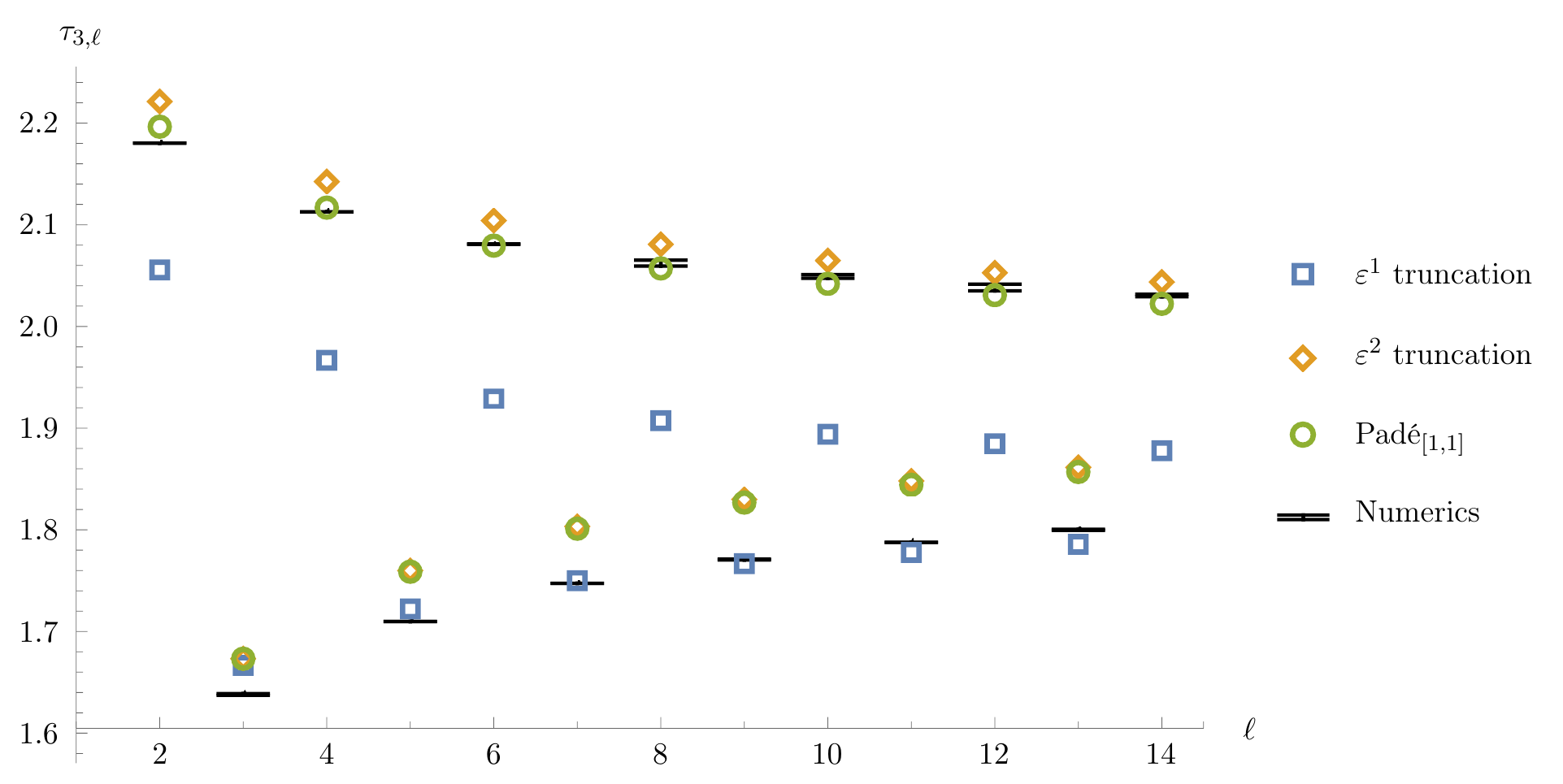}
\caption{Estimates for the twist $\tau_{3,\ell}$ in three dimension, and comparison with the numerical values of \cite{Simmons-Duffin:2016wlq}.}\label{fig:twis3dims}
\end{figure}
\begin{itemize}
\item OPE coefficients $\lambda_{\phi^2\phi^2\mathcal J_\ell}^2$ for the twist-2 operators $\mathcal J_\ell$, identified with the family $[\sigma,\sigma]_{0,\ell}$ of \cite{Simmons-Duffin:2016wlq}. Our data is given by \eqref{eq:OPEcoefsssJ}.
\item Scaling dimensions of operators $\mathcal O_{\ell,i}=\partial^\ell\phi^3$, identified with the family $[\sigma,\epsilon]_{0,\ell}$ of \cite{Simmons-Duffin:2016wlq}. Our data is given by \eqref{eq:gamma3ellres}.
\end{itemize}

In both cases, we compare the following three evaluations: 1) order-$\eps$ truncation, available before this work, 2) order-$\eps^2$ truncation, and 3) a Pad\'e approximant of the form\begin{equation}
\text{Pad\'e}_{[1,1]}(\eps)=\frac{a_0+a_1\eps}{1+b_1\eps}.
\end{equation}
The resulting values for OPE coefficients are displayed in figure~\ref{fig:opecoefs-ssss}, and for dimensions in figure~\ref{fig:twis3dims}. We find that for all OPE coefficients, and for the even-spin scaling dimensions, the new estimates take us closer to the numerical data.

\section{Checks and collected results}
\label{sec:collectedresults}

In this section we will list our results and do some checks with the literature. We start by summarizing the correlators that we have computed in closed form. For each correlator, we refer to the corresponding result:
\begin{align*}
	\expv{\varphi\varphi\varphi\varphi} :\qquad &
	 \quad \eqref{eq:ffffSapp},\,\eqref{eq:ffffTapp},\,\eqref{eq:ffffAapp}
	\nonumber \\
	\expv{\varphi\varphi ss} :\qquad &
	 \quad \eqref{eq:GffssAns}
	\nonumber \\
	\expv{ssss}: \qquad &
	 \quad \eqref{eq:Gssssfinal}
	\nonumber \\
	\expv{\varphi\varphi tt}: \qquad &
	 \quad \eqref{eq:GffttSfinal},\,\eqref{eq:GffttTfinal},\,\eqref{eq:GffttAfinal}
	\nonumber \\
	\expv{tttt} :\qquad &
	\quad \eqref{eq:GttttS},\,\eqref{eq:GttttT},\,\eqref{eq:GttttA},\,\eqref{eq:GttttT4},\,\eqref{eq:GttttH4},\,\eqref{eq:GttttB4}
	\nonumber \\
	\expv{sstt} :\qquad &
	\quad \eqref{eq:Gssttans}
	\nonumber
\end{align*}
In these expressions there may be some constants whose values are specified in the corresponding sections. Apart from $\langle ssss\rangle$ that matches with \cite{Henriksson:2020jwk} and $\langle \varphi \varphi \varphi \varphi\rangle$ that matches with \cite{Bissi:2019kkx} for $n=1$, the other expressions for the correlators have not been determined to the best of our knowledge.\footnote{We computed the expressions for $\langle \varphi \varphi \varphi \varphi\rangle$ purely from conformal data that was known before this paper; see appendix~\ref{app:fffforder2}.}

The knowledge of these correlators gives us access to the anomalous dimensions and OPE coefficients of the non-degenerate operators. The order-$\eps$ anomalous dimensions are, in most cases, known and have been tabulated in \cite{Henriksson:2022rnm} -- we find full agreement. For the OPE coefficients, our results are new except for the operators that appeared in the $\langle \varphi \varphi \varphi \varphi\rangle$ and  $\langle ssss\rangle$ correlators. 
By inputting order-$\eps$ anomalous dimensions, we can also find the leading OPE coefficients for the case of twice-degenerate operators.

\subsection{Operators without mixing}
\label{sec:mixingfreeops}

In table \ref{tab:nondegsIsing} we write part of the non-degenerate operator content for the Ising ($n=1$) case. This is divided into even (+) and odd ($-$) cases. Recall that, for Ising, we denote the basic field by $\phi$.

For twist 2 we have:
\begin{align}
	\label{eq:deltaevenl}
	\Delta_{2,\ell} &=2+\ell-\eps+\frac1{54}\left(1-\frac6{\ell(\ell+1)}\right)\eps^2+\ldots ,
	\\
	\label{eq:phialphaevenl}
	\lambda^2_{\phi \phi \mathcal{J}_\ell} &= \frac{2 \Gamma(\ell+1)^2}{\Gamma(2 \ell + 1)} \Big(1 +\big(-2 S_1(\ell) + S_1(2 \ell) \big) \eps \Big)+\ldots,
\end{align}
known to order $\eps^4$ from the literature, \cite{Derkachov:1997pf} and \cite{Alday:2017zzv} respectively. We determined the OPE coefficients $\lambda^2_{\phi^2 \phi^2 \mathcal{J}_\ell}$ to order $\eps^2$ in \eqref{eq:OPEcoefsssJ}.

\begin{table}
\centering
\caption{Results for non-degenerate operators in the Ising case.}\label{tab:nondegsIsing}
{\small
\renewcommand{\arraystretch}{1.25}
\begin{tabular}{|ll|ccc|}
\hline
$R$ & $(\tau_0,\ell)$ & $\Delta_\O$ & $\lambda^2_{\phi\phi\O}$ & $\lambda^2_{\phi^2\phi^2\O}$
\\\hline
$+$ & $(2,0)$ & $2-\frac23\eps$ & $2(1 - \frac{1}{3} \eps)$ & $8(1-\eps)$
\\
$+$ & $(2,\ell)_{\ell=2,4,\ldots}$ & \eqref{eq:deltaevenl} & \eqref{eq:phialphaevenl} & \eqref{eq:OPEcoefsssJ}
\\
$+$ & $(4,0)$ & $4+0\eps$ & $\frac{1}{54} \eps^2 \left(1 - \frac{47}{27} \eps \right)$ & $6\left(1-\frac{10}9\eps\right)$
\\
$+$ & $(4,2)$ & $6-\frac59\eps$ & $\frac{1}{1440} \eps^2 \left(1 - \eps \right)$ & $\frac85\left(1-\frac{65}{54}\eps\right)$
\\
$+$ & $(6,2)$ & $8 - \frac{10}{9} \eps $ & $\frac{1}{2592000}\eps^4$ & $\frac{1}{5} \left(1 - \frac{1493}{1080} \varepsilon \right)$
\\
$+$ & $(8,0)$ & $8-\frac89\eps$ & $O(\eps^4)$ & $\frac16\left(1-\frac{259}{216}\eps\right)$
\\
$+$ & $(10,0)$ & $10-\frac53\eps$ & $O(\eps^4)$ & $\frac1{175}\left(1-\frac{71}{120}\eps\right)$
\\
$+$ & $(12,0)$ & $10-\frac{16}{15}\eps$ & $O(\eps^4)$ & $\frac1{630}\left(1-\frac{105589}{75600}\eps\right)$
\\\hline\hline
$R$ & $(\tau_0,\ell)$ & $\Delta_\O$ & $\lambda^2_{\phi\phi^2\O}$ &
\\\hline
$-$ & $(1,0)$ & $1-\frac12\eps+\frac1{108}\eps^2$ & $2(1 - \frac{1}{3} \eps)$ & $$
\\
$-$ & $(3,\ell)_{\ell=2,3,4,5,7}$ & \eqref{eq:twist3delta} & \eqref{eq:twist3OPE} &
\\\hline
\end{tabular}
}
\end{table}

In  the mixed channel, we have determined the anomalous dimensions \eqref{eq:gamma3ellres} for the operators $\O_\ell$ of the form $\de^\ell\phi^3$ that have a non-vanishing order-$\eps$ anomalous dimension; see \eqref{eq:gammanonzero}. The corresponding full scaling dimensions take the form
\begin{align}
	\Delta_{\O_\ell} &= 3 + \ell +  \eps\left(-\frac{7}{6} + (-1)^\ell \frac{2}{3(\ell + 1)} \right) 
	+ \eps^2\bigg[\frac{41}{324}-\frac{4S_1(\ell)}{9(\ell-1)(\ell+3)}	
	\nonumber
	+\frac{4(\ell^3+2\ell^2+2\ell+3)}{9(\ell-1)(\ell+1)^3(\ell+3)}
	\nonumber
	\\
	&\quad
	+(-1)^\ell\frac{9(\ell+1)(3\ell^2+6\ell-1)S_1(\ell)+38\ell(1-\ell^2)-69\ell^2-75}{81(\ell-1)(\ell+1)^2(\ell+3)}\bigg]+O(\eps^3).
\label{eq:twist3delta}
\end{align}
The OPE coefficients for these operators are
\begin{align}
	\lambda^2_{\phi \phi^2{\O_\ell}} \ &= \ 2(\ell + 1 + 2 (-1)^\ell) \frac{\Gamma(\ell + 2)^2}{\Gamma(2\ell+ 3)} \Bigg[1-\eps \left(-\frac{7}{6} + (-1)^\ell \frac{2}{3(\ell + 1)} \right) S_1(2 \ell + 2)
	\nonumber \\
	& \quad +\eps \left( \frac{5 \ell^2 + 10 \ell - 3 }{3 (\ell -1 )(\ell + 1)(\ell+3)} (-1)^\ell - \frac{7 \ell^2 + 14 \ell - 15}{3 (\ell - 3)(\ell+3)} \right) S_1(\ell + 1)
	\nonumber\\
	& \quad + \eps\left( -\frac{2 \ell(3 \ell^2 + 8 \ell + 7) }{3(\ell -1)(\ell + 1)^2 (\ell+3)}(-1)^\ell + \frac{8 \ell^3 + 23 \ell^2 + 38 \ell + 3}{12(\ell -1)(\ell + 1) (\ell+3)} \right) \Bigg]+O(\eps^2),
	\label{eq:twist3OPE}
\end{align}
following from \eqref{eq:OPEasffsT3}, \eqref{eq:alphaasffsT3} and \eqref{eq:gammaasffsT3} upon putting $n=1$ and using the observation \eqref{eq:surprisingobservation}.
For twist 5 and above, the squared OPE coefficients are all $\mathcal{O}(\eps^2)$ and above and we have not attempted to determine them.

\begin{table}
\centering
\caption{Results for non-degenerate operators with an even number of fields, part 1. The normalization constants $\mathcal N_R$ are given in \eqref{eq:normVVVV}--\eqref{eq:Norm3}.}\label{tab:nondegseven1}
{\small
\renewcommand{\arraystretch}{1.25}
\begin{tabular}{|ll|cccc|}
\hline
$R$ & $(\tau_0,\ell)$ & $\Delta_\O$ & $\lambda^2_{\varphi\varphi\O}$ & $\lambda^2_{ss\O}$ & $\lambda^2_{tt\O}$
\\\hline
$S$ & $(2,0)$ & $2-\eps+\frac{n+2}{n+8}\eps$ & $\frac2n\left(1-\frac{n+2}{n+8}\eps\right)$ & $\frac 8n\left(1-\frac{n+2}{n+8}\eps\right)$ & $\frac 8n\left( 1 -\frac{n+6}{n+8}\eps \right)$
\\
$S$ & $(2,\ell)_{\ell=2,4,\ldots}$ & \eqref{eq:deltaS2l} & \eqref{eq:ffS2l} & \eqref{eq:sssstwist2inputa0},\eqref{eq:sssstwist2inputalpha} & \eqref{eq:tttta0S},\eqref{eq:ttttalphaS}
\\
$S$ & $(4,0)$ & $4+0\eps$ & $\frac{n+2}{2n(n+8)^2}\eps^2$ & $\frac{2n+4}{n}\left(1-\frac{10}{n+8}\eps\right)$ & $\frac{8}{n(n+2)}\left( 1 -\frac{n+10}{n+8}\eps\right)$
\\
$S$ & $(6,0)$ & $6-\frac{12}{n+8}\eps$ & $O(\eps^4)$ & $\frac{2n-2}{3n}\left( 1-\frac{40-7n}{12(n+8)}\eps \right)$ & $\frac{2}{3n(n-1)}\left( 1 -\frac{41n+40}{12(n+8)}\eps \right)$
\\\hline\hline
$R$ & $(\tau_0,\ell)$ & $\Delta_\O$ & $\lambda^2_{\varphi\varphi\O}/\mathcal N_T$ & $\lambda^2_{tt\O}/\mathcal N_T$ & $\lambda^2_{st\O}$
\\\hline
$T$ & $(2,0)$ & $2-\eps +\frac{2}{n+8}\eps$ & $2\left(1-\frac2{n+8}\eps\right)$ & $\frac{4(n+4)(n-2)}{(n+2)(n-1)}\left( 1 -\frac6{n+8}\eps \right)$ & $\frac8n \left( 1 -\frac{n+6}{n+8}\eps \right)$
\\
$T$ & $(2,\ell)_{\ell=2,4,\ldots}$ & \eqref{eq:deltaT2l} &
 \eqref{eq:ffS2l} &\eqref{eq:tttta0T},\eqref{eq:ttttalphaT} & \eqref{eq:lambdast20}
\\
$T$ & $(4,0)$ & $4-\frac{n}{n+8}\eps$ & $\frac{n+4}{4(n+8)^2}\eps^2$ & $\frac{8(n-2)}{(n+2)(n-1)}\left( 1 -\frac{n+20}{2(n+8)}\eps \right)$ & $\frac{n+4}n\left(1-\frac{10}{n+8}\eps\right)$
\\
$T$ & $(6,0)$ & $6-\frac{n+12}{n+8}\eps$ & $O(\eps^4)$ & $\frac{2(n+4)}{3(n+2)(n-1)}\left( 1 -\frac{25n+40}{12(n+8)}\eps \right)$ & $\frac{n-2}{3n}\left(1-\frac{n+40}{12(n+8)}\eps\right)$
\\
$T$ & $(6,1)$ & $7-\frac{3n+32}{3(n+8)}$ & --- & --- & $\frac14\left(1-\frac{27n+608}{288(n+8)}\eps\right)$
\\\hline\hline
$R$ & $(\tau_0,\ell)$ & $\Delta_\O$ & $\lambda^2_{\varphi\varphi\O}/\mathcal N_A$ & $\lambda^2_{tt\O}/\mathcal N_A$ & 
\\\hline
$A$& $(2,\ell)_{\ell=1,3,\ldots}$ & \eqref{eq:deltaA2l} &  \eqref{eq:ffS2l} & \eqref{eq:tttta0A},\eqref{eq:ttttalphaA} & 
\\
$A$& $(4,1)$ & $5-\frac{n+6}{n+8}\eps$ & $\frac{n+2}{18(n+8)^2}\eps^2$ & $\frac{4n}{(n+2)(n-1)}\left( 1 -\frac{7n+50}{6(n+8)}\eps \right)$ & 
\\\hline
\end{tabular}
}
\end{table}

\begin{table}
\centering
\caption{Results for non-degenerate operators with an even number of fields, part 2. The normalization constants $\mathcal N_R$ are given in \eqref{eq:Norm2}--\eqref{eq:Norm3}.}\label{tab:nondegseven2}
{\small
\renewcommand{\arraystretch}{1.25}
\begin{tabular}{|ll|cc|}
\hline
$R$ & $(\tau_0,\ell)$ & $\Delta_\O$& $\lambda^2_{tt\O}$
\\\hline
$T_4$ & $(4,0)$ & $4-\frac{2n+4}{n+8}\eps$ & $6\left( 1 -\frac8{n+8}\eps \right)\mathcal N_{T_4}$
\\
$T_4$ & $(4,2)$ & $4-\frac{2(3n+11)}{3(n+8)}\eps$ & $\frac85\left( 1-\frac{153n+1696}{180(n+8)} \eps\right)\mathcal N_{T_4}$
\\
$T_4$ & $(6,2)$ & $6-\frac{2(3n+16)}{3(n+8)}\eps$ & $\frac15 \left( 1-\frac{657n+4300}{1800(n+8)} \eps \right)\mathcal N_{T_4}$
\\
$T_4$ & $(8,0)$ & $8-\frac{2(3n+14)}{3(n+8)}\eps$ & $\frac16 \left( 1 -\frac{105n+728}{72(n+8)}\eps \right)\mathcal N_{T_4}$
\\
$T_4$ & $(10,0)$ & $10-\frac{2(n+7)}{n+8}\eps$ & $\frac{1}{175} \left( 1 -\frac{1885 n+6752}{840 (n+8)}\eps \right)\mathcal N_{T_4}$
\\
$T_4$ & $(12,0)$ & $12-\frac{2(5n+26)}{5(n+8)}\eps$ & $\frac{1}{630} \left( 1 -\frac{62655 n+336488}{25200 (n+8)}\eps \right)\mathcal N_{T_4}$
\\\hline
$H_4$ & $(4,1)$ & $5-\frac{2n+8}{n+8}\eps$ & $2\left( 1 -\frac{n+14}{2(n+8)}\eps \right)\mathcal N_{H_4}$
\\\hline
$B_4$ & $(4,2)$ & $6-\frac{6n+34}{3(n+8)}\eps$ & $\left(1 -\frac{9n+62}{9(n+8)}\eps\right)\mathcal N_{B_4}$
\\
$B_4$ & $(6,0)$ & $6-\frac{2n+10}{n+8}\eps$ & $\left(1 -\frac34\eps\right)\mathcal N_{B_4}$
\\\hline
\end{tabular}
}
\end{table}

In table \ref{tab:nondegseven1} and \ref{tab:nondegseven2} we present data for non-degenerate operators with an even number of fields in the $O(n)$ model. The irreps in table \ref{tab:nondegseven1} are common to $\expv{\varphi\varphi\varphi\varphi}$ and $\expv{tttt}$. Likewise, in table~\ref{tab:nondegsodd} we present data for non-degenerate operators with an odd number of fields.

In addition to the properly non-degenerate operators, there are points in the $(\tau_0,\ell)$ plane where there is more than one operator, but only one of $\varphi^4$ type. One example is $\tau_0=6$, $\ell=0$, singlet, where we have $\varphi^6_S$ and $\square\varphi^4_S$, with dimensions $[6-3\eps]+\frac{3(n+14)}{n+8}\eps$ and $2+[4-2\eps]+\frac{2n+4}{n+8}\eps$ respectively. Only the second of these operators has an OPE coefficient at leading order in the $s\times s$ OPE, and we can therefore determine its conformal data. This assertion can be confirmed by the fact that the dimension reported for $(\tau_0,\ell)=(6,0)$ precisely matches with that of $\square\varphi^4_S$. In tables~\ref{tab:nondegsIsing}--\ref{tab:nondegsodd}, we have included many similar cases.

\begin{table}
\centering
\caption{Results for non-degenerate operators with an odd number of fields.}
\label{tab:nondegsodd}
{\small
\renewcommand{\arraystretch}{1.25}
\begin{tabular}{|cc|ccc|}
\hline
$R$ & $(\tau_0,\ell)$ & $\Delta_\O$ & $\lambda^2_{\varphi s\O}$ & $\lambda^2_{\varphi t\O}$
\\\hline
$V$ & $(1,0)$ & $1 - \frac{1}{2} \eps$ & $\frac{2}{n} \left(1 - \frac{n+2}{n+8} \right) \eps$ & $2 \mathcal{N}_T \left(1 - \frac{2}{n+8} \eps \right)$
\\
$V$ & $(3,1)$ & $3-\frac{n+20}{2(n+8)}\eps$ & $\frac{2(n-1)}{3n} \left(1 - \frac{n+11}{3(n+8)} \eps\right)$ & $\frac{2 \mathcal{N}_T}{3(n-1)} \left(1 - \frac{4n+11}{3(n+8)} \eps \right)$
\\\hline
$T_3$ & $(3,0)$ & $3-\frac{3n+12}{2(n+8)}\eps$ & --- & $\frac{3\sqrt{n+4}}{\sqrt{3(n+2)}}\left( 1-\frac{4}{n+8}\eps \right)$
\\
$T_3$ & $(3,2)$ & $5-\frac{9n+52}{6(n+8)}\eps$ & --- & $\frac{\sqrt{n+4}}{2\sqrt{3(n+2)}}\left( 1-\frac{63n+524}{72(n+8)}\eps \right)$
\\
$T_3$ & $(3,3)$ & $6-\frac{3n+22}{2(n+8)}\eps$ & --- & $-\frac{2\sqrt{n+4}}{35\sqrt{3(n+2)}}\left( 1-\frac{589n+3911}{420(n+8)}\eps \right)$
\\
$T_3$ & $(3,4)$ & $7-\frac{15n+92}{10(n+8)}\eps$ & --- & $\frac{\sqrt{n+4}}{18\sqrt{3(n+2)}}\left( 1-\frac{5175n+38236}{3600(n+8)}\eps \right)$
\\
$T_3$ & $(3,5)$ & $8-\frac{9n+64}{6(n+8)}\eps$ & --- & $-\frac{2\sqrt{n+4}}{231\sqrt{3(n+2)}}\left( 1-\frac{1070840n+157257}{83160(n+8)}\eps \right)$
\\
$T_3$ & $(3,7)$ & $10-\frac{3n+21}{2(n+8)}\eps$ & --- & $-\frac{2\sqrt{n+4}}{2145\sqrt{3(n+2)}}\left( 1-\frac{3708339n+541550}{240240(n+8)}\eps \right)$
\\\hline
\end{tabular}
}
\end{table}

\subsection{Cases with two degenerate operators}

When degeneracy occurs one has to unmix the degenerate operators. For twice degenerate operators this procedure has been done in section \ref{sec:fseps2}.

\begin{table}
\centering
\caption{Results for pairs of degenerate operators in the Ising case. Anomalous dimensions taken from \cite{Henriksson:2022rnm}.
}\label{tab:twodegsIsing}
{\small
\renewcommand{\arraystretch}{1.25}
\begin{tabular}{|cc|ccc|}
\hline
$R$ & $(\tau_0,\ell)$ & $(\Delta_{\O_1},\Delta_{\O_2})$ & $(\lambda^2_{\phi\phi\O_1},\lambda^2_{\phi\phi\O_2})$ & $(\lambda^2_{\phi^2\phi^2\O_1},\lambda^2_{\phi^2\phi^2\O_2})$
\\\hline
$+$ & $(4,4)$ & $\left(8-\frac{14}9\eps,8-\frac{11}{15}\eps\right)$ & $\left(\frac{\eps^2}{419580},\frac{\eps^2}{23976}\right)$ & $\left(\frac{125}{2331},\frac{8}{37}\right)$
\\
$+$ & $(6,4)$ & $10-\frac{131\pm\sqrt{697}}{90}\eps$  & $\left(O(\eps^4),O(\eps^4)\right)$
& $\frac{2091\mp  71   \sqrt{697}}{107338}$
\\
$+$ & $(8,2)$ & $\left(10-\frac{17}9\eps,10-\frac{16}{15}\eps\right)$ & $\left(O(\eps^4),O(\eps^4)\right)$  & $\left(\frac1{999},\frac4{111}\right)$
\\\hline\hline
$R$ & $(\tau_0,\ell)$ & $(\Delta_{\O_1},\Delta_{\O_2})$ & \multicolumn{2}{c|}{$(\lambda^2_{\phi\phi^2\O_1},\lambda^2_{\phi\phi^2\O_2})$}
\\\hline
$ -$ & $(3,6)$ & $\left(9-\frac32\eps+O(\eps^2),9-\frac{15}{14}\eps+\frac{48799}{277830}\eps^2\right)$ & \multicolumn{2}{c|}{$\left(O(\eps^2),\frac{3}{572}(1-\frac{8019289 }{5045040}\eps)\right)$}
\\
$ -$ & $(3,8)$ &$\left(11-\frac32\eps+O(\eps^2),11-\frac{59}{54}\eps+\frac{318589}{1837080}\eps^2\right)$ & \multicolumn{2}{c|}{$\left(O(\eps^2),\frac1{2210}(1-\frac{110303617}{60147360}\eps)\right)$ }
\\\hline
\end{tabular}
}
\end{table}

In table \ref{tab:twodegsIsing} we show some degenerate operators and relative CFT data in the Ising ($n=1$) case.
One can perform the same computation in the case of twice-degenerate operators at general $n$. The resulting expressions are quite lengthy and not particularly illuminating. For instance, there are two singlet operators at twist four and spin two, with scaling dimensions
\begin{equation}
\label{eq:S42}
\Delta_{1,2}=6-2\eps+\frac{9 n+44\mp\sqrt{9 n^2-8 n+624}}{6 (n+8)}\eps.
\end{equation}
We find that their OPE coefficients are
\begin{align}
\lambda_{\varphi\varphi\O_{1,2}}^2&=\frac{n+2}{320n(n+8)^2}\frac{\pm n\mp 76+3\sqrt{9n^2-8n+624}}{\sqrt{9n^2-8n+624}}\eps^2+O(\eps^3),
\\
\lambda_{ss\O_{1,2}}^2&=\frac{\mp(9n^2-n+92)+(3n+1)\sqrt{9n^2-8n+624}}{5n\sqrt{9n^2-8n+624}}+O(\eps).
\end{align}

\section{Discussion}
\label{sec:disc}

In this paper, we have applied a combination of large spin perturbation theory and the Lorentzian inversion formula to systems of non-identical correlators. In particular, we have considered the $\eps$-expansion of the $O(n)$ CFT, and studied all four-point correlators of pairwise identical operators drawn from the set $\left\{\varphi,s=\varphi^2_S,t=\varphi^2_T\right\}$. This set of operators is the same as the one considered in the most recent developments in the numerical conformal bootstrap \cite{Chester:2019ifh,Chester:2020iyt}. 

In our approach, we determined the correlators in an iterative procedure using the Lorentzian inversion formula, which is sensitive only to the double-discontinuity of the correlator. The strategy for each correlator was to first determine the double-discontinuity in terms of a few crossed-channel operators, then to compute the inversion integral, and finally to resum the data to produce the correlator. We started from the $\varphi$ four-point function and reviewed the considerations of \cite{Alday:2017zzv}, where the double-discontinuity to order $\eps^3$ is produced purely form the identity operator and the bilinears $s$ and $t$. This provides values of the OPE coefficients $\lambda_{\varphi \varphi s}$ and $\lambda_{\varphi \varphi t}$, which also appear in the mixed correlators of $\varphi,s$ and $\varphi,t$ respectively. These correlators then determine the contribution of leading twist operators, represented by their OPE coefficients $a_{2,\ell}$, which provide the double-discontinuity in the correlators of the bilinears. The final step of determining $\langle sstt\rangle$ required a careful consideration of the inversion into the mixed correlators.
The whole process can be summarized by the following diagram:
\begin{equation}
\begin{tikzcd}
\langle\varphi\varphi\varphi\varphi\rangle \arrow[d,"{\lambda_{\varphi\varphi s}}" left]
\arrow[dr,"{\lambda_{\varphi\varphi s},\,\lambda_{\varphi\varphi t}}"]
&
\\
\langle\varphi\varphi ss\rangle \arrow[d,"{a_{2,\ell}}" left] & \langle\varphi\varphi tt\rangle \arrow[d,"{a_{S,2,\ell},a_{T,2,\ell}}"]
\\
\langle ss ss\rangle \arrow[dr,"{a_{2,\ell}}"] & \langle tt tt\rangle \arrow[d,"{a_{S,2,\ell}}"]
\\
& \langle sstt\rangle
\end{tikzcd}
\end{equation}

Along the way, a few free parameters were introduced, and we were able to fix them at the end using consistency with crossing of the final results.
The result is the determination of all order-$\eps$ correlators of pairwise operators; see the list at the beginning of section~\ref{sec:collectedresults}. From the conformal block decomposition of these correlators, we have been able to confirm the order-$\eps$ anomalous dimension of non-degenerate operators, and computed their OPE coefficients; see tables~\ref{tab:nondegsIsing}, \ref{tab:nondegseven1}, \ref{tab:nondegseven2} and \ref{tab:nondegsodd}. The results for the OPE coefficients appearing in these correlators are new, except for the cases of $\langle\varphi\varphi\varphi\varphi\rangle $ and $\langle ssss\rangle$.

Apart from determining the structure of operators and the interdependence of conformal data, the computations in this paper involved several difficulties, some of which are more technical in nature and have not been emphasized in the main text. This includes manipulations of conformal blocks in general spacetime dimensions (appendix~\ref{app:4-epsblocks}), computing several complicated inversion integrals (appendix~\ref{app:inversions}) and determining the closed-form of the correlator from the CFT data.

In the $n=1$ case, we were able to move to order $\eps^2$ in the correlator $\langle\phi\phi\phi^2\phi^2\rangle$ in section~\ref{sec:fseps2}, thanks to a surprising result that appeared in the order-$\eps$ computation; see \eqref{eq:surprisingobservation}. In short, this result means that, effectively, all operators contributing to the double-discontinuity at order $\eps^2$ are non-degenerate, and their contribution can therefore be evaluated. This yielded some completely new pieces of conformal data, such as the OPE coefficients
\begin{equation}
\lambda^2_{\phi^2\phi^2\mathcal J_\ell}=\frac{8\Gamma(\ell+1)^2}{\Gamma(2\ell+1)}\left(1+\eps \alpha^{(1)}_\ell+\eps^2\alpha^{(2)}_\ell\right)+ O(\eps^3)
\end{equation}
given in \eqref{eq:OPEcoefsssJ}, and the dimensions
\begin{equation}
\Delta_{3,\ell}^{\neq0} = \Delta_\phi+\Delta_{\phi^2}+\ell+\gamma_{3,\ell}, \qquad \gamma_{3,\ell}=\eps\gamma_{3,\ell}^{(1)}+\eps^2\gamma_{3,\ell}^{(2)} + O(\eps^3)
,
\end{equation}
given in \eqref{eq:gamma3ellres}. In most cases, these new determinations take us closer to the numerical data in the 3d Ising CFT, determined in \cite{Simmons-Duffin:2016wlq}; see figures~\ref{fig:opecoefs-ssss} and \ref{fig:twis3dims}.

An assumption in our work is that the inversion formula extends down to spin zero, except at twist $\tau=2$. In section~\ref{sec:finitespincontributions} we considered the possibility of a finite-spin solution contribution in the $\langle ssss\rangle$ correlator based on the considerations in \cite{Heemskerk:2009pn}. However, such contribution must be proportional to a non-analytic contribution at twist $\tau=4$, but we found that the inversion formula result is correct for this dimension (and indeed for all the non-degenerate spin-zero operators at higher-twist operators). Also in the $\langle tttt\rangle$ correlator, we found that the inversion formula gives correct predictions for the scaling dimension at spin zero.

Another way to perform these calculations would be the use of Feynman diagrams, \emph{i.e.} perturbation theory. We would like to stress that in this work we wanted to emphasize the power of the Lorentzian inversion formula and test the limits of its applicability, without using any other method. Also, while order-$\eps$ results can be obtained considering a relatively low number of loop integrals, the evaluation of the integrals tends to get complicated already at second order.

A possible application of our results would be to conformal perturbation theory \cite{Cardy:1996xt,Komargodski:2016auf,Amoretti:2017aze}. In conformal perturbation theory, one considers a controlled perturbation of the CFT with an operator $\O$, which may or may not be relevant:
\begin{equation}
S\mapsto S+\int d^dxg\O(x).
\end{equation}
Quantities away from the CFT, or in a nearby CFT, can then be computed in a series in $g$. At leading order in $g$, the corrections induced by this perturbation involve OPE coefficients with $\O$, evaluated in the CFT. To go to subleading order, order $g^2$, one needs to know four-point functions of the CFT. A concrete setup that could be studied in the $\eps$-expansion with the help of the results in our paper is the perturbation of $N$ copies of the Ising CFT by the operator $\sum^N_{A\neq B}(\phi^2)_A(\phi^2)_B$. This specific setup was considered in \cite{Komargodski:2016auf} in order to study a disordered fixed-point. The determination of order $g^2$ data using conformal perturbation theory involves integrals of the correlators $\langle\phi\phi\phi^2\phi^2\rangle$ and $\langle\phi^2\phi^2\phi^2\phi^2\rangle$.  

The machinery considered in this paper naturally extends to other $\phi^4$ theories, and interesting targets would be theories with symmetry groups that are hypercubic, $O(m)\times O(n)$ and more generally $G^n\rtimes S_n$ for different groups $G$, recently studied in \cite{Kousvos:2021rar}. For these theories, it is difficult to find small isolated islands using the bootstrap, and additional input from the $\eps$-expansion might guide the search.
The algorithm would be completely analogous to the considerations in this paper:
\begin{equation}
\begin{tikzcd}
\langle\phi\phi\phi\phi\rangle \arrow[d]
\\
\langle\phi\phi \phi^2_R\phi^2_R\rangle \arrow[d]
\\
\langle \phi^2_R\phi^2_R\phi^2_R\phi^2_R\rangle \arrow[d]
\\
\langle\phi^2_R\phi^2_R \phi^2_{R'}\phi^2_{R'}\rangle
\end{tikzcd}
\end{equation}
The first step of the $\phi$ four-point function was worked out to order $\eps^3$ in \cite{Henriksson:2020fqi}. This involves all representations $R$ appearing in the tensor product of two fundamental fields, and the only non-trivial step is to determine the crossing matrix $M_{RR'}$. Then one would consider mixed correlators, involving the bilinears $\phi^2_R$, and finally the correlators involving only bilinear operators.

As an alternative approach, one can start directly with the correlator of bilinear operators, and make an ansatz for the contribution of the twist-2 operators which source the double-discontinuity. In this way, the results would apply also to weakly coupled four-dimensional gauge theories containing scalars, where $\varphi$ is not gauge-invariant and therefore not in the spectrum. For flavor-singlet bilinear operators, this was in fact considered in \cite{Henriksson:2017eej}, but can also be extended to non-singlet bilinears.\footnote{In gauge theories, one must allow for a logarithmic scaling of the anomalous dimensions of twist-two operators, $\gamma_{2,\ell}=(c_1 S_1(\ell)+c_0)g_{\mathrm{YM}}^2+O(g_{\mathrm{YM}}^4)+O(1/\ell)$.}

Another potential extension of our work would be to include more correlators, either in the $O(n) $ CFT or in a general $\phi^4$ theory. If one were to include external operators that are cubic or quartic in the field, such as $\varphi^3_{T_3}$ or $\varphi^4_S$, more exchanged operators would fall below the double-twist threshold and the contribution of such operators has to be carefully analyzed. If one instead includes spinning operators, for instance the global symmetry current $J=\mathcal J_{A,1}$ or the stress tensor $T=\mathcal J_{S,2}$, one would need to consider the inversion formula for spinning external operators, perhaps using the technology of \cite{Kravchuk:2018htv}.

\acknowledgments

We thank Ning Su, Aninda Sinha and Aleix Gimenez-Grau for useful discussions and access to unpublished results. We thank Alessandro Vichi for comments on an earlier draft of this manuscript. This project has received its funding from the European Research Council (ERC) under the European Union's Horizon 2020 research and innovation programme (grant agreement no.~758903).

\appendix

\section{Conformal blocks in $\boldsymbol{4 - \varepsilon}$ dimensions}
\label{app:4-epsblocks}

In section \ref{sec:two}, we wrote down the simple expressions for the conformal blocks in the exact $z \to 0$ limit, and in the 4d limit, where $\eps \to 0$. However, to find the OPE coefficients for the operators beyond twist two, we need to go beyond the leading power of $z$ in the collinear expansion for arbitrary dimension. Here we will work with scalar externals which may have different scaling dimensions. We use $\Delta_{12} = \Delta_1 - \Delta_2$ and $\Delta_{34} = \Delta_3 - \Delta_4$.
In the collinear limit, the blocks may be written as a series expansion \cite{Simmons-Duffin:2016wlq}
\begin{align}\label{eq:blocksgenerald}
G^{(d), [\Delta_{12}, \Delta_{34}]}_{\Delta, \ell}(z, \bar z) = z^{h} \sum^\infty_{k = 0} z^k \sum_{m = -k}^k \, c^{(d), [\Delta_{12}, \Delta_{34}]}_{k, m} k_{\bar h + m} (\bar z)
.
\end{align}
The coefficients $c^{(d), [\Delta_{12}, \Delta_{34}]}_{k, m}$ may be determined by solving the Casimir equation order-by-order in $z$. The Casimir equation for this case takes the form
\begin{align}
\mathcal{C}_2 \ G^{(d), [\Delta_{12}, \Delta_{34}]}_{\Delta, \ell}(z, \bar z) = \left( h(h+1- d) + \bar h ( \bar h - 1 ) \right) \, G^{(d), [\Delta_{12}, \Delta_{34}]}_{\Delta, \ell}(z, \bar z)
,
\end{align}
where
\begin{align}
\mathcal{C}_2 = D_z + D_{\bar z} + (d - 2) \frac{z \bar z}{z - \bar z} \left((1 - z) \partial_z - (1 - \bar z) \partial_{\bar z} \right) ,
\end{align}
and
\begin{align}
D_z = z^2 \partial_z (1 - z) \partial_z + (\Delta_{12} - \Delta_{34}) z^2 \partial_z + \Delta_{12} \Delta_{34} z .
\end{align}

In this paper, we have computed all of the coefficients up to $k = 3$ for general scalar externals, and up to $k = 6$ for identical externals. The expressions quickly become rather complicated. We record the first set, $k = 1$, below:
\begin{align}
c_{0,0} \ & = \ 1
,
\\
c_{1,-1}\ & = \ \frac{(\bar h-h) (d - 2)}{d - 4 + 2(\bar h-h)}
,
\\
c_{1,0}\ & = \ \frac{1}{2} \left(-\frac{\Delta _{12} \Delta _{34} \left(d
(h-1)-2 \left(\bar h^2-\bar h+h-1\right)\right)}{(\bar h-1) \hb (d-2
(h+1))}-\Delta _{12}+\Delta _{34}+h\right)
,
\label{eq:c10}
\\
c_{1,1}\ & = \ -\frac{(d-2) \left(\bar h^2-\Delta _{12}^2\right) \left(\bar h^2-\Delta
_{34}^2\right) (\bar h+h-1)}{4 \bar h^2 (2 \bar h-1) (2 \bar h+1) (d-2 (\bar h+h+1))}
.
\end{align}
These expressions are valid for general $d$. In this paper, we are only interested in the order-$\eps$ corrections to the conformal blocks, so to use them we set $d = 4 - \eps$ and expand around $\eps = 0$.

\section{Dimensional shift in OPE coefficients in $\boldsymbol{\langle ssss\rangle }$}
\label{app:dimensionalshift}

Consider the conformal block decomposition of the $\langle ssss \rangle$ correlator

\begin{equation}
\label{eq:CBdimensionalDependence}
\G_{ssss}(u,v)=1+\sum_{\tau,\ell}a^{(0)}_{\tau,\ell}\left(
1+\eps \alpha^{(d)}_{\tau,\ell} \right)G_{\tau +\gamma_{\tau ,\ell},\ell}^{(d)}(u,v)+O(\eps^2)
,
\end{equation}
where we have explicitly written the $d$ dependence of the conformal blocks and the order $\eps$ correction to the OPE coefficients $\alpha^{(d)}_{\tau_0,\ell}$, which, working to order $\eps$, are the only elements in \eqref{eq:CBdimensionalDependence} that depend on dimension.

In the main computations of this paper, we have made use of the $d=4$ dimensional conformal blocks for simplicity. However, the physical CFT data should be expressed in terms of the decomposition in $d=4-\eps$ dimensional blocks. The difference between these two sets of conformal blocks starts at order $\eps$, so it will affect $\alpha_{\tau, \ell}$ but not $a^{(0)}_{\tau, \ell}$. To leading order in $\eps$, this comparison can be done with a linear shift in the corrections to the OPE coefficients:
\begin{equation}\label{eq:4dto4minusepsshift}
\alpha_{\tau,\ell}^{(4-\eps)}=\alpha_{\tau,\ell}^{(4)}+\Delta\alpha_{\tau,\ell}
.
\end{equation}
The quantity $\Delta\alpha_{\tau,\ell}$ can be computed order-by-order in $\tau$ and then the full expression can be determined. It is enough to do this for the free theory: the form of $\Delta \alpha_{\tau, \ell}$ will be the same in the interacting theory to order $\eps$. Setting $c=\frac 4n$, the result takes the form
\begin{align}
\Delta\alpha_{\tau,\ell}&=\frac1{ c (-1)^{\tau/2}+(\ell+1)(\ell+\tau-2)}\bigg[
\frac{(\ell+1)(\ell+\tau-2)}2\bigg(
4S_1(\tau/2-2)-S_1(\ell+1)
\nonumber
\\
&\quad
+2S_1(\tau/2+\ell-1)-2S_1(\tau-4)-S_1(\tau+\ell-3)
\bigg)
\nonumber
\\
&\quad + \frac{c(-1)^{\tau/2}}4\bigg(
S_1\left(\tfrac{\ell+1}2\right)-S_1\left(\tfrac{\ell}2\right)
+2S_1\left(\tfrac{\tau-4}2\right)-2S_1\left(\tfrac{\tau-5}2\right)
\nonumber
\\
&\quad-S_1\left(\tfrac{\tau+\ell-3}2\right)+S_1\left(\tfrac{\tau+\ell-4}2\right)+\delta_{\tau,4}-4\log 2
\bigg)
\bigg].
\end{align}
For $c=\frac4n=0$, this expression reduces to the corresponding expression in the OPE coefficients of the generalized free field of dimension $\Delta=2$:
\begin{equation}
\left.\left(
a^{\mathrm{GFF},d=4-\eps}_{n,\ell}[\Delta=2]
-a^{\mathrm{GFF},d=4}_{n,\ell}[\Delta=2]
\right)\right|_{\eps}=\left.\left(a^{(0)}_{\tau_0,\ell}\,\Delta\alpha_{\tau_0,\ell} \right) \right|_{\tau_0=2n+4,c=0},
\end{equation}
where $a^{\mathrm{GFF},d}_{n,\ell}$ and $a^{(0)}_{\tau_0,\ell}$ are given in \eqref{eq:aGFFgen} and \eqref{eq:ssssa0tauell} respectively.

Although we do not do it in this paper, it is possible to derive similar formulas for $\expv{tttt}$ as well.

\section{Crossing matrices}
\label{app:crossingM}

Here we will recall some expressions for the normalization constants $\mathcal N_R$ and the crossing matrices. The discussion follows closely appendix~A.3 of \cite{Henriksson:2022rnm} and is based on unpublished work by the authors of \cite{He:2021sto}.

For $\langle\varphi\varphi\varphi\varphi\rangle$ we have
{
\begin{equation}
\renewcommand*{\arraystretch}{1.6}
M=\begin{pmatrix}
\dfrac1n &\dfrac{ (n+2)(n-1)}{2n^2} & \dfrac{1-n}{2n}\\
1&\dfrac{n-2}{2n}&\dfrac12\\
-1&\dfrac{n+2}{2n}&\dfrac12
\end{pmatrix}
\end{equation}
}
in the basis $\{S,T,A\}$ and the normalization constants are given by $\mathcal N_S=1$ and $\mathcal N_R=\frac{\sqrt{\dim R}}{\dim V}$ for $R=T,A$; see \eqref{eq:normVVVV}.

For $\langle\varphi\varphi tt\rangle$ we have
\begin{equation}\label{ffttcrossmatrix}
M=
\begin{pmatrix}\sqrt{\frac{2}{(n-1) (n+2)}} & \sqrt{\frac{2(n-2)}{3(n-1)}} & \sqrt{\frac{n+4}{3(n+2)}} \\
\frac{n\sqrt{(n-2) (n+4)}}{(n-1) (n+2)} & -\frac n{n-1}\sqrt{\frac{ n+4}{3(n+2)}} & \frac n{n+2} \sqrt{\frac{2(n-2) }{3(n-1) }} \\
-\frac n{n-1}\sqrt{\frac{n}{ (n+2)}} & - \frac{\sqrt{n(n-2)} }{(n-1)\sqrt3} & \sqrt{\frac{2n (n+4)}{3(n-1) (n+2)}}
\end{pmatrix}
\end{equation}
in the basis $\{S,T,A\}$ and $\{V,H_3,T_3\}$.
For $R=S,T,A$, $\mathcal N_R$ agree with \eqref{eq:normVVVV}, and $\mathcal N_V=\mathcal N_{H_3}=\mathcal N_{T_3}=1$.

For $\langle tt tt\rangle$ we have
\begin{align}
&M=
\nonumber\\
&\begin{pmatrix}
\frac{2}{(n-1) (n+2)} & \frac{1}{n} & -\frac{1}{n+2} & \frac{n (n+1) (n+6)}{6 (n-1) (n+2)^2} & -\frac{(n-2) (n+1) (n+4)}{2 (n-1) (n+2)^2} & \frac{(n-3) n (n+1)}{3 (n-1)^2 (n+2)} \\
\frac{2 n}{(n-1) (n+2)} & \frac{n^2+4 n-24}{2 (n-2) (n+4)} & -\frac{n}{2 (n+2)} & \frac{n^2 (n+1) (n+6)}{3 (n-1) (n+2)^2 (n+4)} & \frac{2 n (n+1)}{(n-1) (n+2)^2} & -\frac{(n-3) n^2 (n+1)}{3 (n-2) (n-1)^2 (n+2)} \\
-\frac{2 n}{(n-1) (n+2)} & -\frac{1}{2} & \frac{1}{2} & \frac{n (n+1) (n+6)}{3 (n-1) (n+2)^2} & 0 & -\frac{(n-3) n (n+1)}{3 (n-1)^2 (n+2)} \\
1 & \frac{(n-1) (n+2)}{n (n+4)} & \frac{n-1}{n} & \frac{(n-2) n}{6 (n+2) (n+4)} & \frac{n-2}{2 (n+2)} & \frac{n-3}{3 (n-1)} \\
-1 & \frac{2 (n-1) (n+2)}{(n-2) n (n+4)} & 0 & \frac{n (n+6)}{6 (n+2) (n+4)} & \frac{1}{2} & \frac{(n-3) n}{3 (n-2) (n-1)} \\
1 & -\frac{(n-1) (n+2)}{2 (n-2) n} & -\frac{n-1}{2 n} & \frac{n+6}{6 (n+2)} & \frac{n+4}{2 (n+2)} & \frac{n^2-2 n+3}{3 (n-2) (n-1)}
\end{pmatrix}
\label{eq:Mtttt}
\end{align}
in the basis $\{S,T,A,T_4,H_4,B_4\}$ and
for $R=S,T,A$, $\mathcal N_R$ agree with \eqref{eq:normVVVV}, and for $R=T_4,H_4,B_4$,
\begin{equation}
\mathcal N_R=\frac{\sqrt{\dim R}}{\dim T}
,
\end{equation}
where $\dim T_4=\frac{n(n+1)(n+6)(n-1)}{24}$, $\dim H_4=\frac{(n+1)(n+4)(n-1)(n-2)}8$ and $\dim B_4=\frac{n(n+1)(n+2)(n-3)}{12}$; see \eqref{eq:Norm2}--\eqref{eq:Norm3}.
The matrix \eqref{eq:Mtttt} appeared previously in \cite{Reehorst:2020phk} in another normalization convention. Our conventions for the normalization differ from those used in that paper by a factor $\dim T=\frac{(n-1)(n+2)}2$ for the last three representations and with a factor $n$ for the $T$ and $A$ representations.

\section{$\boldsymbol{\langle\varphi\varphi\varphi\varphi\rangle}$ at order $\boldsymbol{\varepsilon^2}$}
\label{app:fffforder2}

In this appendix, we determine the correlator $\langle\varphi\varphi\varphi\varphi\rangle$ to order $\eps^2$, by computing an explicit sum of conformal blocks. 
As described in the introduction, the order-$\eps^2$ conformal data can be found by the Lorentzian inversion formula following \cite{Alday:2017zzv}. This computation gives the leading-order anomalous dimensions of $s$ and $t$, as well as the order-$\eps^2$ dimensions and OPE coefficients of the twist-two operators. 
The dimensions take the form
\begin{align}
\label{eq:deltaS2l}
\Delta_{S,2,\ell} &=2+\ell-\eps+\frac{n+2}{2(n+8)^2}\left(1-\frac6{\ell(\ell+1)}\right)\eps^2,\quad \ell=2,4,\ldots,
\\
\label{eq:deltaT2l}
\Delta_{T,2,\ell} &=2+\ell-\eps+\frac1{(n+8)^2}\left(\frac{n+2}2-\frac{n+6}{\ell(\ell+1)}\right)\eps^2,\quad \ell=2,4,\ldots,
\\
\label{eq:deltaA2l}
\Delta_{A,2,\ell} &=2+\ell-\eps+\frac{n+2}{2(n+8)}\left(1-\frac2{\ell(\ell+1)}\right)\eps^2,\quad \ell=1,3,\ldots,
\end{align}
and the corresponding values for the OPE coefficients can be found in \cite{Alday:2017zzv}.\footnote{The dimensions were first determined in \cite{Wilson:1973jj}, and the OPE coefficients in \cite{Dey:2016mcs}. The data is available in computer-readable format in \cite{Henriksson:2022rnm}.} Likewise we need the leading OPE coefficients of twist-four operators, explicitly given in \cite{Henriksson:2018myn}. The inversion problem to this order does not give the finite-spin contributions at spin 0, for which we supplement the following data:
\begin{align}
\gamma_{s}^{(2)}&=\frac{(n+2)(13n+44)}{2(n+8)^3}, & \gamma_{t}^{(2)}&=\frac{(n+4)(22-n)}{2(n+8)^3},
\\
a^{(2)}_s&=-\frac{6(n+2)(3n+14)}{n(n+8)^3}, & a^{(2)}_t&=-\frac{12(3n+14)}{(n+8)^3}.
\end{align}
The order-$\eps$ correction to the OPE coefficients were given in \eqref{eq:ffffas1at1} in the main text.

By computing the sums over conformal blocks, we find the correlators
\begin{align}
\G^S_{\varphi\varphi\varphi\varphi}(u,v)&=1+\frac1n\left(u+\frac uv\right)+\eps\left(-\frac{(n+2)u}{n(n+8)}\Phi(u,v)-\frac{u(1+v)}{2nv}\log u+\frac{u}{2nv}\log v\right)
\nonumber\\&\quad
+
\eps^2\bigg[
\frac{u}{8nv}\left((1+v)\log^2u-2\log u\log v+\log^2v\right)
\nonumber\\&\quad
+\frac{u(n+2)}{4n(n+8)^2v}\left((1+v)\log u-\log v\right)
\nonumber\\&\quad
-\frac{(n+2)(n+11)}{2n(n+8)^2}\frac{\mathrm{Li}_2(z)-\mathrm{Li}_2(\zb)}{z-\zb}\log u
\nonumber\\&\quad
+\frac{u(n+2)}{4n(n+8)^2v}\big(4(n^2+7n+22)+(n+8)(n+17)\log u
\nonumber\\&\quad
-(n+8)(n+11)\log v\big)\Phi(u,v)
\nonumber\\&\quad
-\frac{3u(n+2)}{4n(n+8)^2}\frac{\log^2u\log(\frac{1-z}{1-\zb})}{z-\zb}
+\frac{n+2}{4n(n+8)}\Upsilon(u,v)
\bigg]+O(\eps^3),
\label{eq:ffffSapp}
\end{align}
which was given for $n=1$ already in \cite{Bissi:2019kkx},
\begin{align}
\G^T_{\varphi\varphi\varphi\varphi}(u,v)&=u+\frac uv+\eps\left(-\frac{2u}{n+8}\Phi(u,v)-\frac{u(1+v)}{2v}\log u+\frac{u}{2v}\log v\right)
\nonumber\\&\quad
+
\eps^2\bigg[
\frac{u}{8v}\left((1+v)\log u^2-2\log u\log v+\log v^2\right)
\nonumber\\&\quad
+\frac{u(n+2)}{4(n+8)^2v}\left((1+v)\log u-\log v\right)
\nonumber\\&\quad
+\frac{(3n+22)u}{2(n+8)^2}\frac{\mathrm{Li}_2(z)-\mathrm{Li}_2(\zb)}{z-\zb}\log u
\nonumber\\&\quad
+\frac{u}{4(n+8)^3}\big(8(n^2+7n+22)+(n+8)(5n+34)\log u
\nonumber\\&\quad
-(n+8)(3n+22)\log v\big)\Phi(u,v)
\nonumber\\&\quad
-\frac{(n+6)u}{4(n+8)^2}\frac{\log^2u\log(\frac{1-z}{1-\zb})}{z-\zb}
-\frac{u}{2(n+8)}\Upsilon(u,v)
\bigg]+O(\eps^3)
\label{eq:ffffTapp},
\end{align}
and
\begin{align}
\G^A_{\varphi\varphi\varphi\varphi}(u,v)&=u-\frac uv+\eps\left(\frac{u(1-v)}{2v}\log u-\frac{u}{2v}\log v\right)
\nonumber\\&\quad
+
\eps^2\bigg[\frac{u}{8v}\left((1-v)\log^2u-2\log u\log v+\log^2v\right)
\nonumber\\&\quad
+\frac{(n+2)u}{4(n+8)^2v}\left((1-v)\log u-\log v\right)-\frac{(n+2)u}{4(n+8)^2}\log v\Phi(u,v)
\bigg]+O(\eps^3).
\label{eq:ffffAapp}
\end{align}
We found them by explicitly summing the conformal data and matching with an ansatz based on the structure of the result in \cite{Bissi:2019kkx}.

In these expressions, $\Phi(u,v)$ is given by \eqref{eq:boxfunction}, and
\begin{align}
\Upsilon(u,v)&=\frac1{z-\zb}\bigg[
\left(\log^2\left(\tfrac{1-z}{1-\zb}\right)+\log^2v\right)\log\left(\tfrac z\zb\right)-
\left(\log\left(\tfrac{1-z}{1-\zb}\right)+\log v\right)\log^2\left(\tfrac z\zb\right)
\nonumber\\&\quad
+\log^2\left(\tfrac{1-z}{1-\zb}\right)\log +4\log v\mathrm{Li_2}\left(\tfrac{\zb-z}\zb\right)-4\log u\mathrm{Li_2}\left(\tfrac{\zb-z}{1-\zb}\right)
\nonumber\\&\quad
-\mathrm{Li_3}\left(\tfrac{\zb-z}{\zb}\right)-4\mathrm{Li_3}\left(\tfrac{\zb-z}{1-z}\right)
+4\mathrm{Li_3}\left(\tfrac{\zb-z}{\zb(1-z)}\right)+4\mathrm{Li_3}\left(\tfrac{z-\zb}{1-\zb}\right)
\nonumber\\&\quad
+4\mathrm{Li_3}\left(\tfrac{z-\zb}{z}\right)-4\mathrm{Li_3}\left(\tfrac{z-\zb}{z(1-\zb)}\right)
\bigg]
\end{align}
is another ancillary function, introduced to simplify the expressions.

\section{Inversion dictionary}
\label{app:inversions}

The functions that show up in the inversion integrals in this paper only take a few different forms, so it is possible to determine all of the inversion integrals we may need. The only contribution to the double-discontinuity will come from terms which are singular at $\bar z = 1$. These include $\log(1-\bar z )$ and inverse powers of $(1- \bar z)$. Using $(1 - \bar z)^\eps = 1 + \log(1 - \bar z) \eps + \O(\eps^2)$, the inversion for the log can be determined by the inversion of the inverse powers. The double-discontinuity for inverse powers can be computed using the formula
\begin{align}
\text{dDisc} \left[ \left( \frac{\bar z}{1 - \bar z} \right)^\alpha \right] \ = \ 2 \sin^2(\pi \alpha ) \left( \frac{\bar z}{1 - \bar z} \right)^\alpha.
\end{align}

Plugging this into the inversion integral gives us
\begin{equation}
\kappa_{\hb}\int_0^1\frac{d\zb}{\zb^2}k_\hb(\zb)\dDisc\left[ \left(\frac{\bar z}{1-\bar z} \right)^{\alpha}\right]=
\frac{2\Gamma(\hb)^2\Gamma(\hb-1+\alpha)}{\Gamma(2\hb-1)\Gamma(\alpha)^2\Gamma(\hb+1-\alpha)}.
\label{eq:appInversionInt}
\end{equation}
The end result is 

\begin{equation}
\left(\frac{\bar z}{1-\bar z} \right)^{ \alpha} \to
\frac{2\Gamma(\hb)^2\Gamma(\hb-1+\alpha)}{\Gamma(2\hb-1)\Gamma(\alpha)^2\Gamma(\hb+1-\alpha)}.
\label{eq:inversion_dictionary}
\end{equation}

There is a subtlety that arises for positive integers $\alpha$ -- for us the case $\alpha = 1$ is most relevant. In this case, we see that the double-discontinuity vanishes but the integral diverges. The correct procedure is to keep $\alpha$ free until the end of the calculation. Setting $\alpha \to 1$ in equation (\ref{eq:inversion_dictionary}) will yield the correct answer.

Some inversions are difficult to obtain because they involve an infinite series of poles at $\bar z = 1$. In practice, these can sometimes be computed by inverting a finite number of the terms that appear, guessing the general form, and then resumming the answer. Another useful method is to compute a finite number of terms, which determine the leading powers of $1 / \ell$ after the inversion. With a suitable ansatz, these large-spin asymptotics fix the inversion. We have computed several new inversions using these two tricks. Below is a list of inversions used for this paper,\footnote{The normalization used in this definition may be fixed by the equation \begin{equation*}\frac{2}{J^2} =  \frac{\Gamma(\ell+1)^2}{2 \pi^2 \Gamma(2 \ell + 2) } \int_0^1 \frac{d \bar z}{\bar z^2} k_{\ell+1}(\bar z) \text{dDisc}\left[ \log^2(1 - \bar z) \right].\end{equation*}
} using the normalization defined by~\eqref{eq:appInversionInt} and~\eqref{eq:inversion_dictionary}:
\begin{align}
\log(1-\bar z)^2 \quad & \to \quad \frac{2}{J^2} , \\
\frac{\log(1-\bar z)^2 }{\bar z} \quad & \to \quad \frac{2}{J^2} + \frac{2}{J^2(J^2 - 2)} , \\
\frac{\log(1-\bar z)^2 }{\bar z^2} \quad & \to \quad \frac{2}{J^2} + \frac{4}{J^2(J^2 - 2)} + \frac{8}{J^2(J^2 - 2)(J^2 - 6)} , \\
\log(1-\bar z)^2 \log{\bar z} \quad & \to \quad -\frac{2}{J^4} , \\
\frac{\log(1-\bar z)^2 }{\bar z} \log{\bar z} \quad & \to \quad-\frac{J^2 + 1}{J^4} + \frac{J^2 - 5}{(J^2 - 2)^2} , \\
\frac{\log(1-\bar z)^2 }{\bar z^2} \log{\bar z} \quad & \to \quad -\frac{8J^2 + 6}{9J^4} + \frac{J^2 - 8}{2(J^2 - 2)^2} + \frac{7 J^2 - 72}{18(J^2 - 6)^2} , \\
\log(1-\bar z)^2 \mathrm{Li}_2(1-\bar z) \quad & \to \quad \frac{2 \zeta_2 + 4 S_{-2}(\ell) }{J^2} , \\
\frac{\log(1-\bar z)^2 }{\bar z} \mathrm{Li}_2(1-\bar z) \quad & \to \quad \frac{4- 2 J^2( \zeta_2 + 2 S_{-2}(\ell) -1) }{J^4(J^2-2)} , \\
\frac{\log(1-\bar z)^2 }{\bar z^2} \mathrm{Li}_2(1-\bar z) \quad & \to \quad \frac{2(J^6 + J^4(4\zeta_2 + 8 S_{-2}(\ell)-3) -2J^2(4\zeta_2 + 8 S_{-2}(\ell) -1)+ 24)}{J^4(J^2-2)^2 (J^2 - 6)} , \\
\log(1-\bar z)^2 \frac{\log(\bar z)}{1 - \bar z} \quad & \to \quad 4 S_{3}(\ell) - \zeta_3 - \frac{2}{J^4} , \\
\frac{\log(1-\bar z)^2 }{\bar z} \frac{\log(\bar z)}{1 - \bar z} \quad & \to \quad 4 S_{3}(\ell) - \zeta_3 - \frac{3}{J^4} - \frac{1}{J^2} -\frac{3}{(J^2-2)^2}+\frac{1}{J^2-2} , \\
\nonumber\frac{\log(1-\bar z)^2 }{\bar z^2} \frac{\log(\bar z)}{1 - \bar z} \quad & \to \quad 4 S_{3}(\ell) - \zeta_3 - \frac{11}{3J^4} - \frac{17}{9J^2} -\frac{6}{(J^2-2)^2}+\frac{3}{2(J^2-2)} \\
&\quad\quad -\frac{5}{3(J^2-6)^2}+\frac{7}{18(J^2-6)} .
\end{align}
We have used $J^2 = \ell(\ell+1)$, and $S_{-m}$ is defined as in \cite{Lukowski:2009ce} to be
\begin{align}
S_{a}(\ell) = \sum_{j = 1}^{\ell} \frac{(\mathrm{sgn}(a))^j}{j^{|a|}} .
\end{align}
If $a<0$, this formula is not analytic in spin because the sign of each term alternates. Therefore its even-spin and odd-spin terms each have their own analytic continuation. In the above, we consider the even-spin continuation, which is all we need for our purposes. In particular, we use  the continuation for $S_{-2}$ given by
\begin{align}
  S_{-2}(\ell)  =  \frac{1}{4} \left(  S_2\left( \frac{\ell}{2} \right) -  S_2\left( \frac{\ell-1}{2} \right) - 2\zeta_2 \right),
\end{align}
where $S_2(\ell)$ denotes the (unique) analytic continuation of the second harmonic number.

{
\bibliographystyle{JHEP.bst}
\bibliography{biblio}
}

\end{document}

%% file: preamble.tex
\usepackage[english]{babel}
\usepackage[babel]{csquotes}
\usepackage{amsfonts}
\usepackage{array, tabularx}
\usepackage{mathtools}
\usepackage{amsthm}
\usepackage{url}
\usepackage{multicol, multirow}
\usepackage{emptypage}
\usepackage{verbatim}
\usepackage{dsfont}
\usepackage{hyperref}
\usepackage{float}
\usepackage{amsmath}
\usepackage{amssymb,tikz}
\usepackage{verbatim}
\usepackage{xspace}
\usepackage{youngtab}
\usepackage{tikz-cd}

\newcommand{\hb}{{\bar{h}}}
\newcommand{\zb}{{\bar{z}}}
\newcommand{\1}{{\mathds{1}}}
\renewcommand{\O}{{\mathcal O}}
\newcommand{\G}{{\mathcal G}}
\newcommand{\de}{{\partial}}

\newcommand{\eps}{{\varepsilon}}

\newcommand{\expv}[1]{\left\langle#1\right\rangle}

\usepackage{tensor}

\newcommand{\R}{\mathbb R}

\newcommand{\Z}{\mathbb Z}

\DeclareMathOperator*{\dDisc}{dDisc}

\allowdisplaybreaks[2]

\usepackage[font=small,format=hang,labelfont={sf,bf}]{caption}

\makeatother